\documentclass[12pt,a4paper,twoside]{book}
\usepackage{hyperref}
\usepackage[T1]{fontenc}
\usepackage[utf8]{inputenc}
\usepackage[font=small,labelfont=bf]{caption}
\usepackage{fancyhdr}
\usepackage{xspace}
\usepackage{amsmath}
\usepackage{ifthen} 
\newboolean{uprightparticles}
\setboolean{uprightparticles}{false} 
\usepackage{upgreek}
\usepackage{cancel}
\usepackage{slashed}
\usepackage{mathtools}
\usepackage{graphicx}
\usepackage{caption}
\usepackage{subcaption}
\usepackage{multirow}
\usepackage{tabularx}
\usepackage{upgreek}
\fancypagestyle{plain}{%
                       \fancyhead{}
                       
                      }
\numberwithin{equation}{chapter}
\begin{document}
	\pagenumbering{roman}          
	
	\pagestyle{empty}
	\begin{titlepage}
	\begin{center}
		{\LARGE{University of Ferrara}} \\[0.9ex]
	   	{\large{Physics and Earth Sciences Department}} \\[0.9ex]
	  	{\large{Degree of Doctor of Philosophy in subject of Physics}} \\
		\vspace{0.6cm}
		\includegraphics[height=5.8cm]{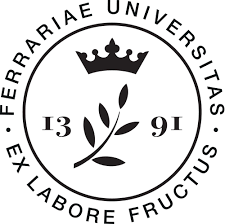}\\[1.2cm] 					
		{\Huge \textbf Research and development in cylindrical triple-GEM detector with $\mu$TPC readout for the BESIII experiment}\\[1.4cm]
	\end{center}
	\begin{flushleft}
		{\em Supervisor}:  \\
		{\em ~~~Dr}. {\sc Gianluigi Cibinetto }\\[1.4ex]
		{\em Examiners}:  \\
		{\em ~~~Dr}. {\sc Erika De Lucia } \\[0.4ex]
		{\em ~~~Prof}. {\sc Theodoros Alexopoulos } \\[1.4ex]
		{\em PhD Coordinator}:  \\
		{\em ~~~Prof}. {\sc Vincenzo Guidi } \\[3.2ex]
	\end{flushleft}
	\begin{flushright}
		{\em Candidate:~~~~~~~~~~~~~~~~~~~} \\
		{\sc ~~~Riccardo Farinelli} 
	\end{flushright}
	\vfill
	\begin{center}
		{Academic years 2015-2018}\\
		{XXXI course - FIS/04}
	\end{center}
\end{titlepage}

	\pagestyle{fancy}

	\chapter*{Abstract}

The third generation of the Beijing Electron Spectrometer, BESIII, is an apparatus for high energy physics research. The hunting of new particles and the measurement of their properties or the research of rare processes are sought to understand if the measurements confirm the Standard Model and to look for physics beyond it. The detectors ensure the reconstruction of events
belonging to the sub-atomic domain. The operation and the efficiency of the BESIII inner tracker is compromised due to the the radiation level of the apparatus. A new  detector is needed to guarantee better performance and to improve the physics research. A cylindrical triple-GEM detector (CGEM) is an answer to this need: it will maintain the excellent performance of the inner tracker while improving the spatial resolution in the beam direction allowing a better reconstruction of secondary vertices. The technological challenge of the CGEM is related in its spatial limitation and the needed cylindrical shape. At the same time the detector has to ensure an efficiency close to 1 and a
stable spatial resolution better than $150 \, \mathrm{\upmu m}$, independently from the track incident angle and the presence of $1 \, \mathrm{T}$ magnetic field.

In the years 2014-2018 the CGEM-IT has been designed and built. Through several test beam and
simulations the optimal configuration from the geometrical and electrical points of view has been
found. A new electronics has been developed to readout charge amplitude and time
information from the detector signal. This allows to measure the position of the charged particle
interacting with the CGEM-IT. Two algorithms have been used for this purpose, the charge centroid
and the $\upmu$TPC, a new technique introduced by ATLAS in MicroMegas and developed here for the
first time for triple-GEM detector. 

A complete triple-GEM simulation software has been developed to improve the knowledge of the detection processes. The software reproduces the CGEM-IT behavior in the BESIII offline software. The simulation of the CGEM-IT in the BESIII apparatus validates the improvements of the detector thought the study of some physics analysis.

	\markboth{Contents}{Contents}
	\baselineskip 5.5mm   
	\tableofcontents
	\baselineskip 5.8mm    
	\markboth{Contents}{Contents}
	\newpage
	
	\listoffigures
	\listoftables
	
	\cleardoublepage
	\pagenumbering{arabic}
	\chapter{The Beijing Electron Positron Collider II and the Beijing Spectrometer III}

High energy collisions between particles create new states of matter as ruled by the nature and described by the physics laws. As the number of collisions increases then it is possible to perform more precise measurements of the particle properties or the transitions from a state to another. Around the world there are several experiments and scientists that approach the knowledge of the physics laws looking at the collision results of high energy particles, one of those experiments is the BEijing Spectrometer (BESIII) at the Institute of High Energy Physics (IHEP) in People's Republic of China (PRC): an apparatus composed of several sub-detectors that measure the properties of the particles such as momentum, energy, mass and path. 
The leptons are provided by the Beijing Electron Positron Collider (BEPCII). They are injected by a linear accelerator in two accumulation rings where electrons and positrons are disposed in bunches. The collision happens in the interaction point (IP). The center of mass energy range varies from $2 \, \mathrm{GeV}$ to $4.6 \, \mathrm{GeV}$ in order to explore the $\tau$-charm region to continue the research about light hadrons, charmonium, exotics and other topics explained in Sect. \ref{cap:BESphysics}

In this Chap. a review about the experimental layout will be reported and the physics research of the BESIII experiment in order to understand the scenario and the background needed by the new particle detector that has been studied in this work.

\section{BEPCII}
BEPCII is the upgrade of the pre-existing BEPC machine and it is a double ring electron positron collider that acts also as a synchrotron radiation (SR) source \cite{ref1:BEPCII_1,ref1:BEPCII_2}. The beam energy ranges from 1 to $2.3 \, \mathrm{GeV}$ with a design luminosity of $10^{33} \, \mathrm{cm^{-2}s^{-1}}$ at $1.89 \, \mathrm{GeV}$ that has been reached in 2016 data taking on April 14$^{th}$. The facility is exploited in high energy physics experiment and synchrotron radiation. The acceleration of $e^+e^-$ is performed by an injector consisting in a $202 \, \mathrm{m}$ electron linac with 16 Radio-Frequency (RF) power sources and 56 S-band RF structures. Then two super-conducting cavities are used with one cavity in each ring to provide the RF voltage. BEPCII requires 363 magnets of different types: bending magnets, quadrupole and sextupole. The details of BEPCII are shown in the Tab. \ref{tab:BEPCII}.
The luminosity of an electron-positron collider is expressed as:
\begin{equation}
L(cm^{-2} s^{-1}) = 2.17 \times 10^{34} (1 + r) \epsilon_y \dfrac{E(GeV)k_bI_b(A)}{\beta^*_y(cm)}
\end{equation}


	\begin{figure}[htbp]
		\centering
   		\includegraphics[width=0.8\textwidth]{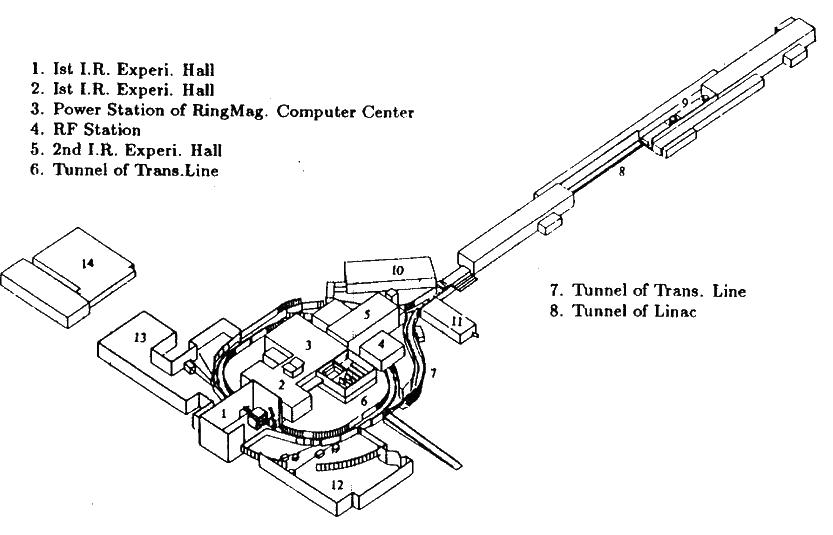}
   		\caption[BEPCII.]{Sketch of the accelerator facility: linac and the two accumulation rings \cite{ref1:BesIIIphysicsbook}.}
	\label{fig:BEPCII}
	\end{figure}

where {\it r} is the beam aspect ratio at the IP, $\epsilon_y$ the vertical beam-beam parameter, $\beta^*_y$ the vertical envelope function at IP, $k_b$ the bunch number in each beam and $I_b$ the bunch current. A double accumulation ring scheme allows to reach a higher bunch number than an single one. The inner and outer rings (represented in Fig. \ref{fig:BEPCII} ) cross each other in the northern and the southern IP. In order to have sufficient separation between the rings without a significant degradation of the luminosity, the horizontal crossing angle between the two beams is 11 mrad $\times$ 2 in the southern IP, where the BESIII experiment is placed. In the northern IP a vertical bump is used to separate the beams. The layout of BEPCII minimizes the machine errors such as misalignment or problems in the multipole field of magnets. This tunes the beta function and reduces the background due to Touschek effect. The mean beam lifetime is around three hours and its limits are due to beam-beam bremsstrahlung, Touschek effect and beam-gas interaction.
\begin{table}[ht!]
\begin{center}
\begin{tabular}{c c c}
Parameter & Unit & Value \\
\hline
Energy & $\mathrm{GeV}$ & 1.89\\
Circumference & $\mathrm{m}$ & 237.53\\
RF frequency & $\mathrm{MHz}$ & 499.8\\
Harmonic & - & 396 \\
RF voltage & $\mathrm{MV}$ & 1.5\\
Transverse tunes & - & 6.53/7.58\\
Damping time & $\mathrm{ms}$ & 25/25/12.5\\
Beam current & $\mathrm{A}$ & 0.91\\
Bunch number & - & 93\\
SR loss per turn & $\mathrm{keV}$ & 121\\
SR power & $\mathrm{kW}$ & 110\\
Energy spread & - & 5.16 $\times$ 10$^{-4}$\\
Compact factor & - & 0.0235\\
Bunch length & $\mathrm{cm}$ & 1.5\\
Emittance & $\mathrm{nm \cdot rad}$ & 144/2.2\\
$\beta$ function at IP & $\mathrm{m}$ & 1/0.015\\
Crossing angle & $\mathrm{mrad}$ & 11 $\times$ 2\\
Bunch spacing & $\mathrm{m}$ & 2.4\\
Beam-beam Parameter & - & 0.04/0.04\\
Luminosity & $\mathrm{cm^{-2}s^{-1}}$ & 1.0$\cdot$10$^{33}$\\
\newline\\
\hline\\
\end{tabular}
\caption[Main parameters of the BEPCII storage ring]{Main parameters of the BEPCII storage ring \cite{ref1:BEPCII_2}.}
\label{tab:BEPCII}
\end{center}
\end{table}

\section{The BESIII detector}
\label{sec:bes}
\begin{figure}[htbp]
\centering
\includegraphics[width=0.9\textwidth]{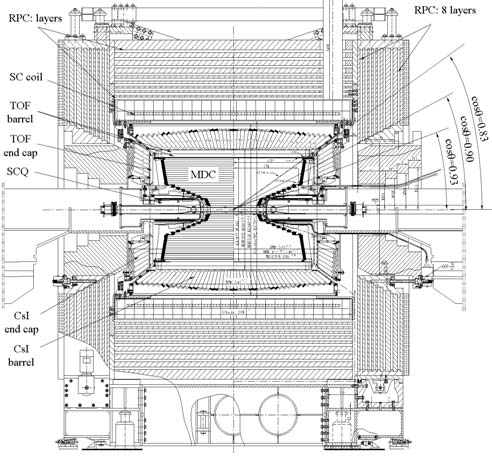}
\caption[BESIII detector]{Schematic drawing of the BESIII detector. The reference frame is define as: z axis along the beam direction, r is the radial coordinate from the beam pipe, $\theta$ is the polar and $\phi$ the azimutal angle, x and y the coordinate in the plane transverse to the beam pipe.}\label{fig:BESIII}
\end{figure}

BESIII measures trajectories and momenta of charged and neutral particles except $\nu$, produced in  $e^+e^-$ collisions in the energy region between 2 and 4.6 GeV by means of a set of detectors and a solenoid producing a 1 T magnetic field \cite{ref1:BESIII}. A schematic drawing of BESIII is shown in Fig. \ref{fig:BESIII}. Around the beryllium beam pipe the Multi-layer Drift Chamber (MDC) is placed and around its conical shape in the end-cap regions there are two superconducting quadrupoles (SCQs). The time-of-flight (TOF) system is located outside the MDC: it is composed by two layers of plastic scintillator counters in the barrel and a multigap resistive plate chamber in the end-caps. The CsI(Tl) ElectroMagnetic Calorimeter (EMC) is placed between the TOF system and the Superconducting Solenoid Magnet (SSM). The MUon Counter (MUC) is built up by several layers of resistive plate chambers (RPCs) placed in the gaps between steel plates of the flux return yoke. The coil of the superconducting magnet is placed outside the electromagnetic calorimeter. It has a mean radius of $1.482 \, \mathrm{m}$ and length of 3.52 m. The covered solid angle is $\Delta\Sigma/4\pi$~=~0.93, where the azimutal angle is totally covered while the polar angle is limited between 21$^\circ$ and 159$^\circ$. The main parameters of BESIII sub-detectors are summarized in Tab. \ref{tab:BESIII}.

\begin{table}[ht!]
\begin{center}
\begin{tabular}{c c c}
Sub-detector & Parameter & Value \\
\hline
MDC & Single wire $\sigma_{r\phi}$ & $130 \, \mathrm{\upmu m}$\\
MDC & $\sigma_{z}$ & $2 \, \mathrm{mm}$\\
MDC & $\sigma_p$/p ($1 \, \mathrm{GeV/c}$) & 0.5\% \\
MDC & $\sigma$ ($\mathrm{d}E/\mathrm{d}x$) & 6\% \\
TOF & $\sigma_t$  barrel & $100 \, \mathrm{ps}$ \\
TOF & $\sigma_t$  end-cap & $110 \, \mathrm{ps}$ \\
EMC & $\sigma_E/E$ ($1 \, \mathrm{GeV}$) & 2.5\% \\
EMC & Position resolution ($1 \, \mathrm{GeV}$) & $0.6 \, \mathrm{cm}$\\
Muon & N$^\circ$ layers barrel & 9 \\
Muon & N$^\circ$ layers end-cap & 8 \\
Muon & Cut-off momentum & $0.5 \, \mathrm{MeV/c}$ \\
Solenoid magnet & Magnetic field & $1 \, \mathrm{T}$\\
\newline\\
\hline\\
\end{tabular}
\caption[Main parameters of the BESIII]{Main parameters of the BESIII spectrometer sub-systems.}
\label{tab:BESIII}
\end{center}
\end{table}	

\subsection{The multi-layer drift chamber}
The design of the MDC has been optimized to detect low momentum particles with  performance shown in Tab. \ref{tab:BESIII}. 
Moreover the MDC provides the level 1 (L1) trigger as described in Sect. \ref{sect:trigger} to select good events and to reject the background. 
The radius of the MDC varies from 59 mm to 810 mm and the entire detector is composed by 43 sense wire layers. The first 8 define the {\it inner} MDC (IDC) and the others the {\it outer}MDC (ODC). The gas mixture is He-C$_3$H$_8$ (60:40) to minimize the multiple scattering effect. 
The measured position in the r-$\phi$\footnote{$r$ and $\phi$ are defined in Fig. \ref{fig:BESIII}.} plane is given by each single wire while the measurement along the beam direction is performed by layers with different stereo angles (-3.4$^{\circ}$ and + 3.9$^{\circ}$). The average gas gain of the
MDC is about $3 \times 10^4$ at the reference operating
high voltage of about $2200 \, V$. \\
Each layer is divided in several cells with an almost square section. The cell size varies from $12$ to $16.2 \, mm$. The cells have a trapezoidal shape with eight field wire on the perimeter and the sense wire in the middle. Each cell of a layer has the same shape and length. The resolution of a single wire is dominated by the electron diffusion and a proper balance is needed between the cell size, the readout channels and the multiple scattering given by the wires \cite{ref1:MDC}. The $25 \, \mathrm{\upmu m}$ wires are made of gold-plated tungsten with 3\% rhenium. The 43 layers of the MDC are divided in 11 superlayers and the field wires within a superlayer are shared between neighboring cells. Track segments reconstructed in superlayers are linked and used in the L1 trigger.\\
The momentum resolution of the MDC is determined by the resolution of a single wire and the multiple scattering.

\begin{equation}		
\label{eq:MDC_momentum_1}
	 	\dfrac{\sigma_{p_t}}{p_t}=\sqrt{\left( \dfrac{\sigma_{p_t}^{wire}}{p_t}\right)^2+\left(\dfrac{\sigma_{p_t}^{ms}}{p_t}\right)^2}
\end{equation}


\begin{figure}[ht!]
\centering
\includegraphics[width=0.6\textwidth]{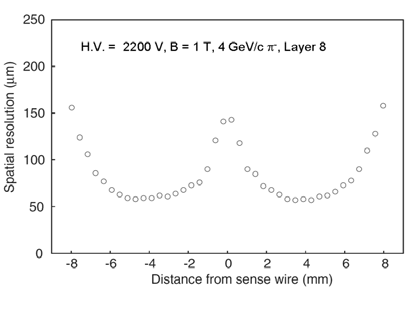}
\caption[MDC performance]{Position resolution of the MDC as a function of the distance from the sense wire \cite{ref1:BESIII}.}
\label{fig:MDC}
\end{figure}


The single wire resolution is a function of the drift distance and includes contribution from primary ionization statistic, longitudinal electron diffusion and the jitter in the time measurement. The distribution has been fitted to two Gaussian functions and the $\sigma$ has been calculated by:

\begin{equation}		
\label{eq:MDC_res}
\sigma=\sqrt{\dfrac{N_1 \sigma_1^2+N_2 \sigma_2^2}{N_1+N_2}}
\end{equation}

where $N_1$ and $N_2$ are the amplitudes and $\sigma_1$ and $\sigma_2$ are the standard deviations of the two Gaussian functions. The average spatial resolution of drift cell with a pion beams at 12 GeV/c is shows in Fig. \ref{fig:MDC}.

The $\mathrm{d}E/\mathrm{d}x$ performance is determined mainly by the fluctuation in the number of primary ionizations along the track. A minimum ionizing particle (MIP) track in helium creates about 8 ion pair per centimeter.  Monte-Carlo simulation shows that the $\mathrm{d}E/\mathrm{d}x$ is about 6\% and the corresponding momentum resolution is better than $\sigma_{p_t}/p_t \, = \, 0.32 \% \, p_t \oplus 0.37 \% \, / \beta$, where the first term is related to the trajectory measurement and the second one to the multiple scattering. It allows to separate $\pi$ and $K$ within 3~$\sigma$ up to momenta of $770 \, \mathrm{MeV/c}$.

The signals from the sense wired are first amplified by fast trans-impedance preamplifiers located close to the wire then output signal is sent outside BESIII thought 18~m cables. Then it is further amplified and split in three branches for timing, charge measurements and the L1 trigger. The wire spatial resolution of $130 \, \mathrm{\upmu m}$ corresponds to a time resolution of $3.5 \, \mathrm{ns}$ and the charge produced by a MIP on a sense wire after the amplification is about $450 \, \mathrm{fC}$.

\subsection{The time of flight}
The time-of-flight system is segmented in a barrel and two end-caps. The barrel is built up by two layers of 88 scintillating bars with a thickness of about $5 \, \mathrm{cm}$ and a trapezoidal cross section. The signal is collected by two photomultiplier tubes (PMT) attached to the bars. The end-caps have been recently upgraded with Multi-gap Resistive Plate Chambers (MRPC) and each end-cap station has 36 trapezoidal shaped modules arranged in a circular double layers \cite{ref1:etof}. 

The PMTs chosen to readout the TOF counters in the barrel are Hamamatsu R5924-70 and they match the size of the scintillator bars. The average quantum efficiency is about 23\%, the rise time is $2.5 \, \mathrm{ns}$ and the signal transition time is $9.5 \, \mathrm{ns}$ with $0.44 \, \mathrm{ns}$ rms spread. A fast preamplifier with a gain factor of 10 is used to boost the signals and to extend the lifetime of the PMT. The signal is received by the front-end electronics that splits the signals into two branches for time and charge measurements. 
The required dynamic range is $60 \, \mathrm{ns}$ and the time resolution of the electronics is less than 25 ps. The charge amplitude signal measured ranges from 200~mV to 4~V.

A different technology is used in the end-cap where the time measurement is performed by MRPC. Each module is divided into 12 readout strips and their length changes from $9.1 \, \mathrm{cm}$ to $14.1 \, \mathrm{cm}$ and the width is $2.4 \, \mathrm{cm}$. The strips are readout from both ends. The internal structure is composed by 14 pieces of thin glass sheet, arranged into two stacks. The stacks are sandwiched in between two layers of readout strips. The time measurement is given by 12 active gas gaps of $0.22 \, \mathrm{mm}$ thick. The MRPC module is placed in a gas-tight aluminum box with a gas mixture of Freon-SF$_6$-iC$_4$H$_{10}$ 90:5:5.

The time resolution is about 100~ps in the barrel and 65 ps in the end-caps and it allows to separate $\pi/K$ within 3~$\sigma$ up to $900 \, \mathrm{MeV/c}$ at 90$^\circ$. The precise time resolution of the TOF is also used in the trigger logic for charged particles. The solid angle coverage of the barrel is $|\cos(\theta)|< 0.83$ while in the end-caps is $0.85<|\cos(\theta)| \, < \, 0.95$. The dead gaps are needed for the mechanical support of the MDC and service line. The inner radius of the first barrel TOF layer is $81 \, \mathrm{cm}$ and the second $86 \, \mathrm{cm}$, while in the end-caps the flight path is about $140 \, \mathrm{cm}$.\\
The time resolution of the TOF is given by several contribution as shown in Tab.~\ref{tab:TOF}, while the intrinsic time resolution is determine by the rise time of the scintillation light, the fluctuation of the photon arrival time at the PMT and the transition time spread of the PMT.


\begin{table}[ht!]
\begin{center}
\begin{tabular}{c c}
Item & Contribution (ps)\\
\hline
Counter intrinsic time resolution & 80-90 \\
Uncertainty from $15 \, \mathrm{mm}$ bunch length & 25\\
Uncertainty from clock system & 20\\
Uncertainty from $\theta$ angle & 25\\
Uncertainty from electronics & 25\\
Uncertainty in expected flight time & 30\\
Uncertainty from time walk & 10\\
Total time resolution, one layer barrel & 100\\
Total time resolution, two layers barrel & 80\\
\newline\\
\hline\\
\end{tabular}
\caption[TOF time resolution]{Analysis of TOF time resolution for 1 GeV/c muon}
\label{tab:TOF}
\end{center}
\end{table}

\begin{figure}[ht!]
\centering
\includegraphics[width=0.8\textwidth]{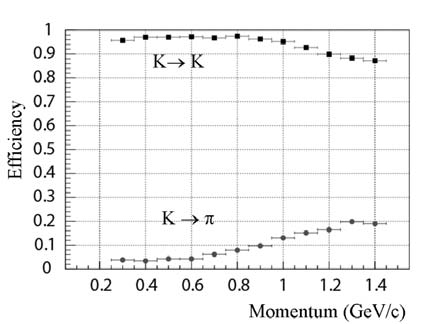}
\caption[TOF performance]{Expected $\pi/K$ separation efficiency and misidentifying rate in TOF detector.}
\label{fig:TOF}
\end{figure}

The $\pi/K$ separation capability depends on the polar angles of the tracks and its limit is reached at momenta of about 0.7 GeV/c at 90 $^\circ$ where the particle flight is the shortest in the barrel, the limit moves to about 1.0 GeV/c in the end-cap systems. The particle identification is calculated by a likelihood analysis. The probability to identify a kaon or to misidentify it as a function of the kaon momentum is shown in Fig. \ref{fig:TOF}. A separation efficiency of 95\% is reached up to 0.9 GeV/c.

\subsection{The electromagnetic calorimeter}
The electromagnetic calorimeter measures the energy of photons above $20 \, \mathrm{MeV}$ and provides a trigger signal. It is mandatory to measure precisely invariant mass of particles with radiative decay, $e.g$ $\pi^0$, $\eta$, $\rho$, \textit{etc}. In radiative decay the photon has to be distinguished from $\pi^0$ and $\eta$ decays. The minimum detectable opening angle between two photons is about 10$^\circ$ and for momenta higher than $200 \, \mathrm{MeV/c}$ it can discriminate $e/\pi$.  

The EMC consists of 6240 CsI(Tl) crystal with a length of $28 \, \mathrm{cm}$, about 15 radiation length (X$_0$), and a surface of $5.2 \, \mathrm{cm}$ $\times$ $5.2 \, \mathrm{cm}$. The crystals are arranged as 56 rings and each one covers an angle of about 3$^\circ$ in both polar and azimuthal directions. The calorimeter is divided in three parts: the barrel and two end-caps. The total weight of the EMC is about 24 tons. The design energy resolution is $\sigma_E/E$ = 2.5\%$\sqrt{E}$ and the position resolution is $\sigma$ = $0.6 \, \mathrm{cm}/\sqrt{E}$ at $1 \, \mathrm{GeV}$. The angular coverage is $|\cos(\theta)|<0.83$ while in the end-caps is $0.85< \cos(\theta)<0.95$. Details of the EMC are shown in Table \ref{tab:EMC}.

\begin{table}[ht!]
\begin{center}
\begin{tabular}{c c}
Parameter & Value\\
\hline
Crystal length & $28 \, \mathrm{cm}$\\
Front size & $5.2 \, \times \, 5.2 \mathrm{cm^2}$\\
Read size & $6.4 \, \times \, 6.4 \mathrm{cm^2}$\\
Number of $\phi$ sectors & 120\\
Barrel number of $\theta$ rings & 44\\
Barrel number of crystals & 5280\\
Barrel inner radius & $94 \, \mathrm{cm}$\\
Barrel $\theta$ coverage & $\cos(\theta)<0.83$\\
Barrel total weight & $21.56 \, \mathrm{tons}$\\
end-caps number of $\theta$ rings & 6\\
end-caps number of crystals & 960\\
end-caps distance to IP & $138 \, \mathrm{cm}$\\
end-caps $\theta$ coverage & $0.84 < \cos(\theta) <  0.93$\\
end-caps total weight & $21.56 \, \mathrm{tons}$\\
\newline\\
\hline\\
\end{tabular}
\caption[Geometrical parameters of EMC]{Geometrical parameters of EMC.}
\label{tab:EMC}
\end{center}
\end{table}	

The energy resolution of EMC depends by the crystal quality, dead material between the IP and the crystal, photodiode and amplifier noise, fluctuation of shower energy, \textit{etc}.

The signal readout is performed by two photodiodes of 1 cm $\times$ 2 cm and a light guide of $2 \, \mathrm{mm}$ thick glued directly at the center of the crystal surface. Only 10\% of the surface is covered but a larger fraction of light is collected thanks to the light reflection performed by reflective material covering the crystal surface. The signal is readout by the electronics that measures the energy deposited and provides a fast energy trigger.

\subsection{The muon counter}

Resistive plate counters are active detector used within the gaps of the flux return steel to identify the muons surviving the EMC with respect other charged particles such as pions. It is divided in three pieces: a barrel and two end-caps. The system is composed by nine layers of steel plates with a total thickness of $41 \, \mathrm{cm}$ and nine layers of RPCs in the barrel. The end-caps have eight layers of counters. The minimum muon momentum at which the RPC starts to become effective is about 0.4 GeV/c due to the energy loss in the EMC and the bending in the magnetic field. The hits reconstructed in the RPC are associated to the tracks reconstructed in the MDC and the energy measured in the EMC. The readout of the RPC is performed by strip of about $4 \, \mathrm{cm}$ in both direction $\theta$ and $\phi$. The space resolution is modest because of the multiple scattering of low momentum muons in the EMC. 

\begin{figure}[ht!]
\centering
\includegraphics[width=0.8\textwidth]{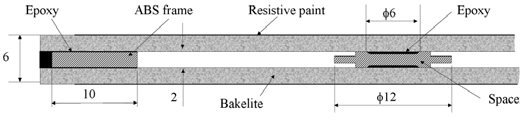}
\caption[RPC internal structure]{The cross sectional view of the RPC gas gap.}
\label{fig:RPC}
\end{figure}

The RPC are constructed by two plates of high resistivity material, the Bakelite, of $2 \, \mathrm{mm}$ thickness around a gas gap of $2 \, \mathrm{mm}$. Figure \ref{fig:RPC} shows the cross-sectional view of the detector. The RPCs work in streamer mode and the signal is collected on the strips outside the gas gap. The gas mixture is Ar-C$_2$F$_4$H$_2$-C$_4$H$_{10}$ 50:42:8 and the working voltage is of 8 kV. This provide a single gap efficiency of about 96\%. In order to improve the detection efficiency a double-gap layout has been used with the readout sandwiched in between the two. Each layer can measure only one coordinate and the orientation of the strips changes layer by layer. Due to the size of the detector, the number of readout channel is about 10000.

\subsection{The trigger system}
\label{sect:trigger}
The trigger, data acquisition and online computing system are designed to sustain an acquisition with multi-beam bunches separated by 8 ns. The trigger system has two levels: a hardware level (L1) and a software level (L3). The L1 trigger signal is generated by the TOF, MDC and EMC. Conditions on the number of minimum tracks are required on the MDC, TOF and EMC. The clock of the trigger system is $41.65 \, \mathrm{MHz}$, it is synchronized with the accelerator RF and it is distributed to the readout electronics crate to synchronize every operation in the BESIII data acquisition. The maximum L1 rate expected is about $4 \, \mathrm{kHz}$ at $3.097 \, \mathrm{GeV}$. The data in the front-end buffers is read once L1 trigger is validated, $6.4 \, \mathrm{\upmu s}$ after the collision. Then the event reconstruction takes place and the background events are suppressed by the L3 trigger.
\section{The BESIII physics}
\label{cap:BESphysics}
The BESIII physics plan is to accumulate a significant data sample in the center of mass energy between 2 and 4.6 GeV: energy point at the mass of charmonium states, above $D\overline{D}$ threshold, energy points at the mass of new resonances and in the continuum region. Up to now it has collected $10^{10}$~$J/\psi$, $10^9$~$\psi(2S)$, $10^6$~$D\overline{D}$.
The aim is to be sensitive to the observation of new physics concerning hot topics like $D^0\overline{D^0}$ mixing, the charmonium-like $XYZ$ states and CP violation in charmed-quark and $\tau$ sectors.
The high precision measurements at BESIII are compared with QCD calculations. The huge data sample allows to search for rare decays like lepton-number violating or flavour violating processes.

\subsection{Hadron Spectroscopy}
BESIII investigates the structure of matter and the nature of the interactions between its constituent components \cite{ref1:BesIIIphysicsbook}. 
To access a small object it is needed to increase the energy scale but increasing the energy the theoretical description of the phenomena changes. 
At momentum transfers around the MeV scale, below $\Lambda_{QCD}$~$\sim$~$200 \, \mathrm{MeV}$, the fundamental scale of Quantum Chromodynamics (QCD), the chiral symmetry theory can provide solutions to study the low-energy dynamics as QCD becomes non-perturbative \cite{ref1:Brambilla}. 
Increasing the energy in a range that is greater than $\Lambda_{QCD}$, the partonic and gluonic behavior is shown in the deep inelastic scattering. In this energy region it is allowed to study QCD processes through perturbative expansion \cite{ref1:Brambilla}. 
Between these two energy regions there is no theory that can describe the phenomena so meson and baryon resonances are studied to know the constituent behavior. The QCD calculations in this region are described with a naive quark model \cite{ref1:Isgur,ref1:KTChao}:
\begin{itemize}
\item spontaneous breaking of chiral symmetry leads to the presence of massive constituent
quarks within a hadron as effective degrees of freedom;
\item hadrons can be viewed as a quark system in which the gluon fields generate effective
potentials that depend on the relative positions and spins of the massive quarks.
\end{itemize}

These pieces compose the QCD puzzle and the hadron spectroscopy plays an important role in its interpretation.
In the light quark sector ($u, d, s$) the mass differences are relatively small ($m_u$~$\sim$~$3 \, \mathrm{MeV}$, $m_d$~$\sim$~$5 \, \mathrm{MeV}$ and $m_s$~$\sim$~$95 \, \mathrm{MeV}$~\cite{ref1:ChinesePhysicsC}) then the strong interaction can be approximated with a global SU(3) flavour symmetry. This approach allows to predict the masses of mesons and baryons composed by the quarks of the symmetry. Different theories and their validity ranges are summarized in Tab. $\ref{table:Teories}$.

\begin{table}[ht!]
\begin{minipage}{\textwidth}
\begin{center}
\begin{tabular}{l|lc}
\hline
Theory Name     & Energy Scale        &       \\
\hline
The chiral-symmetry theory 	& 	E~$\ll~\Lambda_{QCD}$ 	& 	\\
Perturbative QCD		&	E~$\gg~\Lambda_{QCD}$	&	\\
Global SU(3) flavour symmetry	&	E~$\ll~\Lambda_{QCD}$	&	\\
Non-relativistic QCD		&	E~$>$~1 GeV		&	\\
\newline\\
\hline\\
\end{tabular}
\caption[QCD theories and energy scale]{Different theories used and their energy scale validity.}
\label{table:Teories}
\end{center}
\end{minipage}
\end{table}

A non-relativistic approach can be used with the naive method for quarks that have constituent masses comparable or greater than the $\Lambda_{QCD}$. For quarks with an excitation energy of the excited hadron states comparable to their masses this approach is meaningless but it works in a wide range of empirical tests \cite{ref1:Brambilla}. 
Several theories can be found in literature: the semi-relativistic flux-tube model \cite{ref1:Carlson,ref1:Sartor}, the instanton model \cite{ref1:Blask}, the Goldstone boson exchange model \cite{ref1:Glozman}, the diquark model \cite{ref1:Anselmino}, \textit{etc}.

The ultimate goal of the study of the hadron spectroscopy is to understand the dynamics of the constituent interactions. Conventional $q\overline{q}$ meson can be constructed in the quark model and their spectrum is studied to be described in an empirically efficient way. QCD also does not forbid the existence of other bound states that are made completely of gluons ($e.g.$ so-called “glueball”), multiquark mesons, such as $qq\overline{qq}$ and so-called “hybrid” mesons which contain both $q\overline{q}$ and gluon.
Glueballs are bound states of at least 2 or 3 gluons in a color singlet. Gluons inside glueballs could be massive.
The abundance of isoscalar scalars in the 1-2 GeV, mass region, $e.g.$ $f_0$(1370), $f_0$(1500), $f_0$(1710), $f_0$(1790) and $f_0$(1810) should be confirmed in further experiments. They are the natural scalar glueball candidates and they are studied at BESIII as decay products. 
Hybrid mesons are hypothesized to be formed by a $q\overline{q}$ pair plus one explicit gluon field. Evidence for the existence of hybrid mesons would be direct proof of the existence of the gluonic degree of freedom and the validity of QCD. In addition, the lightest hybrid multiplet includes at least one J$^{PC}$ exotic state.
Multiquark are states with a number of quark-antiquark greater than three. They are always colourless. An example of a recent discovery of $c\overline{c}q\overline{q}$ states is the $Y$(4260) \cite{ref1:4260}. Hybrid states are a wide topic and they are connected to Charmonium and Charm physics, further description will be given in next sections.

\subsection{Charmonium physics}
Charmonia are charm-anticharm ($c\overline{c}$) bound states, like the $J/\psi$ resonance or its excitation. Due to their heavy mass ($c$ quark mass is $\sim$ 1.29 GeV) they probe the QCD at different energy scales: from the hard region, where the coupling constant is small and can be expanded, to the soft region, where non-perturbative effects dominate. 

To predict the properties of the charmonia states in these regions, a Non-Relativistic QCD (NRQCD) expansion can be used because $m_c$ is large and the motion velocity of the c quark is smaller than the speed of light ($v^2$ $\sim$ 0.3 by potential model and Lattice simulation \cite{ref1:Brambilla}). This method allows to expand the theory in power of $v$ and describe the mesons annihilation decays. It is named as perturbative NRQCD (pNRQCD) if $mv~\gg~\Lambda_{QCD}$. The system is weakly coupled and the potential of the meson is calculated. 
The next challenge is to interpret the newly discovered charmonia states. 
The spectrum of the charmonium states has been measured and the narrow resonances below the open-charm threshold are experimentally clear. On the other hand above the $c\overline{c}$ threshold many charmonia states have been theorized but only few of them have been observed. 
The spectrum of the charmonium states can be described by a $c\overline{c}$ potential model that combines a Coulomb-like term and a linear one for the confinement: the Cornell potential. 

	\begin{equation}
	\label{eq:strgpot1}
		V_0^{c\overline{c}} (r) = - \dfrac{4}{3} \dfrac{\alpha_s}{r} + br.
	\end{equation}

A spin dependent term is added to better describe the scalar nature of the potential:
	\begin{equation}
	\label{eq:strgpot2}
		V_{spin-dep} = \dfrac{32\pi\alpha_s}{9m_c^2} \mathbf{S_c} \cdot \mathbf{S_{\overline{c}}}  \delta(\mathbf{x}) + \dfrac{1}{m_c^2} \left[\left(\dfrac{2\alpha_s}{r^3} - \dfrac {b}{2r}\right)\mathbf{L} \cdot \mathbf{S} + \dfrac{4\alpha_s}{r^3} T \right]
	\end{equation}



Precision measurements are needed to answer some questions related to the low-mass charmonium spectrum and BESIII is playing its role to determine the masses and the widths of all charmonium states.

Some charmonium-like mesons not fitting in the previous discussed $c\overline{c}$ model were observed, while others, which are foreseen by the theory, like charmonium hybrids, have not been observed. Particles of interest are called <$XYZ$ mesons: multiquarks candidates. On April 26$^{th}$, 2013 BESIII announces a new \textit{mystery} particle: $Z_c^+$(3900) the first observation of a tetra-quark or a charm molecule confirmed later by other experiments. Later on several decays of the $Z$ state has been observed in decay channels with charmonium or open charm state \cite{ref1:Z1,ref1:Z2,ref1:Z3,ref1:Z4}. $Y$ states are well investigated by BESIII thanks to the high precision measurement. $Y$ are connected to the $Z$ states and their nature is still puzzling. An abundance of those states have been observed such as $Y$(4220) and $Y$(4360) in several charmonium decays and open charm mesons \cite{ref1:Y1,ref1:Y2}.

Charmonium spectroscopy is studied with the hadronic and radiative transitions.
Hadronic transitions are decay modes of heavy quarkonia to lighter $c\overline{c}$ state and a light hadrons. The typical mass difference between the initial and final meson is around a few MeV, so that the typical momentum of the emitted hadron is low. The light hadron is produced from gluons and it is emitted in S,D and P-Wave mode. 
Information about the wave-function of the $1^{--}$ charmonium states is accessible from the leptonic partial widths. In the non-relativistic limit of an S-wave and D-wave quarkonium system the coupling $e^+e^-$ through a virtual photon involves the wave function and the partial width is given by \cite{ref1:VanRoyen}:

	\begin{equation}
	\label{eq:Width1}
		\Gamma^{e^+e^-}_{c\overline{c}} \left(^3S_1 \right) = \dfrac{16}{9} \alpha^2 \dfrac{\left|\psi(0)\right|^2}{M^2_{c\overline{c}}}.
	\end{equation}

	\begin{equation}
	\label{eq:Width2}
		\Gamma^{e^+e^-}_{c\overline{c}} \left(^3D_1 \right) = \dfrac{50}{9} \alpha^2 \dfrac{\left|\psi(0)\right|^n}{M^2_{c\overline{c}}m_c^4}.
	\end{equation}

Radiative decay can be used to extract the coupling constant $\alpha_s\left(M_{J/\psi}\right)$ from:
	\begin{equation}
\alpha_s\left(M_{J/\psi}\right)=\dfrac{\Gamma(J/\psi \rightarrow \gamma_{direct} X)} {\Gamma_{strong} (J/\psi \rightarrow X)} 
	\end{equation}
where $X$ stands for light hadrons and $\gamma_{direct}$ is a photon produced by the $J/\psi$.
At BESIII huge data samples of vector charmonium states as $J/\psi$, $\psi'$ and $\psi''$ are produced and it allowed to understand the light hadron spectroscopy and charmonium decay dynamics.
Hadronics decays are the 85-98\% in charmonia mesons, the basic process is the annihilation of the $c\overline{c}$ inside the charmonium to light quarks, gluons or leptons. The energy released in the process is 2m$_c$ and the space-time distance is $\sim$ $\dfrac{1}{2m_c}$. 
$J/\psi$, $\psi$’, $\eta_c$ and $\chi_c$ decays to light hadrons are used to test theories as NPQCD and extract the variables of these or to access the gluonic behavior.

\subsection{Charm physics}
$D$ mesons are the lightest particle containing charm quark and due to Standard Model (SM) prediction the $D$ system is a good environment to test the theory, such as CKM matrix and $D^0\overline{D^0}$ mixing to the CP asymmetries, through their leptonic and semi-leptonic decay. 
The QCD potential used for $D$ meson system needs precise theoretical control to describe branching ratios, spectrum and quantum numbers.
Meanwhile SM predicts small CP-violating asymmetries in charmed particle decay because the mode is Cabibbo suppressed and a large data sample is required with a complex final states to be analyzed.
A $c\overline{c}$ resonance above the $D\overline{D}$ threshold has to be created to produce $D$ mesons pairs via $e^+e^-$. There are four broad resonance states that decay into pair of charmed meson final states: the initial $c\overline{c}$ meson decays via the production of light $q\overline{q}$ pair and forms $c\overline{q}$-$\overline{c}q$ system that separate in two charmed mesons. 
These states are the $\psi$(3770), $\psi$(4040), $\psi$(4160) and $\psi$(4415). They are easily produced at $e^+e^-$ collider because their quantum number J$^{PC}$ 1$^{--}$.
To study this mechanism the charmed cross section is measured
	\begin{equation}
	\label{eq:cross1}
		\sigma_{\mathrm{charm}} = \dfrac{N_{\mathrm{charm}}}{\mathcal{L}}.
	\end{equation}

where $\mathcal{L}$ is the integrated luminosity and $N_{charm}$ is the number of charmed meson obtained using tagging technique.
The cross section of the $c\overline{c}$ states above the open-charm threshold can be understood as the successive onset of the charmed meson channels: $D\overline{D}$,$D^*\overline{D}$,$D^*\overline{D^*}$,$D_s\overline{D}_s$, \textit{etc}.
Each charm decay of these charmonium state is studied to determinate amplitude and width and to compare with the theory that describe these decays \cite{ref1:Brambilla}.
%
Pure leptonic decays $D^+\rightarrow l^+\nu$ are the cleanest decay modes of $D^+$ meson: the $c$ and $\overline{d}$ quarks annihilate into a virtual $W^+$ boson with a decay width given by the SM:
	\begin{equation}
	\label{eq:Width3}
		\Gamma \left (D^+ \rightarrow l^+ \nu \right ) = \dfrac{G^2_F}{8\pi} f_{D^+}^2 m_l^2 M_{D^+} \left (1 - \dfrac{m_l^2}{M^2_{D^+}} \right ) |V_{cd}|^2.
	\end{equation}

where $M_D^+$ is the $D^+$ mass, $m_l$ is the lepton mass, $V_{cd}$ is the CKM matrix element and $f_{D+}$ is the decay constant that contains the information about the dynamics of the quarks inside the meson.
Studying this kind of decay it determines $f_{D+} |V_{cd}|$. In the literature, $f_{D+}$ has been extensively studied in a variety of theoretical approaches \cite{ref1:TWChiu,ref1:Narison}.
In similar way $D_s^+ \rightarrow l^+ \nu$ mode is used to access to $|V_{cs}|$ and $f_{D_s}$.
BESIII physics program include high precision measurement of $f_D$ and $f_{D_s}$ mesons.
Semileptonic decays occur when a charm meson decay via weak interaction by emitting a $l^+\nu$ lepton pair: $c\rightarrow ql^+\nu$, where $q$=$d$,$s$. The $q$ quark is bound to the initial light quark of the charm meson by strong interaction ($X$ meson in Eq. \ref{eq:Amplitude}). In semileptonic decay the leptons do not interfere with the strong interaction then they can factored out the hadronic process. The process amplitude is: 
	\begin{equation}
	\label{eq:Amplitude}
		\mathcal{A} = \dfrac{G_F}{\sqrt{2}}V^*_{cq}\overline{\nu}\gamma_{\upmu}(1-\gamma_5)l\left \langle  X|\overline{q}\gamma^{\upmu}(1-\gamma_5)c|D\right \rangle.
	\end{equation}
The process depends on the quark-mixing parameter $V_{cq}$ so the semileptonic decay is a good system to access to the CKM matrix and test the technique that calculate the hadronic matrix element developed by several method \cite{ref1:Abada,ref1:Shifman,ref1:Khodja,ref1:Wirbel,ref1:HYCheng,ref1:Fajfer}.
These decay are sensitive to $V_{cd}$ and $V_{cs}$ by the determination of the absolute branching fractions, $e.g.$ for $V_{cs}$ in $D \rightarrow \pi l^+ \nu$, $D \rightarrow K l^+\nu$, $D \rightarrow \eta(\eta')l^+\nu$, $D \rightarrow \rho l^+\nu$, $D \rightarrow K^*l^+\nu$, $D_s \rightarrow \varphi l^+\nu$, \textit{etc}.
These parameters are used together with the other CKM element to measure the mixing and the CP violating parameters, to test the self-consistency of the CKM picture and to search possible new physics beyond the CKM mechanism.

Meson-antimeson oscillation is another topic accessible at BESIII experiment. In these oscillation weak interaction effect build up over macroscopic distance and makes tiny mass differences measurable. It is mandatory to probe $D^0-\overline{D^0}$ oscillation as sensitively as possible to provide proof of New Physics and CP violation.

$\Lambda_c$ physics is another branch of interest in charm studies. The lightest charmed baryon $\Lambda_c^+$ ($udc$) decays only through the c weak decay, $c \rightarrow W^+s$,$W^+d$, at the leading order and it is the simpler theoretical environment to describe with non-perturbative models. This particle allows to measure the absolute branching fraction of $\Lambda_c^+ \rightarrow \Lambda l^+\nu$ and 12 Cabibbo-favored decays, studies of singly-Cabibbo-suppressed decays and the observation of the modes with a neutron in the final state.


\section{The needs of a new inner tracker for BESIII}
\label{sec:MDCaging}

\begin{figure}[htbp]
		\centering
   		\includegraphics[width=0.8\textwidth]{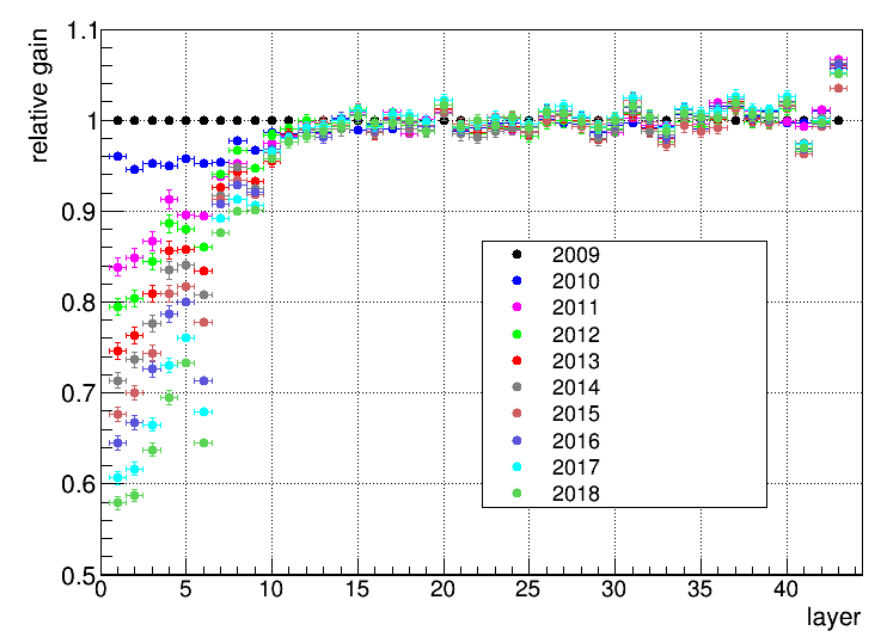}
   		\caption[MDC aging effect in hit efficiency]{Hit efficiency versus the layer number for the past 10 years. A drop of efficiency is clear for the first layers due aging effects.}
	\label{fig:aging}
	\end{figure}

BESIII started to operate in 2009. In about the past 10 years the detectors suffered of aging due to radiation effect. The integrated charge on the innermost layer is about $170 \, \mathrm{mC}/\mathrm{cm}$. The drift chamber is the system closest to the beam pipe. After several years of operation the performance of the drift chamber decreased due to the integrated charge. The aging in the anode is given by gas polymers that condense on the sense wire surface as thin films, whiskers and powder due to the high electric field. Those accumulation could be conductive or insulating; both contaminations cause a gain loss due to the increase of the effective diameter of the sense wire. Reduction of the gain is due also to the reduction of the electric field caused by the accumulation of charges in the insulating layer.
The cathode suffers the aging too. A polymer deposits on the cathode surface and insulate it. This effect prevents the neutralization of the positive charge that modifies the detector electric field around the cathode up to a value that extracts electrons from the electrode. Those electrons can drift to the sense wire and create avalanches. The avalanche positive ions come back to the cathode, enhance the electric field around the cathode and thus feeds continuously this self-sustaining local discharge. This effect is called the Malter effect \cite{ref1:malter} and some water vapor has been added to the MDC gas mixture to slow down this issue. In order to reduce the dark currents of the sense wires and to use the detector in a safe configuration the voltage of the first four layers has been decreased with respect to the normal values, which result in the decrease of the performance of the detector. The gain with respect to 2009 status has been reduced up to 60\% and this affects the event reconstruction worsening the detection efficiency up to 50\% \cite{ref1:aging}. The gain of the cells closer to the IP shows a sizable decrease, as shown in Fig. \ref{fig:aging}. This worsening does not allow the MDC to maintain the required performance in the next years. The BESIII collaboration decided to replace the inner part of the MDC with a new and more performing inner tracker\cite{ref1:CDR} with required performance described in Tab. \ref{tab:requirement}. 

A three layers cylindrical triple-GEM inner tracker has been proposed to replace the first eight layers of the MDC. A GEM (Gas Electron Multiplier) is a Micro-Pattern Gas Detector, a particular class of gas detectors that profits from modern photolitography and deposition of polyamide thin-layer. This allows to reach unprecedented performance with gas detectors as it will be shown in Chap. \ref{sec:graal}.

\begin{table}[htbp]
\begin{center}
\begin{tabular}{c c}
Parameter & Value\\
\hline
Rate capability & $10^4 \, \mathrm{Hz/cm^2}$ \\
$\sigma_{r-\phi}$ & $150 \, \mathrm{\upmu m}$ \\
$\sigma_z$ & $1 \, \mathrm{mm}$ \\
$\sigma_{p_t}/p_t$ & 0.5\% $@$ 1GeV \\
Efficiency & 98\% \\
Material budget & $\leq$ 1.5 X$_0$ \\
$\Delta \Omega$ & 93\% 4$\pi$ \\
Inner radius & $78 \, \mathrm{mm}$\\
Outer radius & $178 \, \mathrm{mm}$\\
\newline\\
\hline\\
\end{tabular}
\caption[Inner tracker requirements]{Inner tracker requirements}
\label{tab:requirement}
\end{center}
\end{table}

	\chapter{CGEM-IT}
\label{sec:cgem}
Despite various improvements, the detection technique based on wire structures suffers by limitation due to diffusion processes, space charge effects and aging of the anode and cathode electrodes. Modern construction technique allows to solve many of those problems, $e.g.$ by using pitch size of a few hundred of $\mathrm{\upmu m}$ to increase the granularity, or by photolithographic processes and the polyamide depositions of a layers of few $\mathrm{\upmu m}$ to define new amplification geometries and detection techniques as reported in App. \ref{sec:mpgd}. The GEM technology is the one chosen to replace the inner MDC: thanks to a thin pierced foil of copper-insulator-copper can magnify the primary ionization and measure the particle position within approximately $1 \, \mathrm{cm}$ of gas volume. The resolution achieved depends on gas mixture, gain and front-end electronics. In literature spatial resolution below hundred $\mathrm{\upmu m}$ are reported \cite{ref2:sauli2016} for orthogonal tracks without magnetic field. The GEM foils can be shaped to the needed surface and thanks to their thickness and formability are suitable to be used as inner trackers with cylindrical surface. In this Chap. a general overview about the components, the construction technique and the electronics of CGEM-IT will be provided. The details of GEM technology will be explained in detail in Chap. \ref{sec:GEM}.
\section{The KLOE-2 IT}
The KLOE-2 inner tracker is an innovative detector built up by four coaxial layers of cylindrical triple-GEM: each layer exploits three GEM foils to amplify the signal from the particles that interact with it and it is shaped around the beam pipe. The inner radius of the layers ranges from 130 to $205 \, \mathrm{mm}$ and the length is about $700 \, \mathrm{mm}$. Each layer is made by five foils: three GEMs, a cathode and a segmented anode. The layers provide a bi-dimensional measurement due to longitudinal and stereo strips of $650 \, \mathrm{\upmu m}$ pitch.

GEM foils of the needed dimension can not be produced then they have been obtained through a gluing process with vacuum bag technique of two or three GEM foils together by means of epoxy glue on a $3 \, \mathrm{mm}$ peripheral region of kapton, the insulator material between the copper foils of the GEM. Then the large GEM foils are wrapped on an aluminum mould coated with Teflon film and glued in a $3 \, \mathrm{mm}$ overlap region. During the curing of the epoxy, a vacuum bag system is used to constrain the foil to the cylindrical shape given by the mould \cite{ref2:kloe2_built}. Anode and cathode electrodes are shaped with the same technique then on a Honeycomb structure and, together with two permaglass rings glued at the edges of those two electrodes provide the structural support to the detector. The five electrodes are inserted one onto each other through a precise machine named Vertical Insertion System. 

The front-end electronics used by KLOE-2 IT is GASTONE-ASIC \cite{ref2:gastone}: an analogue preamplifier integrates the signal that it is filtered from noise and then discriminated with a threshold range from 0 to $200 \, \mathrm{fC}$. A digital output from each strip is sent out for the position measurement.
\section{Innovation of the BESIII CGEM-IT}
\label{sec:BESIII_CGEM} 
The BESIII IT starts from the KLOE-2 experience to develop a new tracker with GEM technology and cylindrical shape. The configuration used by KLOE-2 could not match the BESIII requirements due to its dimension, its total material budget and its spatial resolution (see Sect. \ref{sec:MDCaging}) then a series of improvement have been introduced. Due to the dimension of the inner MDC in BESIII, the volume left free from its removal permits to insert three cylindrical layers instead of four. To keep reasonable the cost of the IT and for mechanical constrains, a pitch of $650 \, \mathrm{\upmu m}$ has been chosen despite it would not reach the needed resolution if the digital readout was kept. It was mandatory to develop a new ASIC to measure charge and time information needed in the reconstruction algorithms described in Sect. \ref{sec:reconstrution_method}. More details about the new ASIC will be discussed in Sect. \ref{sec:tiger}. This is one of the most important innovation with respect to KLOE-2.

\begin{figure*}[ht!]
\centering
\begin{tabular}{lcr}
\includegraphics[width=0.25\textwidth]{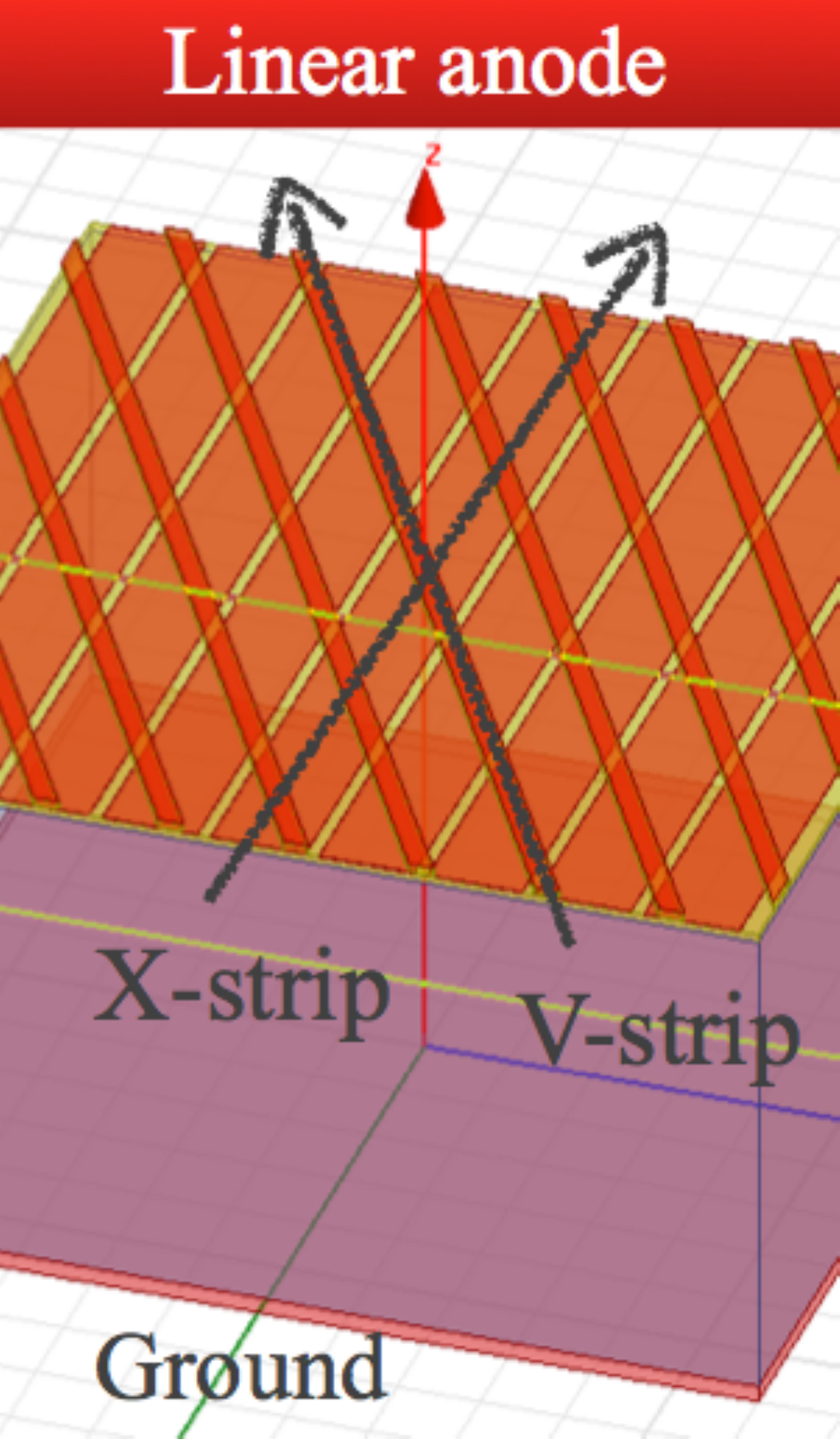} & ~~~~~~~~
\includegraphics[width=0.25\textwidth]{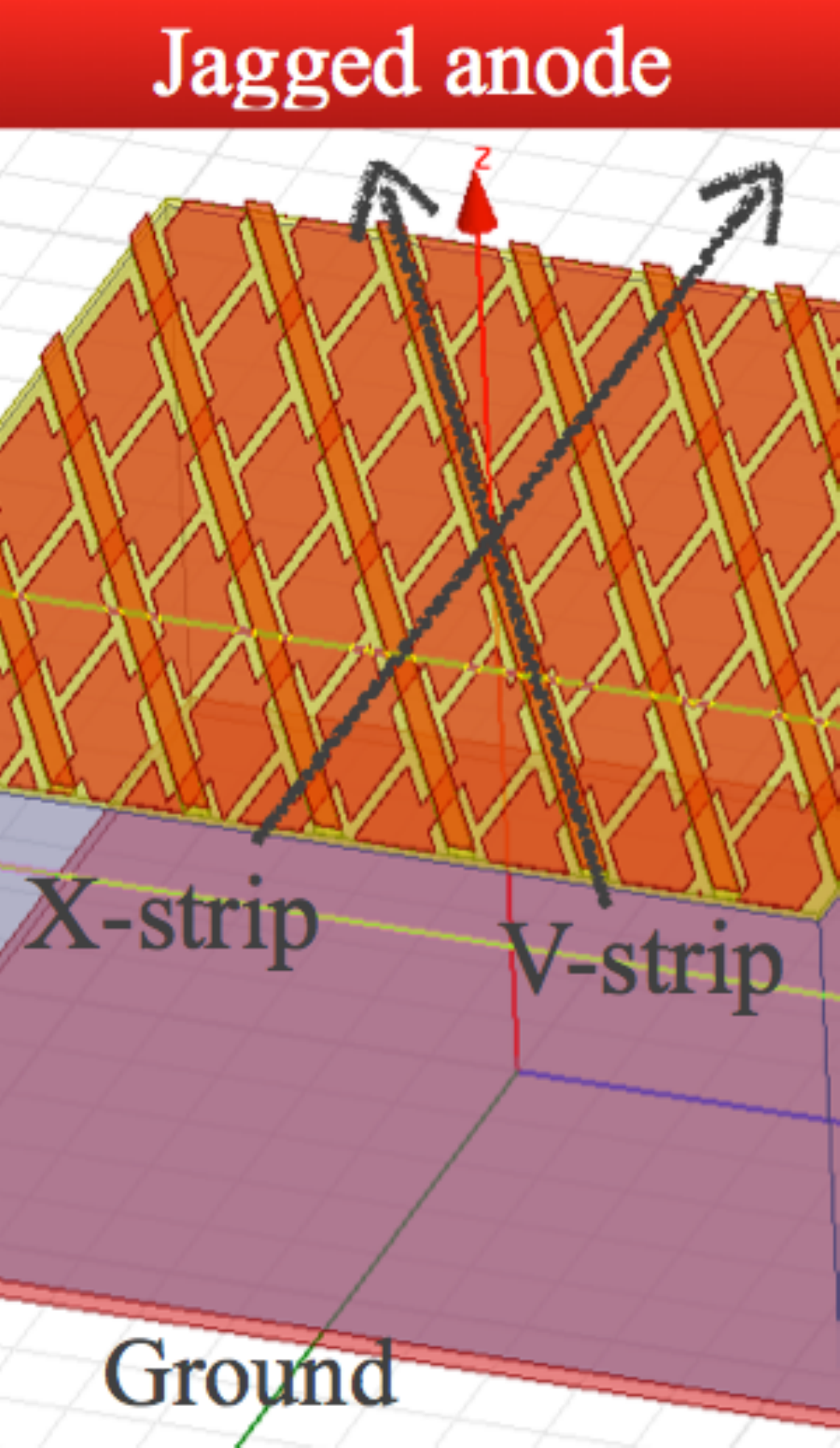} \\
\end{tabular}
\caption[Jagged anode layout]{Anode layouts are shown in the figure with (left) the linear layout where the width of the strips are constraint along their length and (right) the jagged readout where the width of the strips is reduced in the overlap region \ref{fig:jagged}.}
\label{fig:jagged}
\end{figure*}

A new anode design has been implemented to reduce the inter-strip capacitance: longitudinal and stereo strips cross each other many times in the anode and each overlap of the two surfaces introduce a capacitance between the two strips that increases the noise in the readout channel. A jagged layout has been used in the BESIII IT to reduce the overlapping area. Simulations \cite{ref2:jagged} show that it allows to reduce the inter-strip capacitance of about 30\% with respect to the regular layout used by KLOE-2. Figure \ref{fig:jagged} shows the differences between the two design.

The sensitive volume in a triple-GEM detector has about $1 \, \mathrm{cm}$ of thickness. A signal bigger than 1 fC is collected mostly from a fraction of this sensitive volume: the region between the cathode and the first GEM because only the primary electrons generated in this region cross three timesthe GEM foils. In KLOE-2 this region, named $drift~gap$, has $3 \, \mathrm{mm}$ thickness while in the BESIII IT has been increased to $5 \, \mathrm{mm}$ to increase the number of primary ionizations and to improves the reconstruction performances that will be explained in Sect. \ref{Analysis_results}: larger gap correspond to better spatial resolution for the $\upmu$TPC technique.

The mechanical structure of the CGEM-IT in the KLOE-2 experiment is given by a Honeycomb foil on the anode and the cathode electrodes and the permaglass ring at the edges of the detector. BESIII uses a different material: the Rohacell  that it allows to reduce the thickness of the structure, keeping the same mechanical properties. This reduces the material budged of the detector then its interaction length $X_0$. The last design innovation is in the reduction of the copper thickness on the faces of the GEM foil from 5 to $3 \, \mathrm{\upmu m}$. This reduces the radiation length of the entire detector. The entire CGEM-IT will have a total material budget below 1.5\% of $X_0$.

In Tab. \ref{tab:CGEM_config} are reported the values of the inner radius of each CGEM-IT layer, the active area length and the number of strips and the angle between longitudinal and stereo strips. 

\begin{table}[ht]
\begin{center}
\begin{tabular}{lccccc}
~ & Inner radius & Length & Stereo angle & N$^\circ$ strips $\phi$ & N$^\circ$ strips $V$ \\
\hline
\newline\\
Layer 1 & $76.9 \, \mathrm{mm}$ & $532 \, \mathrm{mm}$ & 43.3$^\circ$ & 846 & 1177\\
Layer 2 & $121.4 \, \mathrm{mm}$ & $690 \, \mathrm{mm}$ & -31.1$^\circ$ & 1282 & 2194\\
Layer 3 & $161.9 \, \mathrm{mm}$ & $847 \, \mathrm{mm}$ & 33.0$^\circ$ & 1692 & 2838\\
\newline\\
\hline\\
\end{tabular}
\caption[CGEM-IT geometrical details]{CGEM-IT geometrical details.}
\label{tab:CGEM_config}
\end{center}
\end{table}


\section{The new ASIC for the BESIII CGEM-IT}
\label{sec:tiger}
\begin{figure*}[ht!]
\centering
\begin{tabular}{c}
\includegraphics[width=0.7\textwidth]{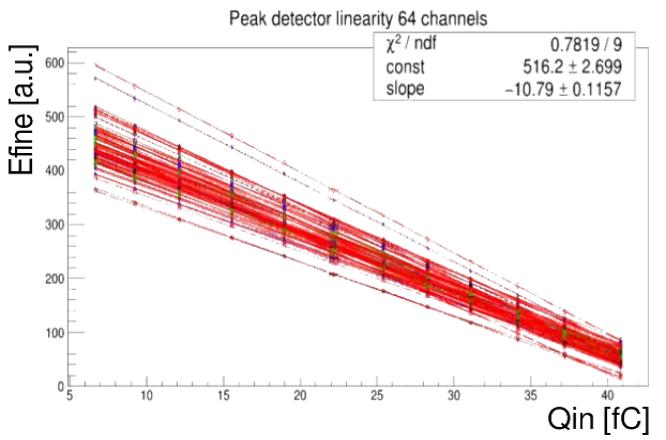} \\
\includegraphics[width=0.7\textwidth]{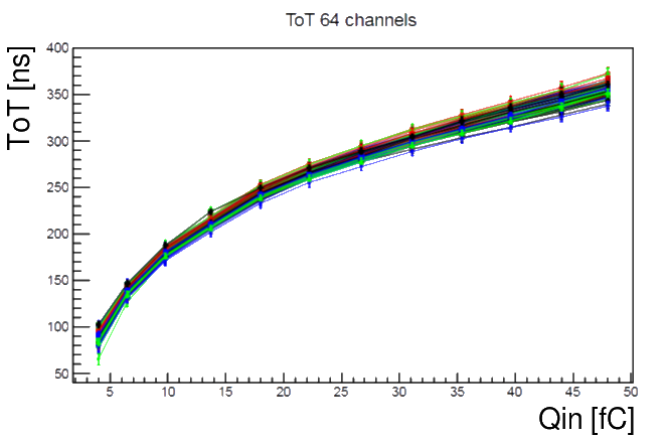} \\
\end{tabular}
\caption[TIGER calibration curves]{TIGER calibration curves with (top) S/H method where the correlation between the measured charge and the injected one is linear up to the saturation value around above $50 \, \mathrm{fC}$ and (bottom) the ToT mode with a linear slope above $25 \, \mathrm{fC}$ and a bended one in the region below \cite{ref2:calib}.}
\label{fig:calib}
\end{figure*}

TIGER (Torino Integrated Gem Electronics for Readout) electronics, is a mixed-signal 64-channel ASIC developed to readout the CGEM detector of the BESIII Experiment \cite{ref2:tiger}. The ASIC is installed on a Front-End Board (FEB) with two chips to measure the signals from 128 strips. The ASIC measures time and charge of the signal. The chip uses two threshold, a first one to measure the rise up edge of the signal and a second one to discriminate signal from noise. The second threshold depends on the strip capacitance and chip temperature. Due to the variable length of the stereo strips, different threshold are used for each channel. A chiller stabilizes the chip temperature to avoid threshold variation. The time is measured with a lower threshold that send out the time information when the signal overstep the discrimination one. The charge can be measured with two methods: sampling and hold (S/H) mode and time-over-threshold (ToT) mode. S/H measures the signal amplitude in a dynamic range from 0 to $50-60 \, \mathrm{fC}$. The ToT measures the leading edge of the signal and it convert the time length of the signal into charge information. Both methods need a calibration that has to be performed for each channel of the TIGER. Figure \ref{fig:calib} shows the calibration curve of a TIGER for S/H and ToT methods \cite{ref2:calib}. The output of the chip is fully digital. The advantage of S/H is the linearity between the charge and the output but the operating range is limited while ToT has an unlimited range but the output is not linear and it needs calibrations.

	\chapter{The GEM technology}
\label{sec:GEM}
Modern physics experiments use gaseous detector to reveal and measure particles properties and as the power of the accelerating machine increases a more sensitive and precise detectors are needed.

Gas detector exploits the physical processes between the particles, gas mixture and the high electric field. Charged particles interact with the gas and ionize it. As a function of the cross section, some gases can create more primary electrons with respect to others, then they have a larger amplitude in the collected signal on the readout. The probability to have a certain number of primary electrons is ruled by the Poisson distribution and the energy loss is described by the Bethe-Block formula. The electric field can drift the electrons: transverse and longitudinal diffusion contribute the time and spatial resolution of the detector. The electrons can trigger an avalanche if the electric field is higher and the electrons excite new gas molecule. Several working regimes are defined: if the electric field is too low the electrons and ions recombine; increasing the electric field the electrons generate an avalanche proportionally to the electric field; increasing more the electric field the electron can generate sparks. Once the avalanche is generated, it can induce a signal on the readout plane such as pads or strips and it is described by the Ramo theorem. More details are reported in the App. \ref{sec:mpgd} and in the bibliography there in.

A gas detector with a large diffusion is the Multi-Wire Proportional Chamber (MWPC) and its evolution: this is suitable detectors from different application but it suffers limitation such as the creation of a large amount of positive ions and the electric field distortion due to their back-flow, large drift volume that introduces limitations in rate capability higher than $10 \, \mathrm{KHz/mm^2}$ \cite{ref3:pc_rate}, low multi-track resolution due to the wire granularity limited at $1-2 \, \mathrm{mm}$ and its operability limitation for aging problems. Modern photolitography and thin-layer polymide deposition introduced by the Micro Strip Gas Chamber (MSGC) \cite{ref3:oed} overcome several limitation of the previous technology but it resulted too fragile in case of sparks. The new MPGD focused their design to increase the gain up to the Raether limit \cite{ref3:sauli_book} of 10$^7$-10$^8$ electrons in the avalanche, in order to avoid the rupture of the gas dielectric rigidity and discharges.

In this Chap. a detailed description of the detector working setting, the mechanical design and the operability configuration will be reported. The modern application profits of the goals achieved step by step by this technology. A description of the advantages and weakness will be shown.
\section{Gaseous electron multiplier}
The Gaseous Electron Multipliers (GEM) is an electron amplification technique invented by F.Sauli in 1996 \cite{ref3:sauli1996}. It consist of a kapton foil with copper coated on the two faces and a high density holes. The foil is placed between two electrodes defining two regions: the drift gap where the primary electrons are generated collected to the GEM foil; the induction gap where the electron avalanche generated in the holes is driven from the GEM to the readout plane. The GEM operates with a high voltage difference between the two coppered electrodes on the two sides of the GEM: this creates an electric field up to $100 \, \mathrm{kV/cm}$ inside each holes that accelerates the upcoming electrons and generates a cascade in a localized position given by the hole dimension. The GEM thickness is less than $100 \, \mathrm{\upmu m}$ and this lets high gain regime without exceeds the critical Raether limit between the two GEM foil faces.
\section{Design and construction}
\begin{figure}[t]
\centering
\includegraphics[width=0.4\textwidth]{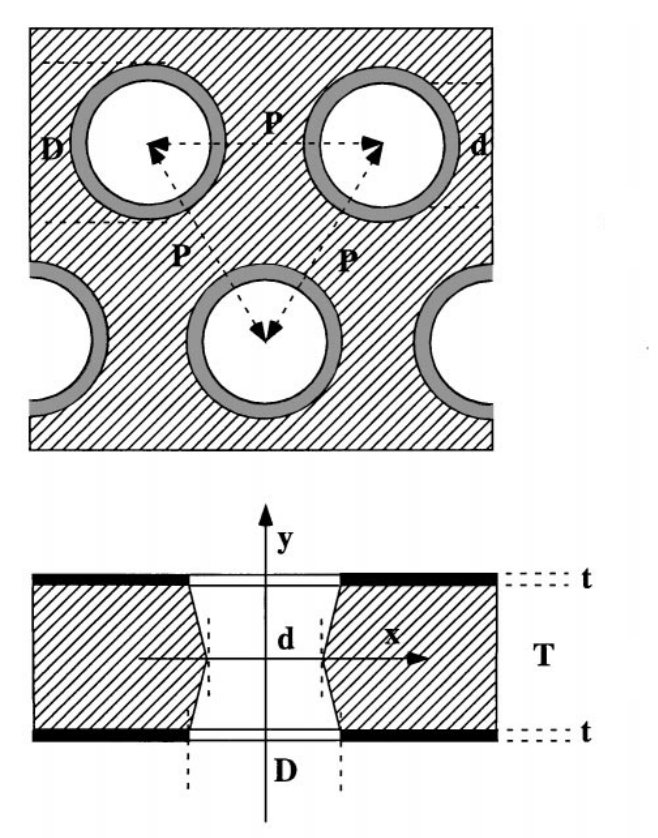}
\caption[Gem Holes]{Schematics of the standard bi-conical GEM holes geometry\cite{ref3:bachmann1999}.}
\label{fig:gem_hole}
\end{figure}

The shape, dimension and pitch of the holes are topics extensively studied by S. Bachamann and collaborators \cite{ref3:bachmann1999} in 1999 to optimize the GEM design. The typical GEM foil is composed by $50 \, \mathrm{\upmu m}$ kapton and $5 \, \mathrm{\upmu m}$ copper on both sides. The holes are produced with the photolitography on the metal and a specific solvent for the kapton. A bi-conical shape of the holes is produced applying the solvent on both sides: this technique is named $double-mask$ process. In 2009 a new process, $single-mask$ \cite{ref3:single_mask}, has been developed to overcome the alignment needed by the two mask. The holes are equidistant and circular as shows Fig. \ref{fig:gem_hole}. The study \cite{ref3:bachmann1999} ranged hole size from $40$ to $140 \, \mathrm{\upmu m}$ and pitch between $90$ and $200 \, \mathrm{\upmu m}$. The optimal values have been found to be a diameter $D$ in the basis of the cone of $70 \, \mathrm{\upmu m}$ and a diameter $d$ in the internal region of $50 \, \mathrm{\upmu m}$. The pitch chosen is $140 \, \mathrm{\upmu m}$. This configuration gives an optical transparency of 46\%.

\section{Electrical configuration}
\begin{figure}[t]
\centering
\includegraphics[width=0.5\textwidth]{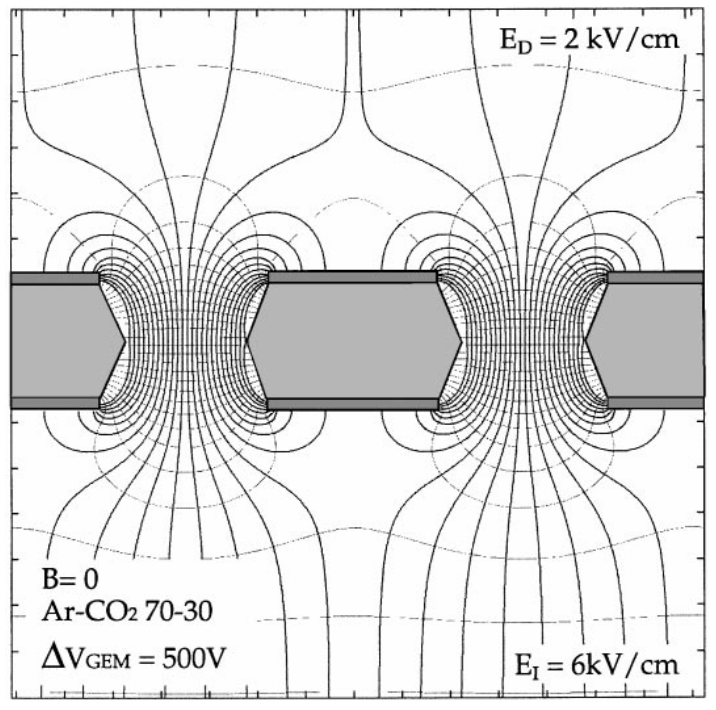}
\caption[Gem electric field]{Electric field lines around the GEM hole \cite{ref3:bachmann1999}.}
\label{fig:gem_field}
\end{figure}

\begin{figure}[tp]
\centering
\includegraphics[width=0.7\textwidth]{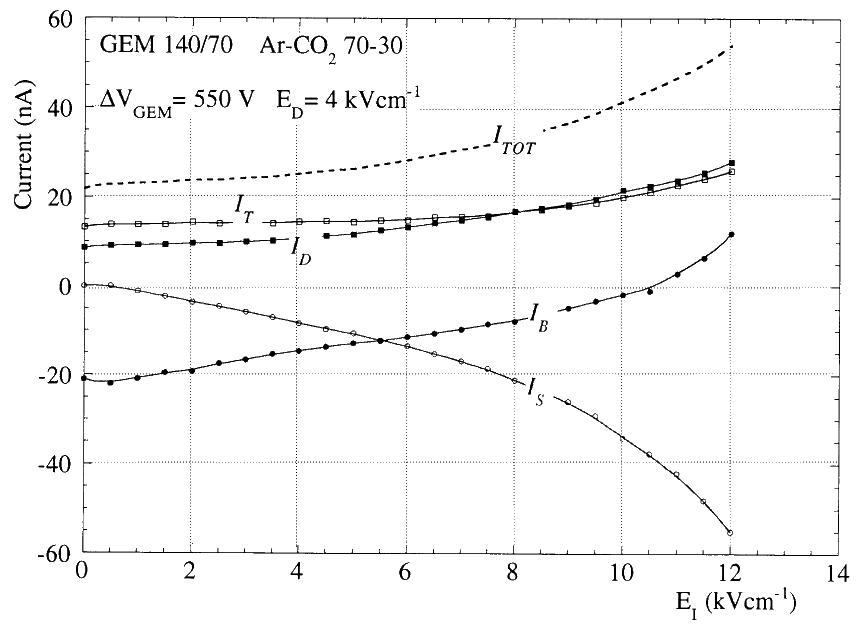}
\caption[Induction field in a GEM]{Current sharing between the electrodes in single GEM as a function of the induction field. $I_S$ is the current collected on the anode, $I_B$ on the bottom of the GEM, $I_{\mathrm{TOT}}$ is given by the sum of the previous two and is related to the GEM gain \cite{ref3:bachmann1999}.}
\label{fig:gainEI}
\end{figure}

\begin{figure}[tp]
\centering
\includegraphics[width=0.7\textwidth]{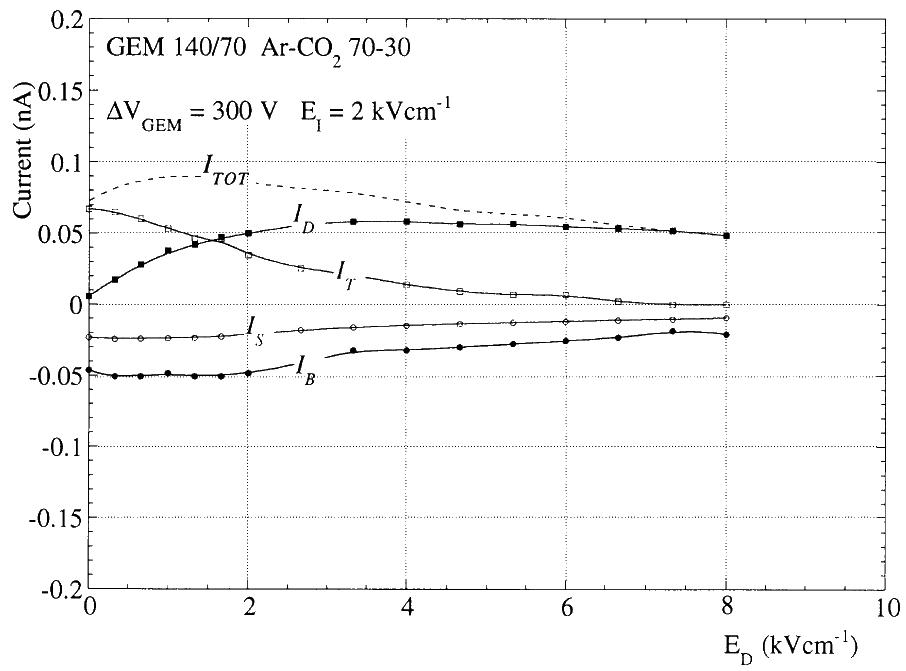}
\caption[Drift field in a GEM]{Current sharing between the electrodes in single GEM as a function of the induction field. $I_D$ is the current collected on the cathode, $I_T$ on the top of the GEM, $I_{\mathrm{TOT}}$ is given by the sum of the previous two and is related to the GEM gain \cite{ref3:bachmann1999}.}
\label{fig:gainED}
\end{figure}

In a GEM detector three electric fields are needed: 
\begin{enumerate}
\item between the cathode above the GEM foil and the electrode on top the GEM to generate the drift field $E_D$;
\item inside the GEM holes to multiply the electron number with a gain related to the high voltage difference $\Delta V_{\mathrm{GEM}}$;
\item between the bottom electrode of the GEM and the anode below the GEM to generate the induction field $E_I$.
\end{enumerate}
Computation of the electric field has been done by \cite{ref3:bachmann1999} as shown in Fig. \ref{fig:gem_field}. 
The gain in a single GEM detector depends on those three components: $\Delta V_{\mathrm{GEM}}$, $E_D$ and $E_I$. The $\Delta V_{\mathrm{GEM}}$ is related to the gain with an exponential relationship. Figure \ref{fig:discharge_gain} shows the exponential dependency of the gain from the $\Delta V_{\mathrm{GEM}}$.  The dependencies on the $E_D$ and $E_I$ have been studied measuring the currents on the electrodes varying those fields. As the induction field $E_I$ increases then the total gain increases: a larger number of electrons are extracted from the holes and the current on the bottom face of the GEM decreases. See Fig. \ref{fig:gainEI}. The current as a function of the drift field $E_D$ has a different behavior as shown in Fig. \ref{fig:gainED}. The total gain is maximum if $E_D \approx 1.5 \, kV/cm$. If $\Delta V_{\mathrm{GEM}}$ is increased then a higher $E_D$ is needed to maximize the total gain.


\section{Multistage multiplication}
\begin{figure}[ht]
\centering
\includegraphics[width=0.7\textwidth]{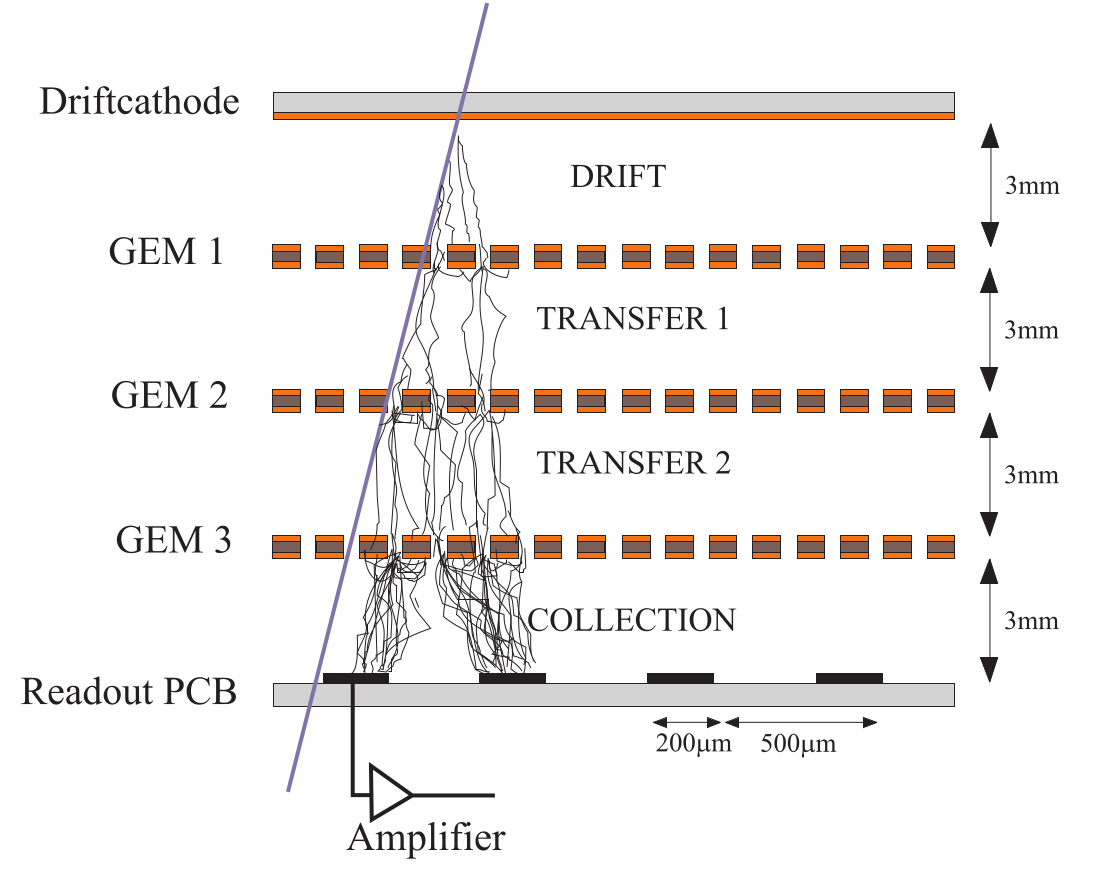}
\caption[Triple-GEM representations]{Representation of a triple-GEM detector \cite{ref3:tripleLHCb}.}
\label{fig:tripleGEM}
\end{figure}
\begin{figure}[tp]
\centering
\includegraphics[width=0.7\textwidth]{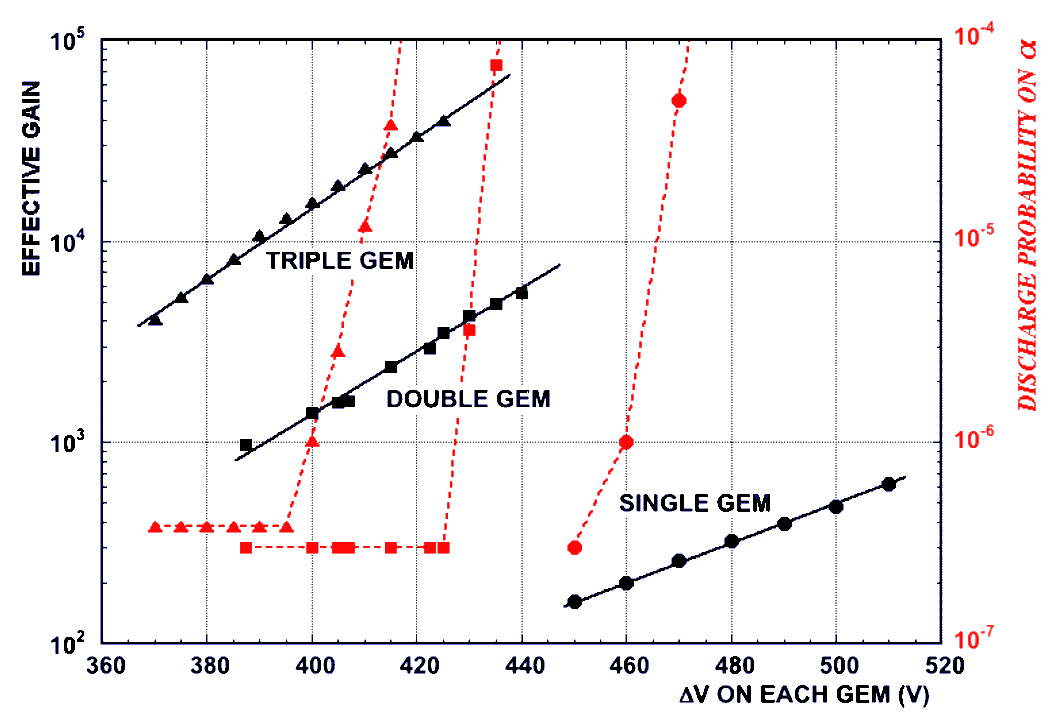}
\caption[Gain and discharge probability in GEM]{Discharge probability and gain as a function of the high voltage on each GEM for single-, double- and triple-GEM \cite{ref3:bachmann1999}.}
\label{fig:discharge_gain}
\end{figure}

Several GEM amplification foils can be used to reach higher values of gain such as bi-GEM or even quadruple-GEM structures reducing the discharge probability. An example is shown in Fig. \ref{fig:tripleGEM}. Configuration with two foils of GEM has extensively studied in test beam by \cite{ref3:bachmann1999,ref3:bigem}, triple-GEM by \cite{ref3:tripleLHCb,ref3:triple2D} and quadruple-GEM by \cite{ref3:quadruple1,ref3:quadruple2}. Those configuration needs another field, named transfer field $E_{T}$, that has to extract the electron from the GEM above and transport them to the next GEM. Optimization of $E_{T}$ is needed to improve the gain of the multiple-GEM detectors. The most significant improvement of multiple amplification stages is to reduce $\Delta V_{\mathrm{GEM}}$ then the discharge probability while the gain is even higher. This grants higher electrical stability and long living detector. Figure \ref{fig:discharge_gain} shows the differences between single-, double- and triple-GEM.

\section{Innovations and performance of triple-GEM detectors}
In the past 20 years triple-GEM detectors have been extensively studied and several applications have been developed. The usual configuration in reported in Fig. \ref{fig:tripleGEM}. Best performance report a spatial resolution of tens of $\mathrm{\upmu m}$ and time resolution below $10 \, \mathrm{ns}$; it has been used in low and high pressure environments \cite{ref3:sauli2016}; gain up to 10$^5$ and discharge probability below 10$^{-6}$ at a gain lower than 20000 \cite{ref3:bachmann1999}. It can be shaped to to non-planar geometry such as cylindrical \cite{ref3:cgem} or spherical \cite{ref3:spherical} shapes. Moreover the application of the triple-GEM varies from high precision tracking detector, large TPC, single photon or single electron detection and even neutron detection \cite{ref3:sauli_book}.

	\chapter{Characterization and reconstruction of a triple-GEM signal}
\label{sec:graal}
A full characterization of the triple-GEM signal is needed to optimize the geometrical and electrical configurations of the detector in order to fulfill the BESIII requirements defined in Sect. \ref{tab:requirement}. 
The challenge is to achieve this requirement with a limited number of electronic channels, around 10000, over the whole detector and to build a system that fits the geometrical space left from the removal of the current Inner Drift Chamber. These details have been discussed in Sect. \ref{sec:BESIII_CGEM}.
The detector under study has a bi-dimensional readout with $X$ axial and $V$ stereo strips. The strip pitch chosen is $650 \, \mathrm{\upmu m}$ in order to limit the number of channels in the final design. A smaller pitch would have led to a better measurement of the time and charge profile of the signal but it would be impossible to instrument and readout the signal from the detector.

Between 2014 and 2018 different triple-GEM detectors have been studied with test beam to measure their performance as a function of the gas mixture, the drift gap thickness, the geometry, the dimension and the readout electronics. In detail  Ar+30\%CO$_2$ and Ar+10\%iC$_4$H$_{10}$ gas mixtures has been used, drift gaps of $3$ and $5 \, \mathrm{mm}$ thickness, planar and cylindrical shape. Transfer and induction gaps have $2 \, \mathrm{mm}$ thickness. The tested planar triple-GEM detectors have 10 $\times$ $10 \, \mathrm{cm^2}$ active area and bi-dimensional readout while the cylindrical triple-GEM chambers have the dimensions of the first two layers of the CGEM-IT.

The bi-dimensional readout differs from chamber to chamber. Some chambers have $XY$ readout with orthogonal strips for the two views, others have $XV$ readout with an angle of 60$^\circ$ between the strips of different views, while the CGEM have $\phi V$ readout with cylindrical shape and a stereo angle of 43.3$^\circ$ and 31.1$^\circ$.

\begin{figure}[ht]
\centering
\includegraphics[width=0.7\textwidth]{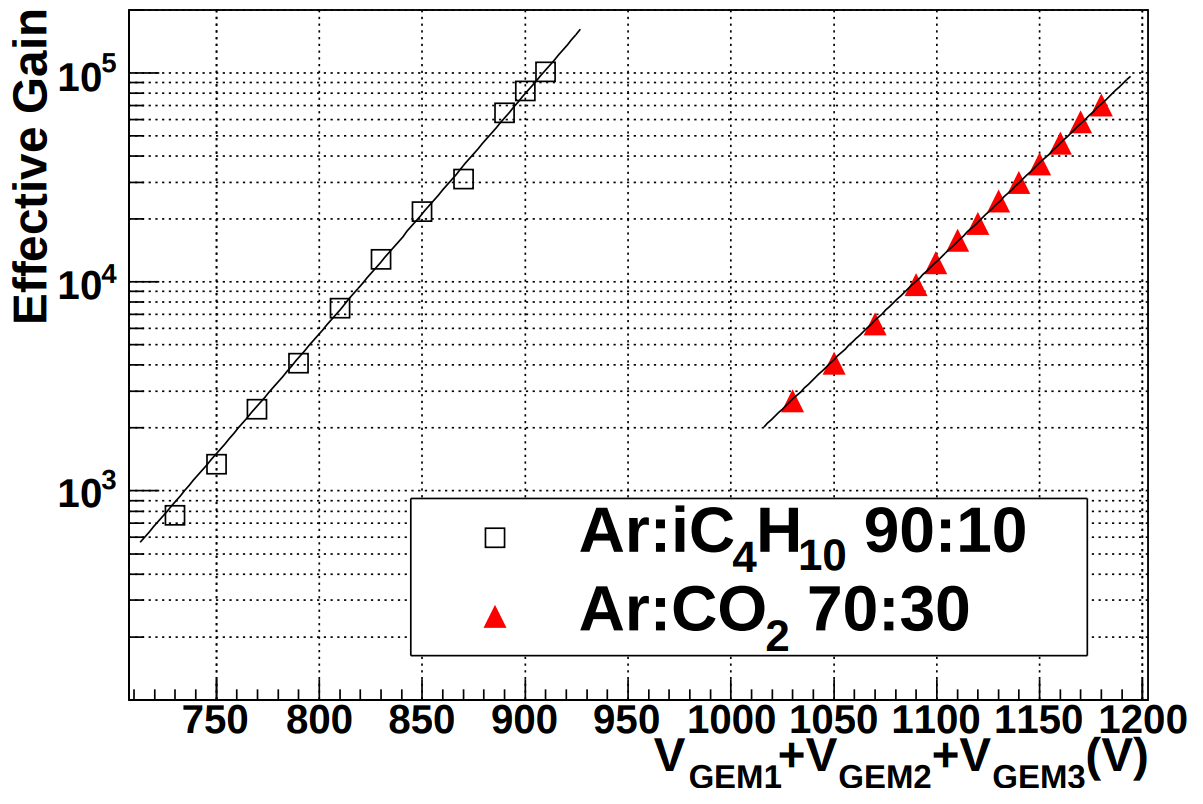}
\caption[Gas gain in a triple-GEM]{Ar+30\%CO$_2$ and Ar+10\%iC$_4$H$_{10}$ gas mixtures gain as a function of the $\Delta V_{\mathrm{GEM}}$ applied on each GEM in a triple-GEM detector \cite{ref4:kloe_gasgain}.}
\label{fig:gasgain}
\end{figure}

As starting point the used high voltage settings are the ones used by KLOE-2 \cite{ref4:kloe_tb} ($E_D \, = \, 1.5 \, \mathrm{kV/cm}$, $E_{T_1} \, = \, E_{T_2} \, 3 \, \mathrm{kV/cm}$ and $E_I\, = \, 5 \, \mathrm{kV/cm}$) and then an optimization is performed. The electric fields define the transport properties of the electrons in the triple-GEM geometry. The gain dependency on the $\Delta V_{\mathrm{GEM}}$ is shown in Fig. \ref{fig:gasgain}.

The information on the used electronics, the offline software and a selection of results will be provided in this Chap. Offline software includes reconstruction of the data, its alignment and the analysis. Results of interest will show the configuration that fulfills the BESIII requirements.

\section{Experimental setup}
\label{sec:setup}
\subsection{The facilities}

The various detectors have been characterized at two test beam facilities:
\begin{enumerate}
\item CERN - H4 beam line in North-Area in Prevessin with muon and pion beams up to 150 GeV/c momentum;
\item MAMI - MAinz MIcrotron facility in Mainz with electron beam up to $855 \, \mathrm{MeV/c}$ momentum.
\end{enumerate}

\begin{figure}[htp]
\centering
\begin{tabular}{c}
\includegraphics[width=0.7\textwidth]{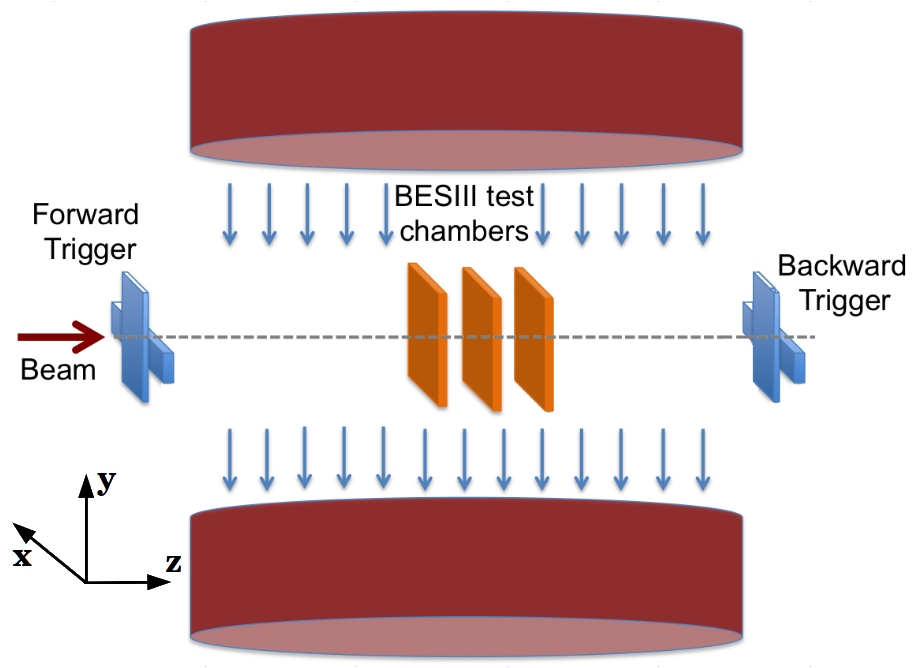}\\
\end{tabular}
\begin{tabular}{lr}
\includegraphics[width=0.34\textwidth]{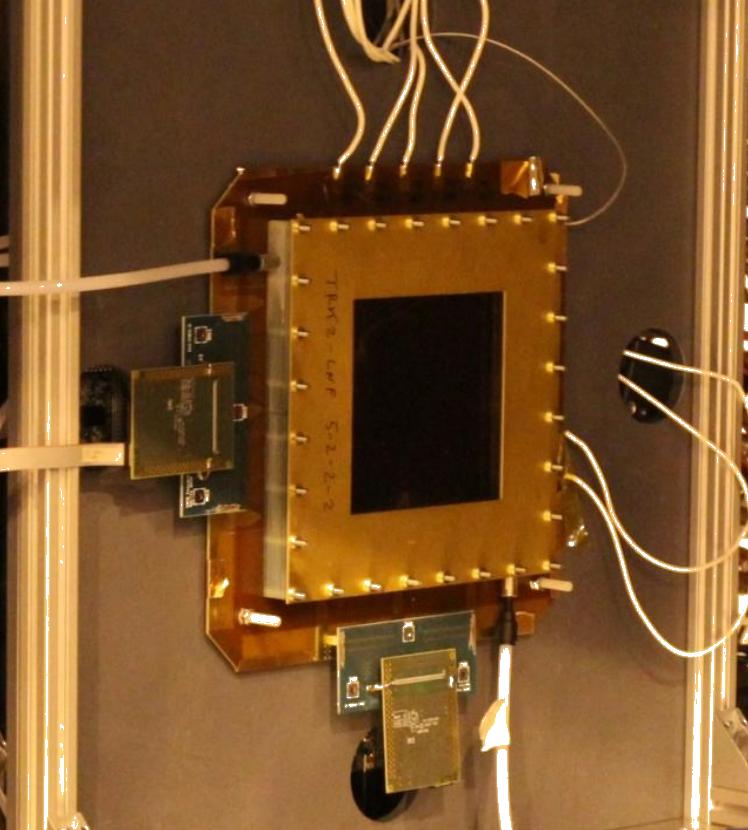}&
\includegraphics[width=0.48\textwidth]{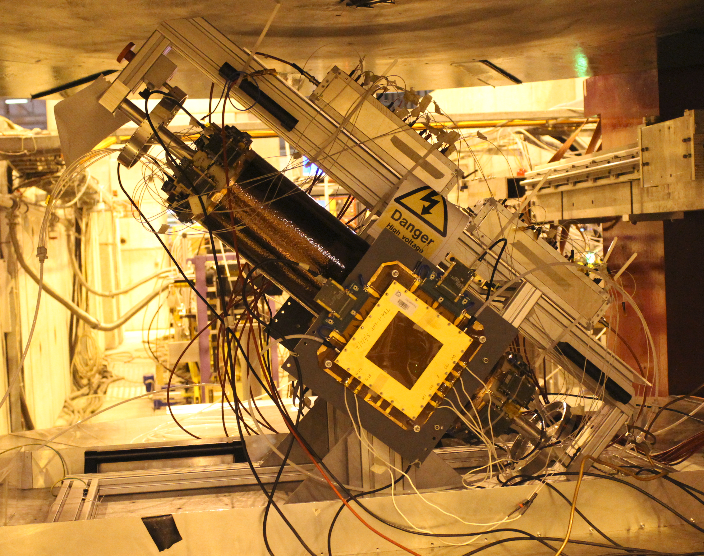}
\end{tabular}
\caption[Test beam setup]{A representation of a test beam setup is shown (top). A picture of the planar triple-GEM detector instrumented (bottom left).  A picture of the CGEM detector in a test beam are shown (bottom right).}
\label{fig:setup}
\end{figure}

The CERN facility allows to test the chambers in magnetic field thanks to Goliath \cite{ref4:goliath}, the largest ferromagnetic dipole magnet in the world providing a magnetic field up to 1.5 T in both polarities that can recreate an environment similar to BESIII. Let's define the $z$ axis the one along the beam direction, $x$ axis the one perpendicular to $\mathbf{B}$ and $y$ axis parallel to $\mathbf{B}$. The beam direction is orthogonal to the GEM detector plane and the magnetic field $\mathbf{B}$ is perpendicular to the electric field $\mathbf{E}$ direction and parallel to the GEM plane. Figure \ref{fig:setup} represents the setup. Some measurements have been performed with a rotation of the chambers around the vertical axis in order to characterize the chamber with inclined tracks.  A trigger system composed of two scintillator bars readout by photomultiplier tubes has been placed outside the magnet.

A similar setup has been used in the MAMI facility, where the magnet was not present and the setup was more compact.

\subsection{Readout electronics}
The design of the TIGER chip has been carried out from 2014 to 2018, therefore all the detector characterization has been performed using available commercial electronics, the APV25 developed by CMS collaboration. Only in 2017-2018 the test and integration of TIGER and triple-GEM started.

\subsubsection{APV25}
\label{sec:apv}
Some details about the APV25 \cite{ref4:apv25,ref4:apv25_2} and TIGER \cite{ref4:tiger1,ref4:tiger2} can be found in the references but some details has to be provided to understand the reconstruction technique. The APV25 chip has 128 channels connected to strips with a maximum capacitance of $100 \, \mathrm{pF}$ and it samples the charge up to $50-60 \, \mathrm{fC}$ each $25 \, \mathrm{ns}$ for 27 times after the trigger signal. The time is referred to the trigger time. The charge amplitude is digitized in 1800 bins. A conversion factor of $30 \, \mathrm{ADC}$ counts per fC is used \cite{ref4:apv_cf}. The acquisition is performed with SRS \cite{ref4:SRS} technology and mmDAQ software has been used to acquire the data from the Front-End Card (FEC). This allowed to record the information with an online pedestal subtraction.

\subsubsection{TIGER}
The TIGER chip has 64 channels and it is mounted on a Front-End Board (FEB). Each FEB hosts two chips in order to measure the information of 128 channels. It has two threshold levels: the first one used to measure the time once the signal overstep it; the second to discriminate the signal from the noise. The chip has an analogue readout and it can measure the charge in two different modes: sampling and hold (S/H) mode and time-over-threshold (ToT) mode. The output is sent out digitally and decoded in 10 bits. Other details have been reported in Sect. \ref{sec:tiger}.

\section{The offline software: GRAAL} 
\label{sec:reconstrution_method}
\begin{figure}[ht]
\centering
\includegraphics[width=0.6\textwidth]{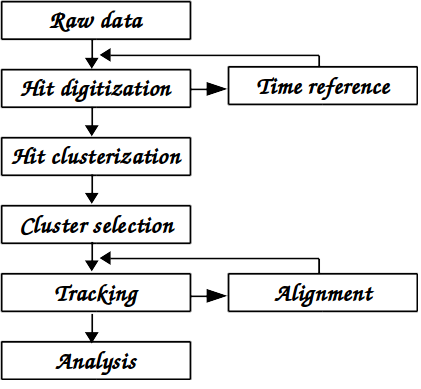}
\caption[Offline software block diagram]{Offline software block diagram.}
\label{fig:block}
\end{figure}

\footnotetext{Gem Reconstruction And Analysis Libraries}
The test beam data have been reconstructed and analyzed through a software developed for this purpose. A scheme of the offline software is shown in Fig. \ref{fig:block}. Each trigger signal defines an {\it event}, for each event the signal on a strip is recorded. This signal is labelled {\it hit}. Hit information is reconstructed in order to obtain the time and charge information. The hit digitization depends on the front-end electronics.

Hits with charge greater than $1.5 \, \mathrm{fC}$ have been used in the analysis. Contiguous hits in the same detector and same view have been used to define a {\it cluster} of strips. If dead strips are present the clusterization algorithm takes them into account and it clusterizes the hits even if there is an empty strip between strips with signal. The charge of the cluster $Q_{\mathrm{cluster}}$ is defined by the sum of the charges of the hits and the information of each hit is used to evaluate the position measurement as described in Sect. \ref{sec:cc_tpc}.

\subsection{Hit digitization with APV25}
The APV25 chip acquires the signal shape sampling every $25 \, \mathrm{ns}$. The total charge induced on the strip is the maximum charge measured in the 27 time-bins, $Q_{\mathrm{hit}}~=~Q_{\mathrm{max}}$, while the time needs to be extracted from the time profile of the charge through a fit with a Fermi-Dirac (FD) function on the rising edge of the signal, as described in Eq. \ref{eq:FD_equation}. 
\begin{equation}		
\label{eq:FD_equation}
Q(t)=Q_0+\dfrac{Q_{\mathrm{max}}}{1+\exp \left(- \dfrac{t-t_{\mathrm{FD}}}{\sigma_{\mathrm{FD}}} \right)}
\end{equation}

\begin{figure}[tp]
\centering
\includegraphics[width=\textwidth]{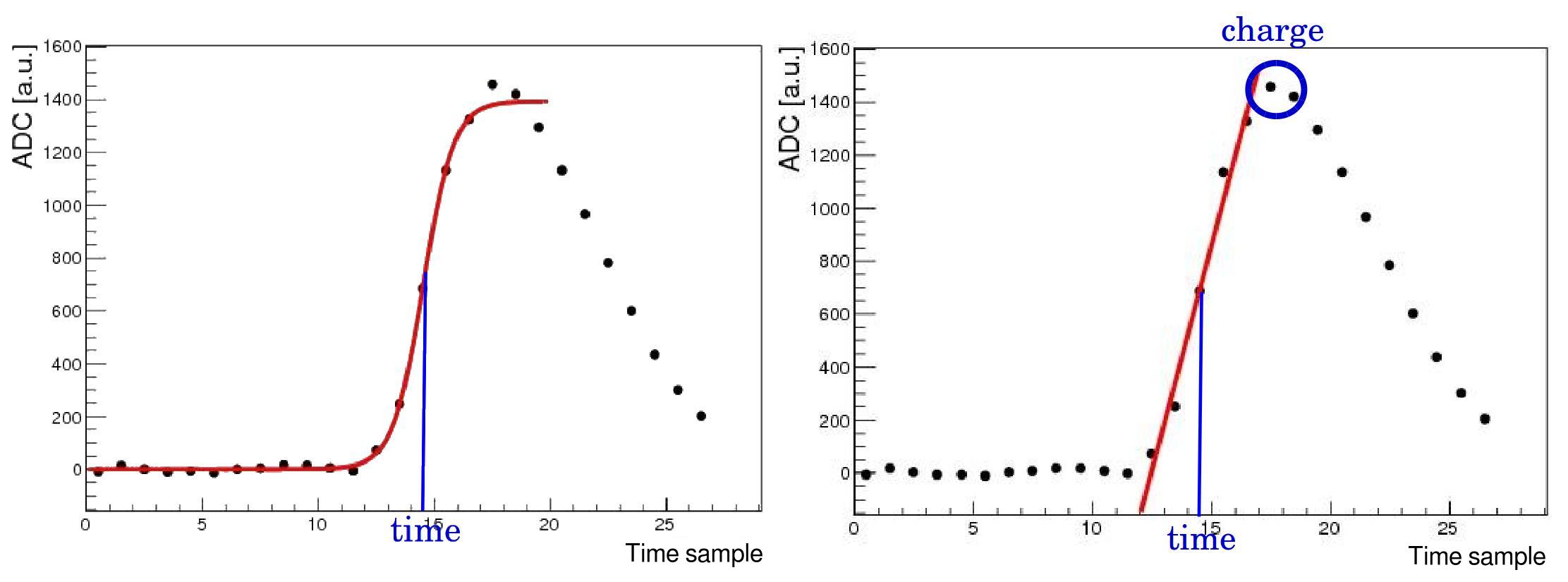}
\caption[Fermi-Dirac and linear fit in time measurement]{Fermi-Dirac (left) and linear (right) fits of a hit signal.}
\label{fig:FD_linear}
\end{figure}

First a pre-analysis in needed to measure the noise level $Q_0$ and the time-bin where the signal is recorded: this is necessary to initialize the FD fit that otherwise would not converge. $Q_0$ is initialized to the charge mean value of the first three time-bins, $t_{\mathrm{FD}}$ to the time mean value between the time-bin associated to $Q_{\mathrm{max}}$ and the time-bin with 10\% of $Q_{\mathrm{max}}$, $\sigma_{\mathrm{FD}}$ to $12.5 \, \mathrm{ns}$. The time related to the signal is assumed to be the the inflexion point, the value in the middle of the signal rise: $t_{\mathrm{hit}}=t_{\mathrm{FD}}$. If the FD fit fails or if converge to anomalous values then a linear fit is used to extract the time information. The hit time $t_{\mathrm{hit}}$ is evaluated with the fit in the middle of the rising edge. Examples of those two fit are shown in Fig. \ref{fig:FD_linear}.
The time-bin amplitude of $25 \, \mathrm{ns}$ introduces uncertainties in the time measurement. In the standard APV25 setup the trigger signal is used to start the data acquisition of each event. The setup has been upgraded to measure with higher precision the trigger time through the injection of this signal into an APV25 channel with a proper circuit \footnote{Using an attenuator (10:1) in series with a 1 pF capacitance.}. Then the trigger signal has been reconstructed in the same way as the standard hits and it has been used as reference for the hit time. 

\subsection{Hit digitization with TIGER}
\label{sec:tiger_rec}
\begin{figure}[htp]
\centering
\includegraphics[width=0.7\textwidth]{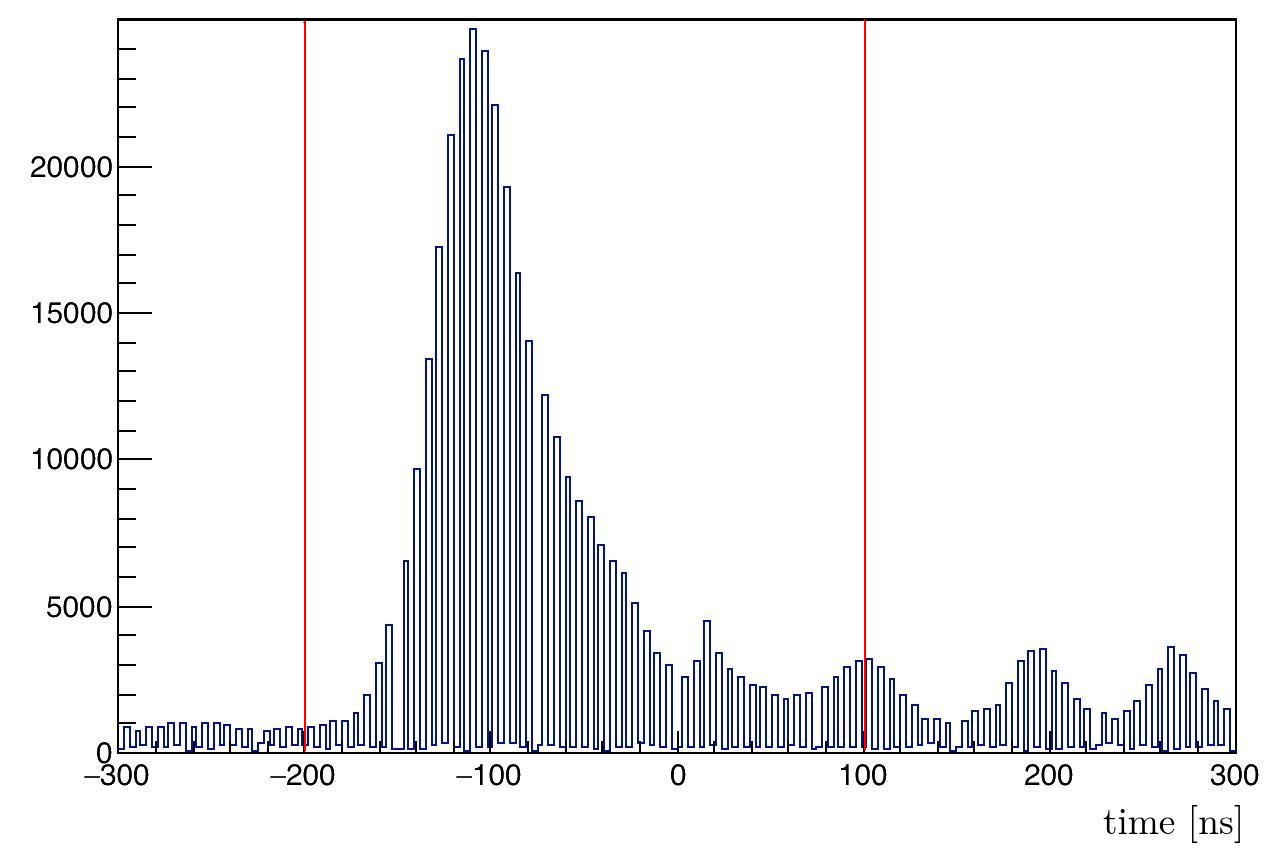}
\caption[Time distribution with TIGER]{Hit time distribution of a triple-GEM with $5 \, \mathrm{mm}$ drift gap and a beam incident angle of 0$^\circ$ with TIGER electronics.}
\label{fig:time_distr_tiger}
\end{figure}

TIGER chip is a technology that can provide the hit information with and without an external trigger signal, named $trigger-match$ and $trigger-less$ modes respectively. Every time that a signal oversteps the threshold the TIGER measures the charge and time informations. In trigger-match mode an external trigger is used as signal for the front-end board to collect in a packet the information from each channel, in trigger-less mode a continuous stream of data is sent out from the TIGER to the front-end board and then to the acquisition computer. At the time of the test beam the trigger-match mode was not ready yet then the data acquisition was performed in trigger-less mode. To reconstruct the event was mandatory to inject the trigger inside a channel of the TIGER. Offline reconstruction has been performed to build each event with respect to the trigger time. The time window around the trigger time spans $300 \, \mathrm{ns}$. Figure $\ref{fig:time_distr_tiger}$ shows the hit time distribution with respect to the trigger time. 

The APV25 can not be used in BESIII because it is an IBM technology and it can not be exported to China but the custom design of TIGER grants additional features: it reduces the capacitance between the detector and the ASIC and moreover it has been designed with the needed minimal dimension.

\subsection{Hit clusterization}
\label{sec:cc_tpc}

The strip position is defined as the coordinate in the middle of the strip, orthogonal to the strip length direction.
To measure the incident particle position two different algorithms have been deployed, the charge centroid (CC), also known as center of gravity, that uses the strip charge and position information and the micro-Time Projection Chamber ($\upmu$TPC), that uses also the time information.

\subsubsection{The charge centroid}

The CC method measures the position of the particles with a weighted average as described in the Eq. \ref{eq:CC}.
\begin{equation}		
\label{eq:CC}
x_{\mathrm{CC}}=\dfrac{\sum_{i}^{N_{\mathrm{hit}}}Q_{\mathrm{hit},i} \, x_{\mathrm{hit},i}}{\sum_{i}^{N_{\mathrm{hit}}} Q_{\mathrm{hit},i}}
\end{equation}
where $N_{\mathrm{hit}}$ is the number of hits in the cluster, referred as {\it cluster size}, $x_{\mathrm{hit},i}$ and $Q_{\mathrm{hit},i}$ the hit position and charge. 

\subsubsection{The micro-Time Projection Chamber}

The time information is used to reconstruct the particle path since the invention of drift chambers but in the last decade the ATLAS collaboration introduced the idea of using the same technique in MPGD technology, $e.g.$ the MicroMegas detector. The few millimeters of drift gap are used to reconstruct the particle path. Very important contributions in this field come from \cite{ref4:kostas,ref4:kostas2} that define the state of the art. A detailed study has been performed in this thesis work to develop a similar technique for triple-GEM detectors that shares some features with MicroMegas. 

\begin{figure}[tp]
\centering
\includegraphics[width=\textwidth]{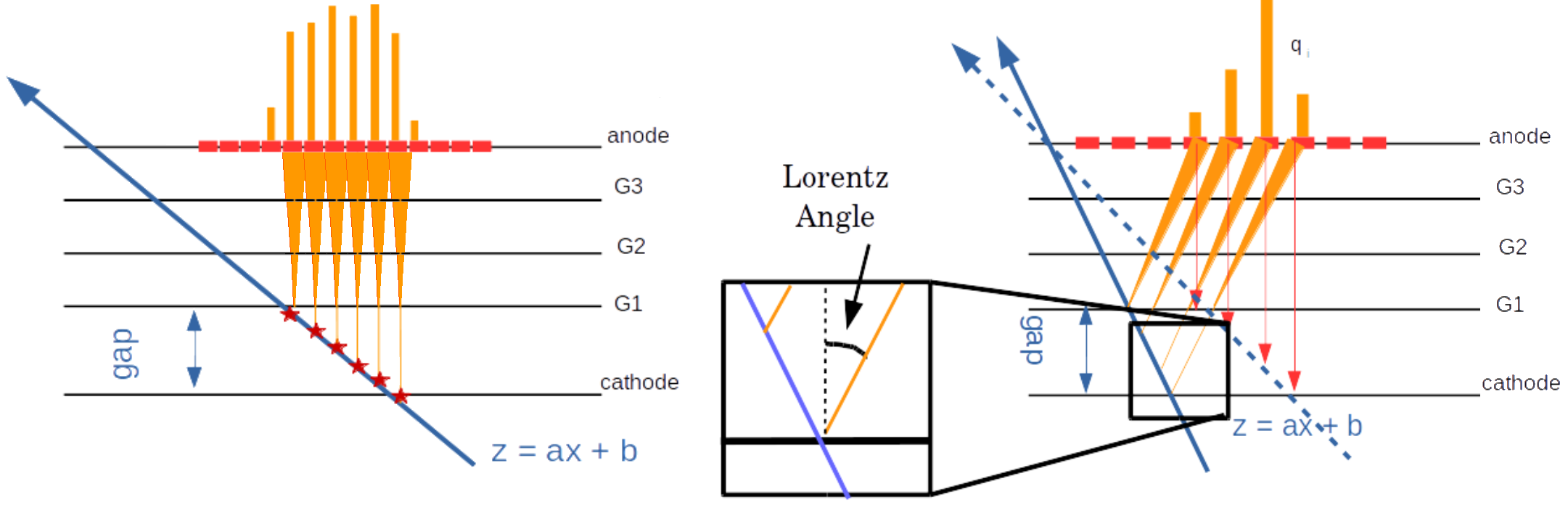}
\caption[$\upmu$TPC reconstruction]{Representations of the signal amplification in a triple-GEM and the $\upmu$TPC reconstruction with magnetic field (right) and without (left). Continuous blue arrow represents the charge particle, stars are the primary ionization, orange areas are the electrons avalanches, red rectangles are the strips and the orange bars are the charge collected on each strip and dashed arrow is the reconstructed line without magnetic field correction.}
\label{fig:TPC_noB_yesB}
\end{figure}

The $\upmu$TPC algorithm reconstructs the particle path inside the drift gap of the triple-GEM: strips are associated to points in x:z plane, named $\upmu TPC~point$, where $x_{\mathrm{hit}}$ is the strip position and $z_{\mathrm{hit}}$ is the product of the hit time $t_{\mathrm{hit}}$ times the electron drift velocity. The drift velocity $v_{\mathrm{drift}}$ is computed by from Garfield++ \cite{ref4:garfield} simulation\footnote{Magboltz program is the one related to the simulation of motion of the electron in gas sensitive in Garfield++.}. It is a quantity that depends on the electro-magnetic field in the drift gap and on the gas mixture. If the cluster size is at least two, a linear fit is performed with the $\upmu$TPC points and the value at the middle of the gap is chosen as the position measured by the $\upmu$TPC algorithm.
\begin{equation}		
\begin{tabular}{cc}
$z_{\mathrm{hit}}=t_{\mathrm{hit}} \, v_{\mathrm{drift}}$~;~$x_{\mathrm{\upmu TPC}}=\dfrac{gap/2 - b}{a}$
\end{tabular}
\label{eq:TPC}
\end{equation}

where $gap$ is the drift gap thickness, $a$ and b the linear fit parameters as shown in Fig. \ref{fig:TPC_noB_yesB}.

To improve the $\upmu$TPC reconstruction it is important to assign to the $\upmu$TPC points proper uncertainty. Along the $z$ direction the fit uncertainty on the $t_{\mathrm{FD}}$ measurement has been considered. In the $x$ direction the pitch over $\sqrt{12}$ is considered as uncertainty because the primary ionization that generates the main signal could start in any position within the strip pitch. Studies from ATLAS \cite{ref4:kostas} show that the position distribution of the primary electrons in not flat but this will be discussed in App. \ref{sec:capacitive}. Moreover, strips with lower $Q_{\mathrm{hit}}$ have a higher probability to have a cross-talk contribution from the neighboring strips with higher charge. As it will be described in Sect. \ref{sec:result_noB} and discussed in App. \ref{sec:capacitive}, the signal generated from a primary electron is mainly collected on the strip below its position (if no magnetic field is present) but a certain fraction of the signal is collected on the neighboring strips: the higher is the fraction of charge carried by a strip, the smaller is the error associated on the $x$ coordinate. Due to this behavior an charge-dependent error is used on $x_{hit}$. Eq. \ref{eq:TPC_errors} describes the errors associated to each $\upmu$TPC point.

\begin{equation}		
\begin{tabular}{ccc}
$dz_{\mathrm{hit}}=dt_{\mathrm{FD}} \, v_{\mathrm{drift}}$ & $dx_{\mathrm{\upmu TPC}}=\sqrt{  \left(\dfrac{\mathrm{pitch}}{\sqrt{12}}\right)^2 + \left(\dfrac{\mathrm{pitch}}{\sqrt{12}} \, \dfrac{Q_{\mathrm{cluster}}}{N_{\mathrm{hit}} \, Q_{\mathrm{hit}}} \right)^2  }$
\end{tabular}
\label{eq:TPC_errors}
\end{equation}

\subsection{Time reference for $\upmu$TPC reconstruction}
\label{sec:time_ref}
\begin{figure}[htp]
\centering
\includegraphics[width=0.7\textwidth]{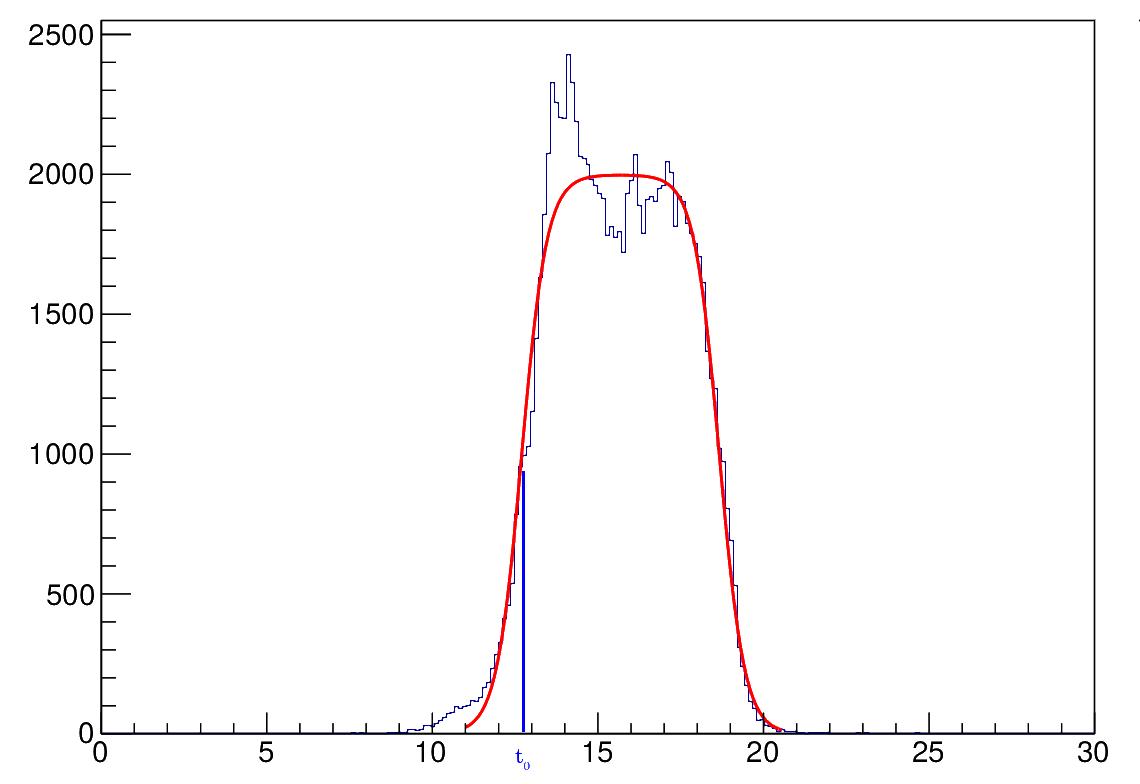}
\caption[Time distribution with APV25]{Hit time distribution of a triple-GEM with $5 \, \mathrm{mm}$ drift gap and a beam incident angle of 45$^\circ$ with APV25 electronics.}
\label{fig:time_distr}
\end{figure}

As it will described in Sect. \ref{sec:cc_tpc}, the $\upmu$TPC algorithm reconstructs the particles path in the drift volume. The larger signal in a triple-GEM comes from the electrons amplified three times, bi-GEM effects are less than 2\% \cite{ref4:poli}.
The hit time is the interval between the primary ionization and the signal induction on the anode. The time distribution of the hits is studied to evaluate the drift time needed from the first GEM to the anode.
Figure \ref{fig:time_distr} shows the time distribution of the hits in a run. The histogram is fitted with two FD function to describe the rising and falling edges. The time in the middle of the rise is associated to the fastest electrons: the closest to the first GEM. This is chosen as $t_0$. The time in the middle of the falling edge corresponds to the slowest electron generated close the cathode. The time distribution shows peaks on the top due to capacitive effects. This will be discussed in App. \ref{sec:capacitive}. The $t_0$ measurement is performed with a sub-sample of the run entries, 10\% of the total number of events, to avoid bias.

\subsection{Residual measurement}

The test beam setup is composed by several triple-GEM detectors and they are divided in two groups: the {\it trackers} used to measure the particle path along the setup and the {\it test detectors} that are used to determine the performance of this technology. From three to seven triple-GEM detectors have been used in the tests beam. The outer detectors are usually used as trackers, the more trackers are used the more precise is the track position measurement. A linear fit is performed from the tracker position measurement, even if the magnetic field is present and the particle path is bent. 
Once the track is defined, the expected position $x_{\mathrm{expected}}$ at the test detector plane $x_{\mathrm{expected}}$ is evaluated and it is used to measure the residual $\Delta x_{\mathrm{tracker}}$ as shown in Eq. \ref{eq:residual_trk}.

\begin{equation}		
\Delta x_{\mathrm{tracker}} = x_{\mathrm{detector}} - x_{\mathrm{expected}}
\label{eq:residual_trk}
\end{equation}

The residual distribution has a Gaussian shape and a Gaussian fit is used to describe it. The $\sigma_{\mathrm{tracking}}$ of the Gaussian is related to the spatial resolution of the detector but $\Delta x_{\mathrm{tracker}}$ distribution is not used for this purpose because it contains the contributions of the tracking system. $\Delta x_{\mathrm{tracker}}$ is used in the alignment procedures. Another technique has also been used to characterize and extract the detector performance.

The residual distribution of the two detectors in the middle of the setup is used to measure their spatial resolution and efficiency as shown in Eq. \ref{eq:efficiency}: 

\begin{equation}		
\Delta x_{1,2} = x_{\mathrm{detector},1} - x_{\mathrm{detector},2}
\label{eq:residual_enemy}
\end{equation}

This technique allows to remove in a easier way the systematic error due to tracking system on the spatial resolution thought the assumption that the two detectors are equal and they have the same performance. Then the spatial resolution of the two detectors can be evaluated with a Gaussian fit of the $\Delta x_{1,2}$ distribution and the $\sigma_{\mathrm{Gaussian}}$ is divided by $\sqrt{2}$ as shown in Eq. \ref{eq:resolution_enemy}:

\begin{equation}		
\begin{tabular}{ccc}
$\sigma_{\mathrm{residual}}^2 = \sigma_{\mathrm{detector},1}^2 + \sigma_{\mathrm{detector},2}^2$ & \, if \, &$ \sigma_{\mathrm{detector},1} = \sigma_{\mathrm{detector},2}=\sigma_{\mathrm{detector}}$\\ & \\$ \rightarrow \sigma_{\mathrm{detector}} = \dfrac{\sigma_{\mathrm{residual}}}{\sqrt{2}}$
\end{tabular}
\label{eq:resolution_enemy}
\end{equation}

\subsection{The alignment procedures}

\begin{figure*}[ht!]
\begin{tabular}{ll}
\includegraphics[width=0.5\textwidth]{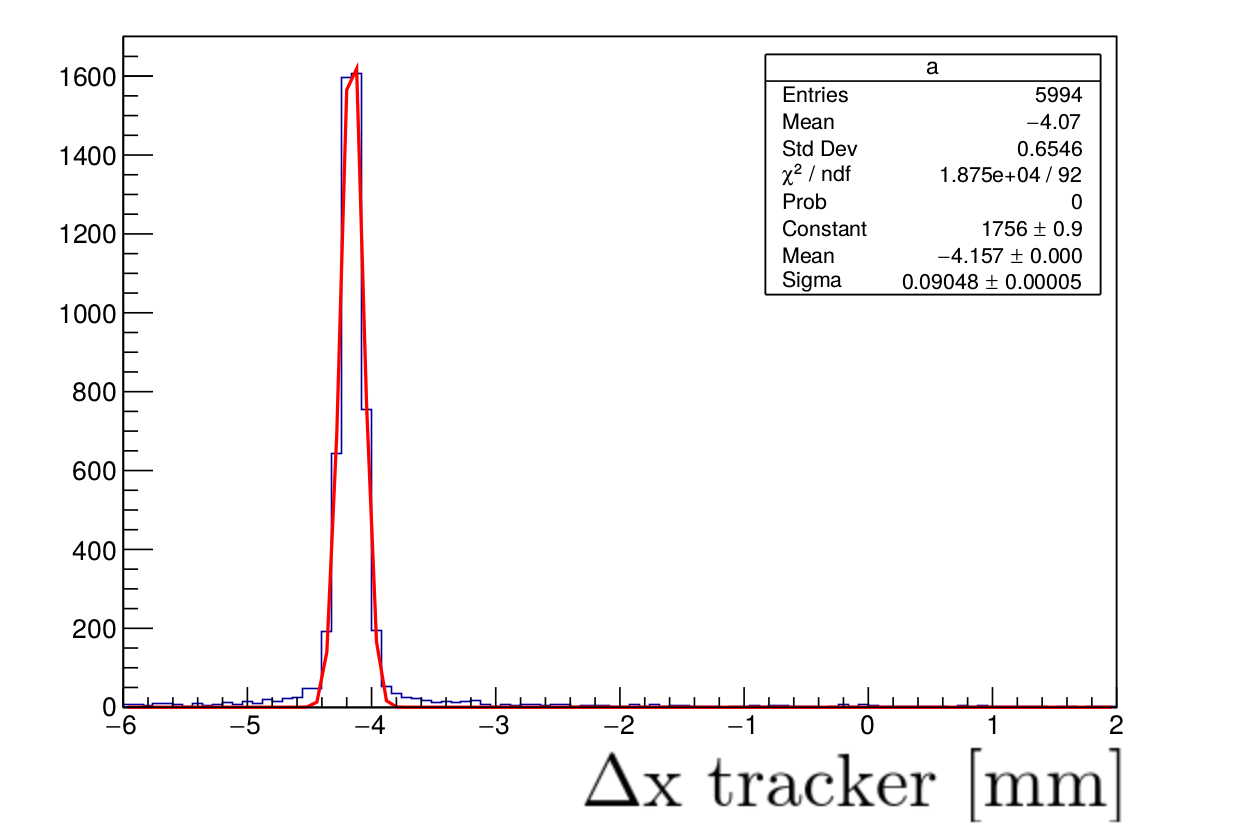} & 
\includegraphics[width=0.5\textwidth]{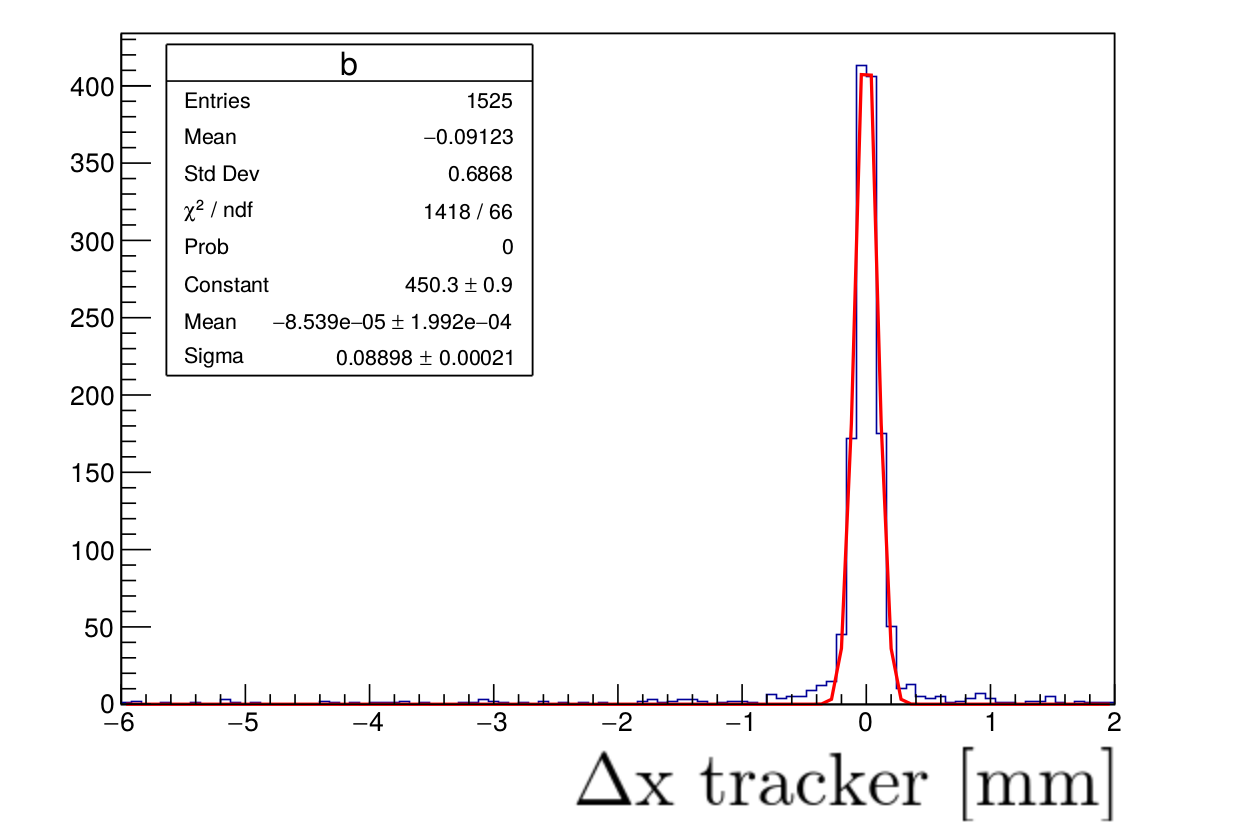} \\
\includegraphics[width=0.5\textwidth]{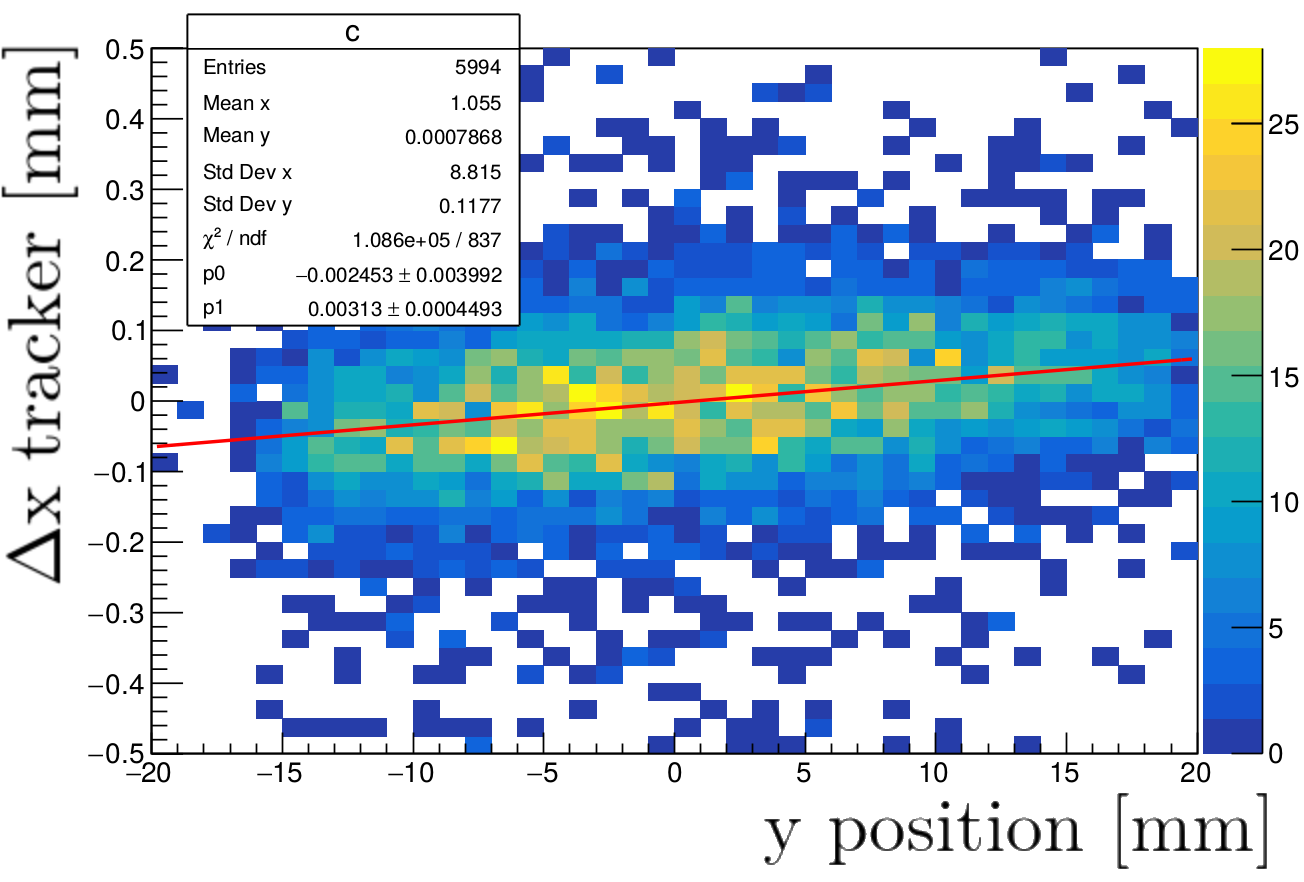} & 
\includegraphics[width=0.5\textwidth]{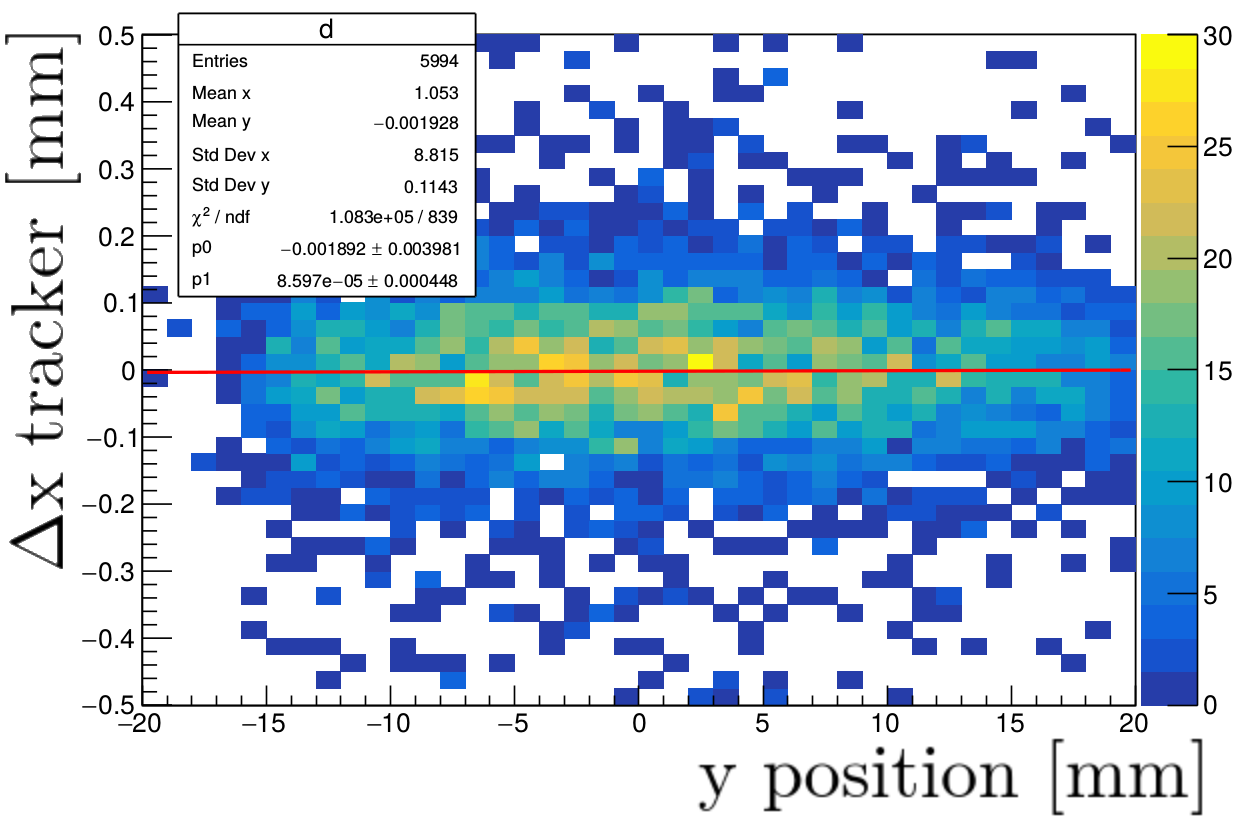} \\
\includegraphics[width=0.5\textwidth]{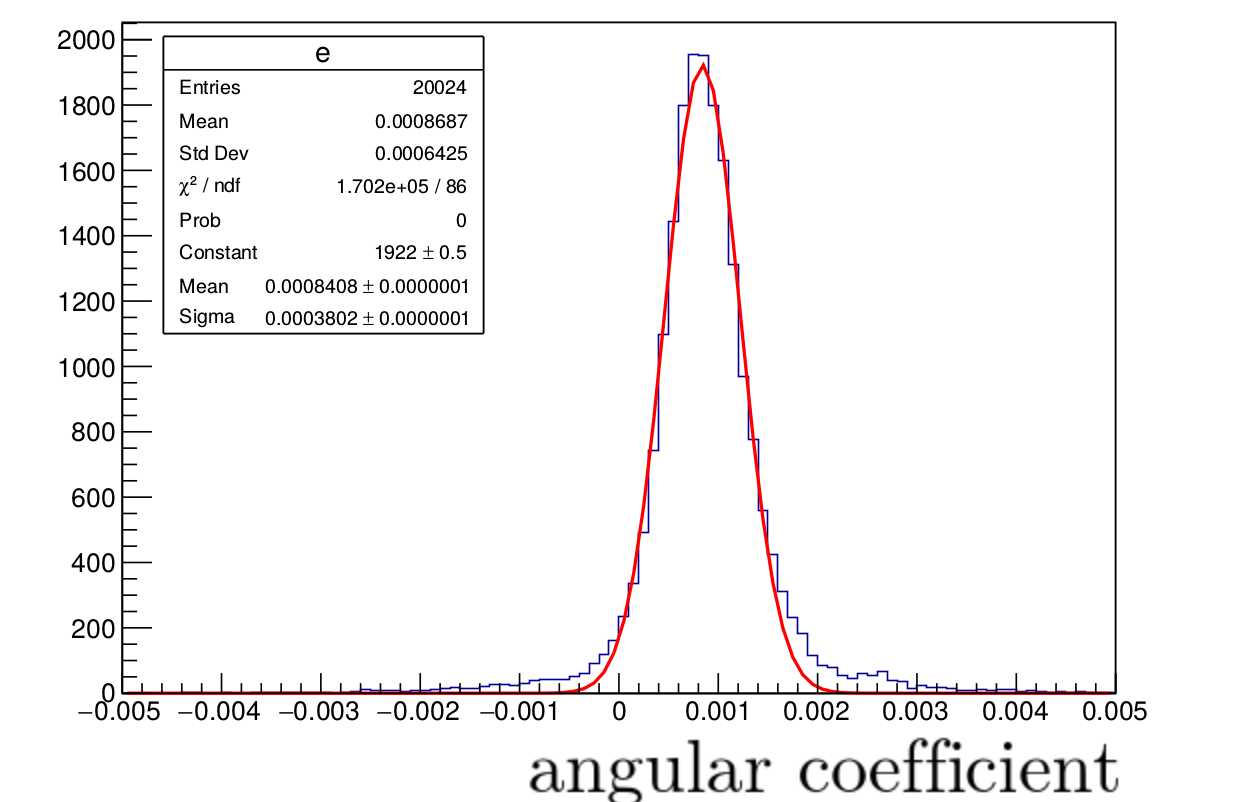} &
\includegraphics[width=0.5\textwidth]{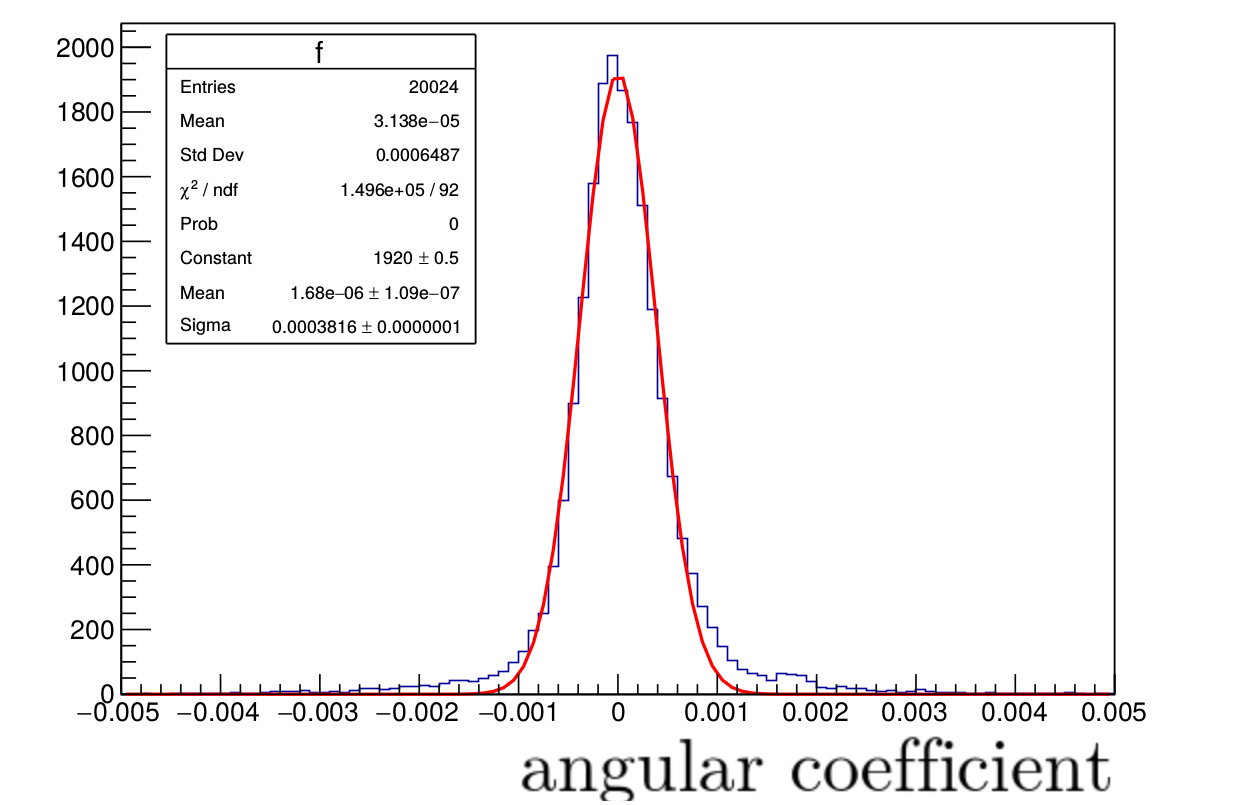}  
\end{tabular}
\caption[Alignment plots]{Alignments plots examples before the alignment (left) and after it (right). The first row shows the $\Delta x_{\mathrm{tracker}}$ to remove the shift between the real and reconstructed position. The second row shows the tilt in $xy$ plane. The third row shows the track angular coefficient measured by the tracking system.}
\label{fig:align}
\end{figure*}

Alignment procedures are needed to remove the translations and rotations of the real detector position with respect to the one reconstructed by the software. The detectors are installed perpendicular to the beam line (or rotated with respect this position in certain runs) with a precision within the millimeter. This can affect the resolution measurements. A sub-sample of the event corresponding to the 10\% of the total event number is used in the alignment procedures. Three iteration cycle are needed to the alignment procedures before the analysis of the entire run. The quantities under study are the shifts along the $x$ and $y$ coordinate, tilts in the $xy$ plane and rotation in the $xz$ plane.

The shift alignments are the easiest to be performed: once the residual distribution $\Delta x_{\mathrm{tracker}}$ is evaluated and the Gaussian fit is performed then the central value of the Gaussian is used to shift the detector reference frame and remove the difference between the real position of the detector and the reconstructed one.
The tracking system itself needs an alignment, the angular coefficient of the track with the entire setup has to be perfectly orthogonal and being the resolution of the tracking system below the mrad then it is possible to rotate the entire setup in order to reconstruct precisely the track incident angle.
The tracking system needs also an alignment: each detector is referred to the first tracker that has been fixed in (0,0,0) reference frame.
 
Those three are the macroscopic alignment. Now fine alignments, the ones related to the detector rotations, are needed. The first one is the evaluation of the tilt in the $xy$ plane. A bi-dimensional plot $\Delta x_{\mathrm{tracker}}:y_{\mathrm{detector}}$ is performed to do this alignment where $y_{\mathrm{detector}}$ is the position measurement\footnote{$y$ measurement is performed in the same way as the $x$ measurement.}. A linear fit of the bi-dimensional distribution is performed and the angular coefficient of the line is used to rotate the detector in the $xy$ plane.
Similarly, it is evaluated the tilt of the x coordinate of the tested detector with respect to the x coordinate of the first tracker. Similarly to the the previous measurement, a bi-dimensional plot is drawn for $\Delta x_{\mathrm{tracker}}:x_{\mathrm{tracker}}$ and a linear fit is used to measure the rotation between the two detectors.

\section{Analysis results}
\label{Analysis_results}
A wide range of configurations have been studied for this PhD. Several hundreds of runs have been collected in order to measure the behavior of triple-GEMs in different conditions and to perform calibrations of the reconstruction algorithms to cross-check the results. The main goal is to find the optimal parameters to fulfill the BESIII requirements with a triple-GEM operated in $1 \, \mathrm{T}$ magnetic field. A selection of the obtained results will be shown in the next sections: the detector efficiency at different gain; signal information ($e.g$ cluster charge and size) as a function of the gain, the beam incident angle and particle rate; spatial resolution in different conditions ($e.g.$ gain, beam incident angle and magnetic field).
\subsection{Detector efficiency and signal characterization}
It is important that a detector used in high energy physics has an efficiency higher as possible to detect almost every particle. In Argon gas the number of electrons per millimeter is about 8, if CO$_2$ or iC$_4$H$_{10}$ are mixed in the Argon this quantities increases. The probability to have at least one ionizing interaction is almost 100\% from Eq. \ref{eq:det_eff}. Once the ionization occurs the signal has to be amplified depending by many factors. Those detail will be discussed in Sect. \ref{sec:digi}.

To evaluate the efficiency the $\Delta x_{1,2}$ distribution is fitted with a Gaussian: the number of events within 5$\sigma_{\mathrm{Gaussian}}$ from the mean value is the number of times with a successful reconstruction in the two test detectors. This number is related to the number of event where the tracking system has a good event and each tracker has an inclusive residual distribution within 5$\sigma_{\mathrm{tracking}}$. If the two detectors have the same behavior then the detector efficiency can be evaluated from Eq. \ref{eq:efficiency}:
\begin{equation}		
\begin{tabular}{c}
$\varepsilon_{1\& 2}=\varepsilon_1 \, \varepsilon_2=\dfrac{N_\varepsilon}{D_\varepsilon}$\\
~\\
if $\varepsilon_1=\varepsilon_2=\varepsilon$\\
~\\
$\varepsilon=\dfrac{N_\varepsilon/D_\varepsilon}{\sqrt{2}}$
\end{tabular}
\label{eq:efficiency}
\end{equation}

where N$_\varepsilon$ and D$_\varepsilon$ are:
\begin{itemize}
\item $D_\varepsilon$ = N$^\circ$ of events with a good tracking system and the tracker events are within 5$\sigma_{\mathrm{tracking}}$ in their inclusive residual distribution $\Delta x_{\mathrm{tracker}}$;
\item $N_\varepsilon$ = N$^\circ$ of event with a good tracking system and $\Delta x_{1,2}$ within 5$\sigma_{\mathrm{Gaussian}}$ where $\Delta x_{1,2}$ is measured with both test detectors working properly.
\end{itemize}

\begin{figure}[ht!]
\centering
\includegraphics[width=0.7\textwidth]{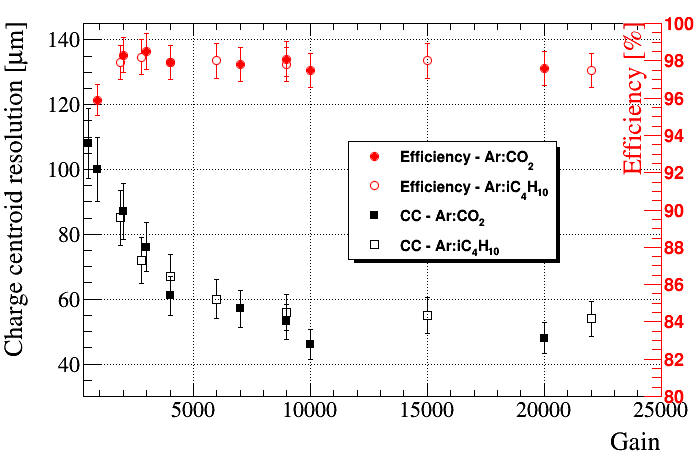}
\caption[Efficiency and CC resolution results]{Efficiency and CC resolution as a function of the effective detector gain in Ar+30\%CO$_2$ and Ar+10\%iC$_4$H$_{10}$ gas mixtures with $5 \, \mathrm{mm}$ drift gap. The incident particle are orthogonal to the detector. No magnetic field is present.}
\label{fig:eff_res}
\end{figure}

Figure \ref{fig:eff_res} shows the efficiency and the resolution of the CC algorithm as a function of the gain for both the studied gas mixtures. As expected from \ref{eq:det_eff}, the highest efficiency is reached almost immediately at a gain of 2000 while the CC spatial resolution achieves $50 \, \mathrm{\upmu m}$ at a gain of 10000. No $\upmu$TPC spatial resolution is shown right now because the algorithm does not work properly with orthogonal tracks and no magnetic field. 

\begin{figure}[ht!]
\centering
\includegraphics[width=0.7\textwidth]{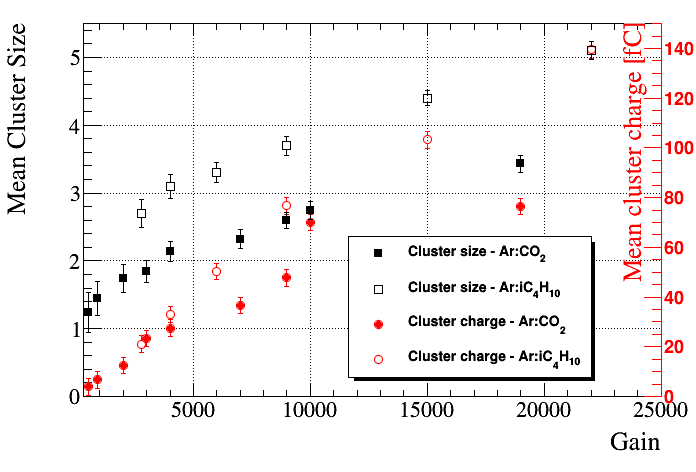}
\caption[Cluster size and charge results]{Cluster size and charge as a function of the effective detector gain in Ar+30\%CO$_2$ and Ar+10\%iC$_4$H$_{10}$ gas mixtures with $5 \, \mathrm{mm}$ drift gap. The incident particle is orthogonal to the detector. No magnetic field is present.}
\label{fig:q_nHit_noB}
\end{figure}

Figure \ref{fig:q_nHit_noB} shows the behavior of the cluster size and charge as a function of the gain. The charge has linear dependency on the gain, with a slope different between the two gas mixtures since the number of electrons generated in the ionization is about 55.1 in Ar+10\%iC$_4$H$_{10}$ and 38.4 in Ar+30\%CO$_2$, evaluated from Eq. \ref{eq:n_ele_ariso}. The cluster size, the number of fired strip, is related to the diffusion properties of the electrons in the gas. The higher the transverse diffusion the larger the cluster size. Both gas mixtures increase the cluster size with the gain but the Ar+10\%iC$_4$H$_{10}$ is larger because its transverse diffusion in around $350 \, \mathrm{\upmu m }$ in a cm while Ar+30\%CO$_2$ $250 \, \mathrm{\upmu m}$ \cite{ref4:sharma_gas}.

The results in Fig. \ref{fig:q_nHit_noB} have been measured with orthogonal tracks and no magnetic field: if the primary electrons are generated on the same line when they drift to the first GEM  they reach it in the same point, within diffusion effects. If the detector is set at high voltage value around 10000 then the electrons starting the avalanche in the same point of the first GEM generate their signal on three strips. This behavior has to be underlined to understand the further results.

Smaller cluster size and charge values have been obtained with $3 \, \mathrm{mm}$ drift gap detectors since the number of primary electrons and the path to diffuse are smaller.

\subsection{Performance of a triple-GEM  in magnetic field}
The triple-GEM detector has been proposed as inner tracker for the BESIII experiment and it has to operate in 1 T magnetic field. A characterization as a function of the magnetic field has been performed. If the magnetic field is present then the Lorentz force acts on the electrons and it bends their path. The reconstruction algorithms has to take this effect into account because the magnetic field changes the diffusion properties of the electron and, more important, their path. The $Lorentz~angle$ ($\theta_{\mathrm{Lorentz}}$) is the angle between the electric field and the electrons path in the gas and it is used to determine the reconstruction properties as shown below.

\begin{figure*}[ht!]
\begin{tabular}{ll}
\includegraphics[width=0.5\textwidth]{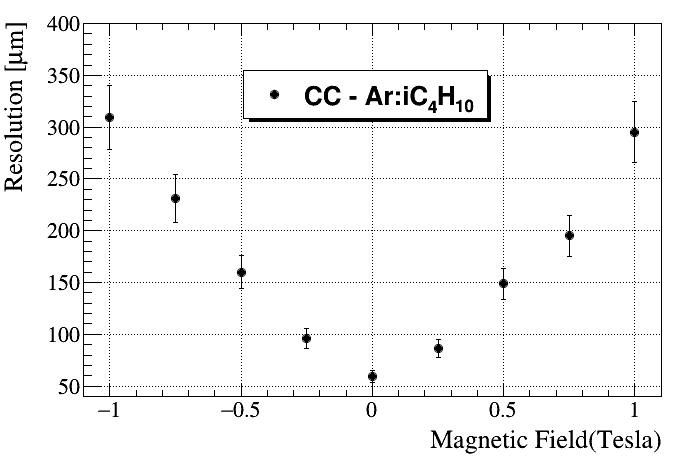} & 
\includegraphics[width=0.5\textwidth]{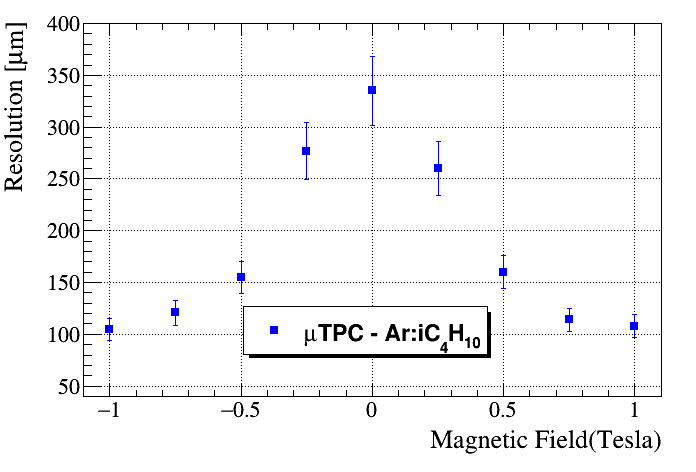} \\
\end{tabular}
\caption[CC and $\upmu$TPC in magnetic field]{Triple-GEM spatial resolution as a function of the magnetic field for the two algorithms: CC (left) and $\upmu$TPC (right).}
\label{fig:res_yesB}
\end{figure*}

In the orthogonal track configuration, the spatial distribution of the signal is larger if the magnetic field is present because the primary electrons position reaching the first GEM is spread. Similarly the case without magnetic field and non orthogonal tracks, the position width where the electrons arrive on the first GEM is $gap \, \tan(\theta_{\mathrm{Lorentz}})$. This effect explains the behavior of the CC and $\upmu$TPC shown in Fig. \ref{fig:res_yesB}. 

The $\mathrm{\upmu}$TPC reconstruction in magnetic field, as shown in Fig. \ref{fig:TPC_noB_yesB}, needs a correction that takes into account the shift due to the Lorentz force as shown in Eq. \ref{eq:TPC_yesB}:

\begin{equation}
\label{eq:TPC_yesB}
x_{\mathrm{\upmu TPC}}^{B_{\mathrm{field}}} = \dfrac{gap/2-b}{a} \pm \sqrt{ \left( \dfrac{gap}{2 \, \cos(\theta_{\mathrm{Lorentz}})} \right)^2 - \left( \dfrac{gap}{2} \right)^2}
\end{equation}

where $\pm$ depends on the magnetic field direction.

\subsection{Performance of a triple-GEM with sloped tracks}
\label{sec:result_noB}

\begin{figure}[htp]
\centering
\includegraphics[width=0.7\textwidth]{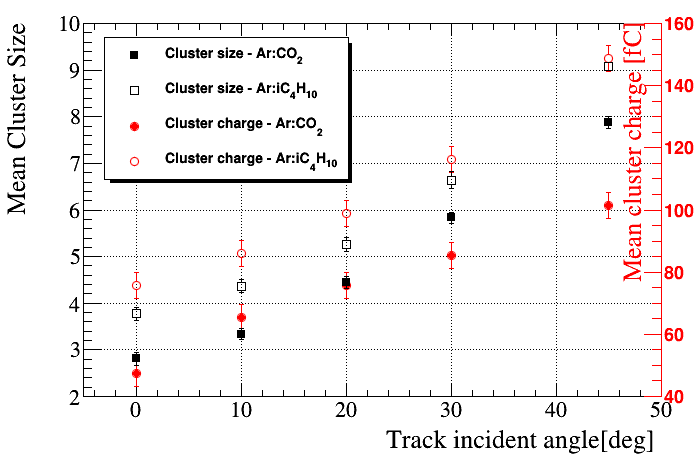}
\caption[Cluster size and charge results]{Cluster size and charge as a function of the incident angle in Ar+30\%CO$_2$ and Ar+10\%iC$_4$H$_{10}$ gas mixtures with $5 \, \mathrm{mm}$ drift gap. No magnetic field is present.}
\label{fig:q_nHit_yesRot}
\end{figure}

At BESIII low momenta particles interact with the detector with an angle different from zero. It is important to guarantee an efficient reconstruction in this case. A characterization of the triple-GEM detector as a function of the incident angle has been performed to determine the signal shape ($Q_{cluster}$ and $N_{hit}$ as shown in Fig. \ref{fig:q_nHit_yesRot}. 

\begin{figure}[tp]
\centering
\begin{tabular}{cc}
\includegraphics[width=0.5\textwidth]{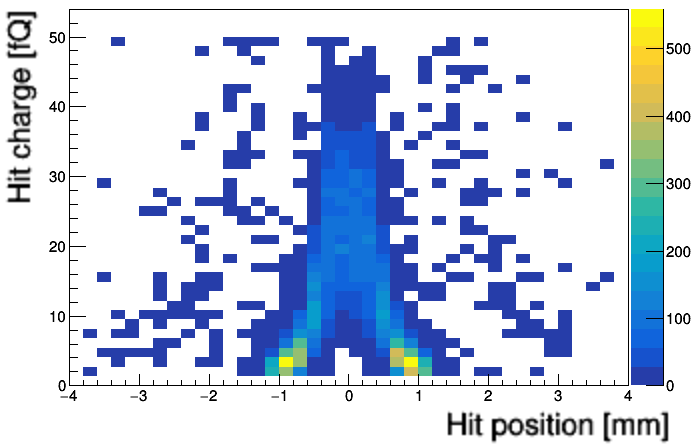} &
\includegraphics[width=0.5\textwidth]{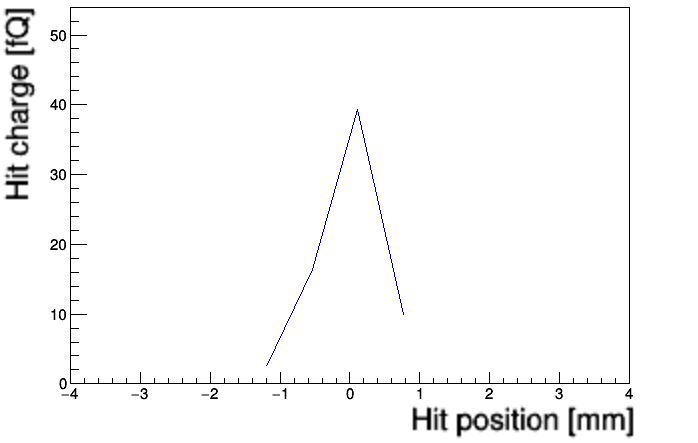} \\
\includegraphics[width=0.5\textwidth]{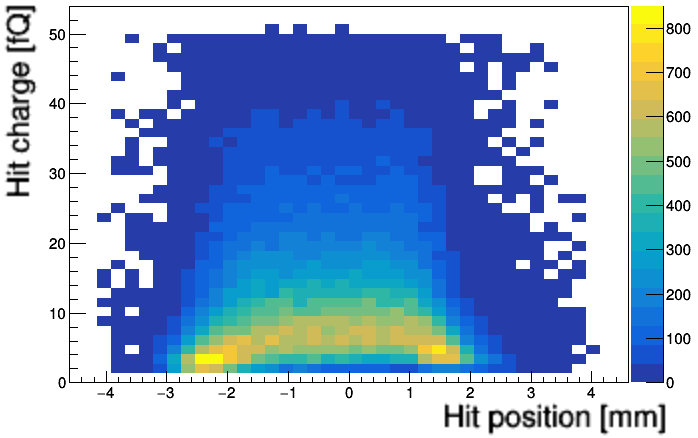} &
\includegraphics[width=0.5\textwidth]{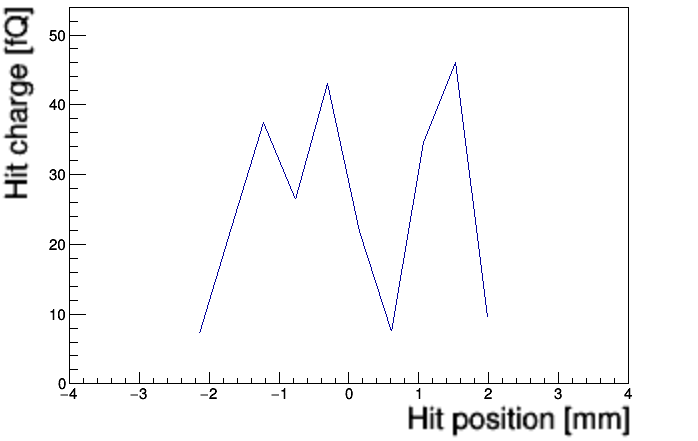} 
\end{tabular}
\caption[Cluster charge distribution]{The mean charge distribution of the entire run (left) is compared to the charge distribution of a single event (right). In the first row the incident angle is zero and both mean and single event distributions show a Gaussian-like distribution. In the second row  the incident angle in 45$^\circ$ and a broaden shape describes the data (bottom left) while on the single event shows a multi-peak distribution (bottom right).}
\label{fig:charge_distr}
\end{figure}

Let's consider the angle $\theta$ between the track and the normal to the detector surface. As this angle increases then the path length $l$ of the particle in the drift gap $d$ is defined by $l=d/\cos(\theta)$. The number of primary electron and the cluster charge are proportional to the path $l$. If $\theta \neq 0$ then the projection on the first GEM plane of the path is not point-like such as in the case $\theta$ = 0$^\circ$ but it is equal to $d \, \tan(\theta)$. Each electron reaching the first GEM generates an avalanche that induces a signal on the three strips below. On average the total signal generated by a particle in this configuration has a broaden shape given by a convolution of a box of a width $d \, \tan(\theta)$ and a Gaussian function similar to the charge distribution in the $\theta$ = 0$^\circ$ configuration. The single event is very different from the average: the charge distribution is not flat due to the large spread in the number of amplified electrons from each primary. This quantity can vary up to the 50\% between each amplification. Charge distributions with $\theta = 0^\circ$ and $\theta \neq 0^\circ$ are shown in Fig. \ref{fig:charge_distr}. This is the reason why the cluster size increases with $\theta$ and the charge distribution is no more Gaussian.

\begin{figure}[tp]
\centering
\begin{tabular}{cc}
\includegraphics[width=0.4\textwidth]{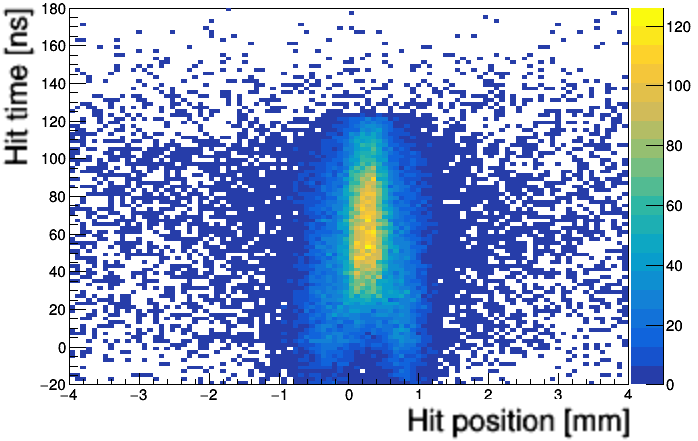} &
\includegraphics[width=0.4\textwidth]{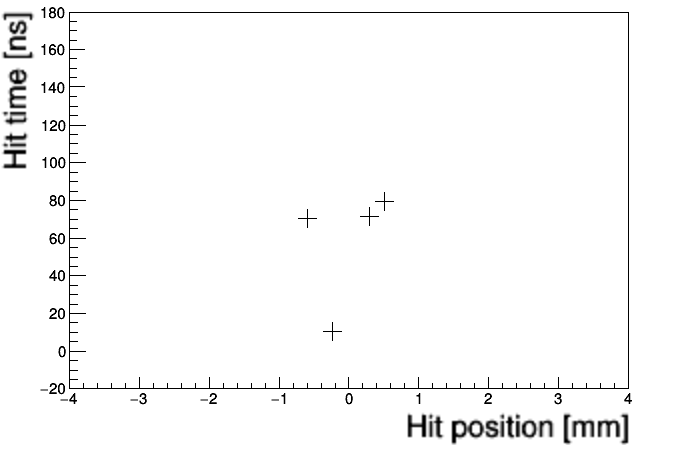} \\
\includegraphics[width=0.4\textwidth]{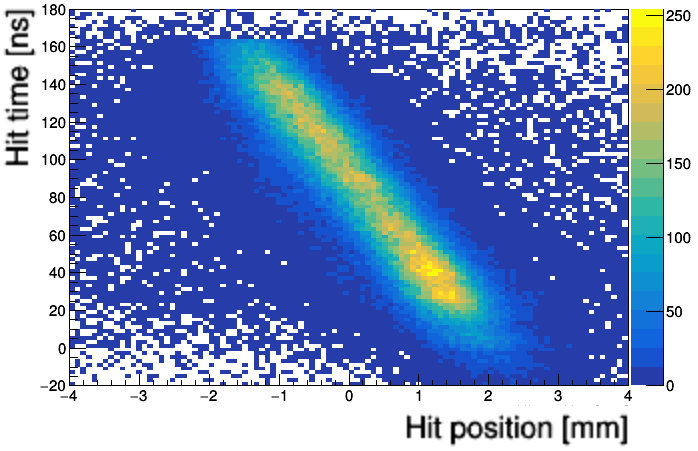} &
\includegraphics[width=0.4\textwidth]{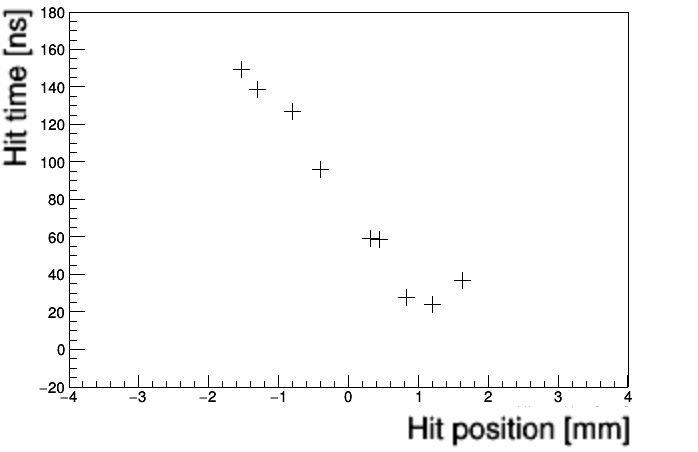} 
\end{tabular}
\caption[Cluster time distribution]{The mean time distributions of the entire run (right) is compared with the time distribution of a single event. In the first row the incident angle is zero and both mean distribution and single event show a compact distribution. In the second row the incident angle is 45$^\circ$ and a linear shape describe the data in both cases.}
\label{fig:time_distr2}
\end{figure}

\begin{figure}[htp]
\centering
\includegraphics[width=0.7\textwidth]{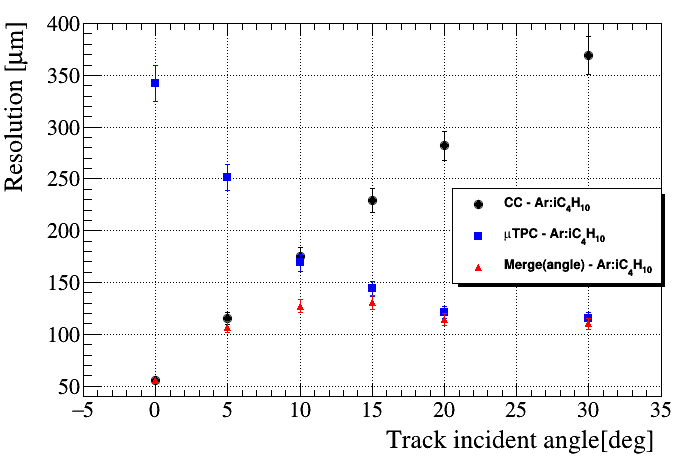}
\caption[CC, $\upmu$TPC and merge resolution with sloped tracks]{The triple-GEM spatial resolution as a function of the incident angle and no magnetic field for the three algorithms: CC, $\upmu$TPC and the merging procedure described in Sect. \ref{sec:merge}.}
\label{fig:res_yesRot}
\end{figure}

The CC algorithm is no more efficient as soon as the charge distribution is not Gaussian but $\upmu$TPC does because cluster size and the time difference between nigh strips are larger as shown in Fig. \ref{fig:time_distr2}. Spatial resolution of the algorithms is shown in Fig. \ref{fig:res_yesRot}.

The $\upmu$TPC spatial distribution shows a Gaussian-like behavior with broad tails and its shape is fitter with a double-Gaussian fit and the resolution is measured with a weighted average of the two Gaussian like in Eq. \ref{eq:MDC_res}. The events in the wider Gaussian are related to pathological $\upmu$TPC event and the cluster properties are not optimal, $e.g$ the cluster size is too small or too large. The position measurement through a fit make this algorithm less stable with respect to the CC. It is sufficient one bad $\upmu$TPC point to worsen the $\upmu$TPC-linear fit. Moreover, the shape of the reconstructed path is not properly a line due to the presence of diffusion and capacitive effects. Further studies are needed to improve more the $\upmu$TPC. Some of those study are present in App. \ref{sec:capacitive}.


\subsection{Temporal studies with sloped tracks}
\label{sec:driftvelocity}
In Sect. \ref{sec:time_ref} it has been shown how to measure the time needed to an electron to drift from the first GEM to the anode from the time distribution of the hits, named $t_0$. See Fig. \ref{fig:time_distr}. Similarly $t_{\mathrm{last}}$ is defined  as the inflexion point of the FD used to describe the leading edge of the time distribution. The time value $t_{last}$ is associated to the time need by an electron to  drift from the cathode to the anode. If the gap is know, then it is possible to measure the drift velocity of the electrons in the drift gap from the Eq.:
\begin{equation}
v_{drift}^{\mathrm{measured}}=\dfrac{gap}{t_{\mathrm{last}}-t_0}
\label{eq:drift_velocity_meas}
\end{equation} 

\begin{figure}[htp]
\centering
\includegraphics[width=0.7\textwidth]{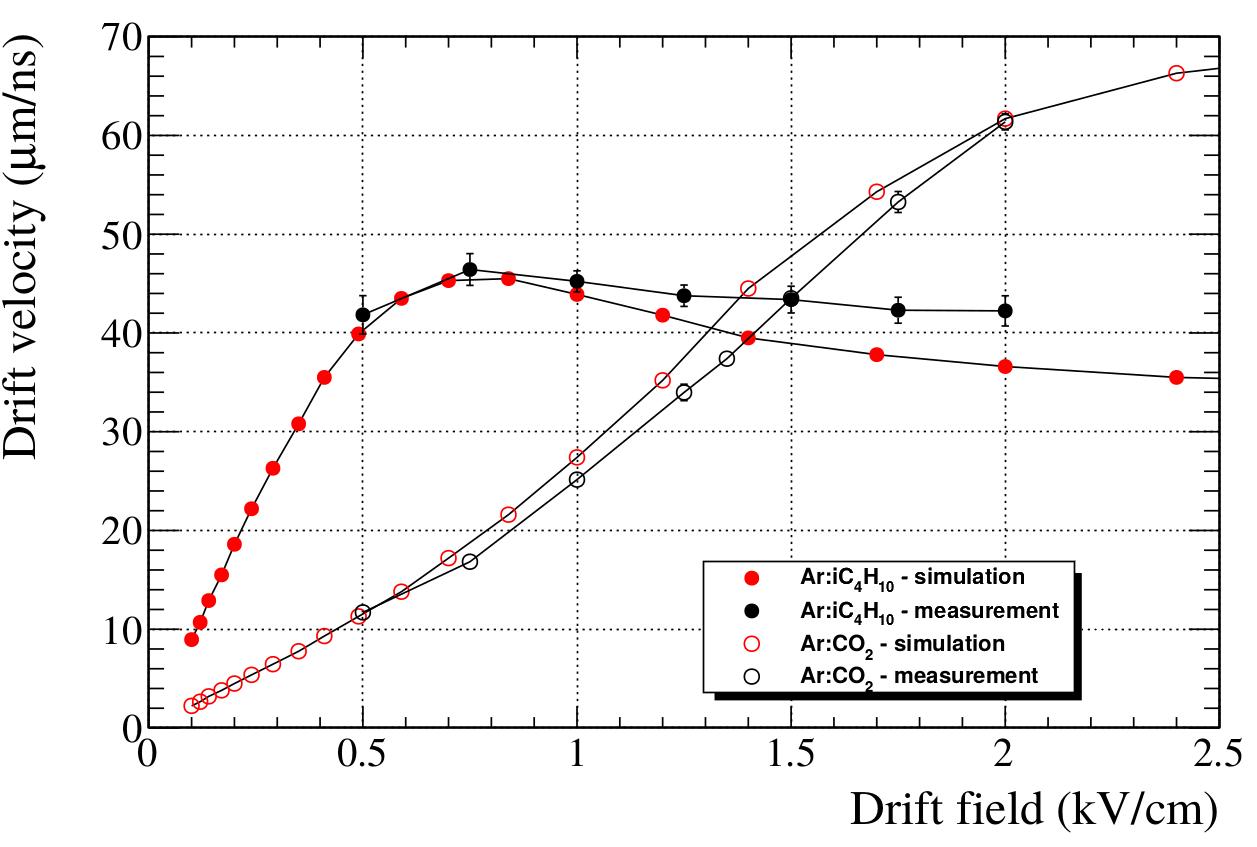}
\caption[Simulated and measured drift velocity]{Simulated drift velocity evaluated with Garfield++ compared to the measured drift velocity with a triple-GEM with $5 \, \mathrm{mm}$ drift gap and an incident angle of 30$^\circ$ in Ar+30\%CO$_2$ and Ar+10\%iC$_4$H$_{10}$ gas mixtures. No magnetic field is present.}
\label{fig:drift_sim_meas}
\end{figure}

The measurement accuracy of the drift velocity improves with larger incident angle, then this measure can not be performed if $\theta$ = 0$^\circ$ where the time distribution is squeezed to $t_0$, see Fig. \ref{fig:time_distr2}. The drift velocity depends on the electric field and it can be computed with programs such as Garfield++ \cite{ref4:garfield}. A comparison of the simulated and measured value of the drift velocity has been performed and the results, as shown in Fig. \ref{fig:drift_sim_meas} agree within 10\%.

Similarly to the drift velocity, the time resolution of the detector is a measurement that improves with larger angles. The time difference between the trigger time and the faster hit of the cluster is the time distribution used to evaluate the detector time resolution. The distribution is fitted with Gaussian function and a $\sigma_t$ of $15 \, \mathrm{ns}$ has been measured. The contribution of the electronics is about $7 \, \mathrm{ns}$ then the time resolution of the detector with both gas mixtures is about $12 \, \mathrm{ns}$.



\subsection{Merging algorithm}
\label{sec:merge}
The CC and $\upmu$TPC are two algorithms anti-correlated: if the first performs properly then the second is not efficient and vice-versa. Once the detector will be installed in BESIII it has to provide a single measurement. Another algorithm to weight properly the two is needed. To achieve this purpose two methods have been developed: the first one uses the cluster size information $x_{merge}(\mathrm{nHit})$, and the other one uses the track incident angle information $x_{merge}(\theta)$. Both methods use the Eq. \ref{eq:merge}:
\begin{equation}
x_{\mathrm{merge}} = w_{\mathrm{cc}} \, (x_{\mathrm{cc}}-\Delta_{\mathrm{cc}})+(1-w_{\mathrm{cc}}) \, x_{\mathrm{tpc}}
\label{eq:merge}
\end{equation}
where $w_{cc}$ is the CC weight and $\Delta_{cc}$ is the shift due to the magnetic field, if it is present. A difference between CC and $\upmu$TPC is the shift $\Delta_{cc}$: the CC reconstructs the position in the anode plane while the $\upmu$TPC in the middle of the drift gap. This shift has to be taken into account.

\begin{figure}[htp]
\centering
\begin{tabular}{cc}
\includegraphics[width=0.4\textwidth]{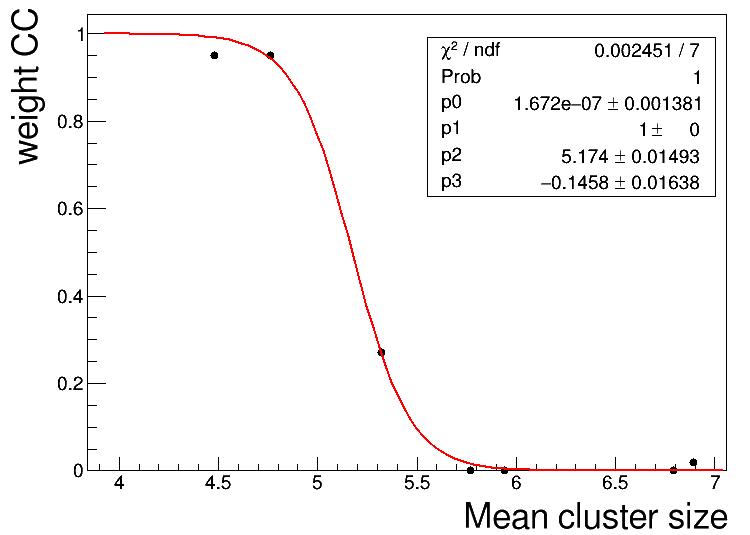} &
\includegraphics[width=0.4\textwidth]{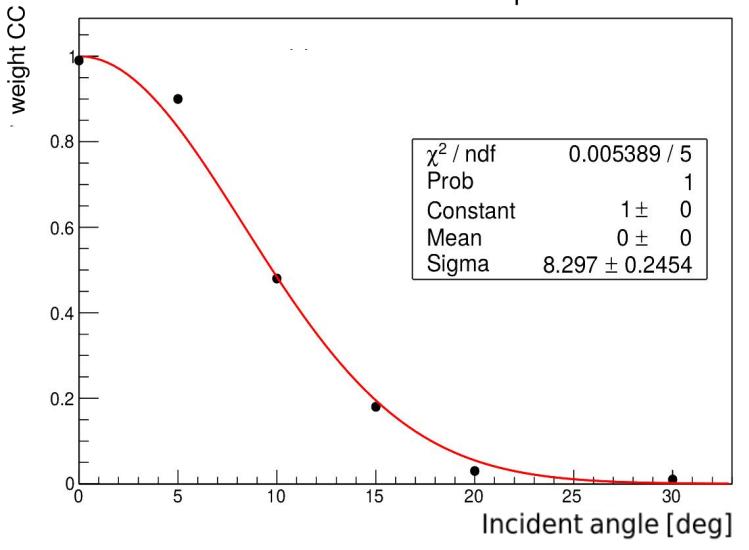}
\end{tabular}
\caption[CC weighting functions]{CC weighting functions as a function of the cluster size (left) and the incident angle (right).}
\label{fig:merge_weight}
\end{figure}

The CC weight $w_{cc}$ used in both methods has been calibrated from a data-set acquired in a test beam, these are data-driven procedures, and the results have been applied to different data-set. The weights have been evaluated from the optimization of the $x_{\mathrm{merge}}$ as a function of the weights. $E.g.$, at 0$^\circ$ the algorithm to use is the CC while at 20$^\circ$ only the $\upmu$TPC. The CC resolution crosses the $\upmu$TPC between 10$^\circ$ and 20$^\circ$ as shown in Fig. \ref{fig:res_yesRot}. Between 0$^\circ$ and 20$^\circ$ the mean cluster size vary from 4 to 6. The calibrated weights are shown in Fig. \ref{fig:merge_weight}. A FD function has been used to evaluate $w_{\mathrm{cc}}$ in $x_{\mathrm{merge}}(\mathrm{nHit})$ and a Gaussian function to fit $w_{\mathrm{cc}}$ in $x_{\mathrm{merge}}(\theta)$. The functions are used to evaluate the weights to use in each situation. In the BESIII experiment the first method will be used to measure the position with a spatial resolution between $100$ and $200 \, \mathrm{\upmu m}$. Once the particle track will be reconstructed then the track incident angle information will be known and used to improve the precision with the second method whit a spatial resolution almost flat between $100$ and $130 \, \mathrm{\upmu m}$.

\begin{figure}[htp]
\centering
\includegraphics[width=0.7\textwidth]{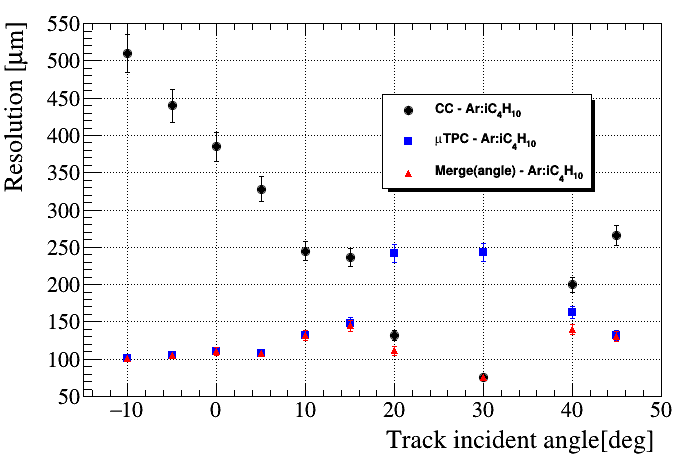}
\caption[Spatial resolution for different incident angle with magnetic field]{Spatial resolution for different incident angle with magnetic field: black dot is the CC resolution, blue squares the $\upmu$TPC and the red triangle the merging algorithm.}
\label{fig:res_yesB_yesRot}
\end{figure}

The performance of the triple-GEM in magnetic field for different incident angles are summarized in Fig. \ref{fig:res_yesB_yesRot}. The points are similar to Fig. \ref{fig:res_yesRot} but they are centered around the Lorentz angle at about 27$^\circ$. If the incident angle coincides with the Lorentz angle then the charge distribution at the anode is close to the one without magnetic field and orthogonal tracks. This is named {\it focusing} effect and this configuration has an efficient CC. Once the angle differs from the Lorentz angle then the $\upmu$TPC becomes the most precise algorithm. The combination of the two through the $x_{\mathrm{merge}}(\theta)$ allows to reach a uniform spatial resolution in the interested angular range.

If the magnetic field is present then $x_{\mathrm{merge}}(\theta)$ depends on the Lorentz angle too. The focusing configuration corresponds to the $\theta$ = 0$^\circ$ configuration without magnetic field. Then $x_{\mathrm{merge}}(\theta)$ is evaluated with $\theta~\rightarrow~\theta - \theta_{\mathrm{Lorentz}}$. 

\section{Performance at high rate}
$\upmu$TPC algorithm shows promising performance in a wide range of configuration, $e.g.$ large angle track and high magnetic field. It is mandatory to understand the application limits of this technique in order to exploit it as much as possible in future detectors. Triple-GEM in high rate environments shows an elevate robustness and electrical stability and it keeps a stable gain up to few $\mathrm{MHz/cm^2}$ \cite{ref4:sauli_review}.

\begin{figure}[htp]
\centering
\begin{tabular}{cc}
\includegraphics[width=0.4\textwidth]{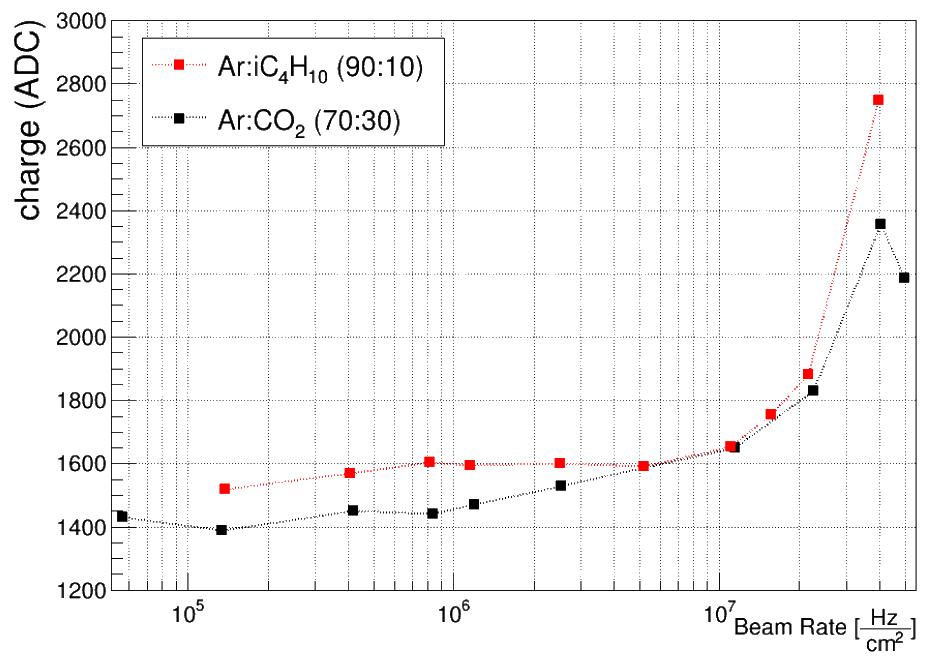} &
\includegraphics[width=0.47\textwidth]{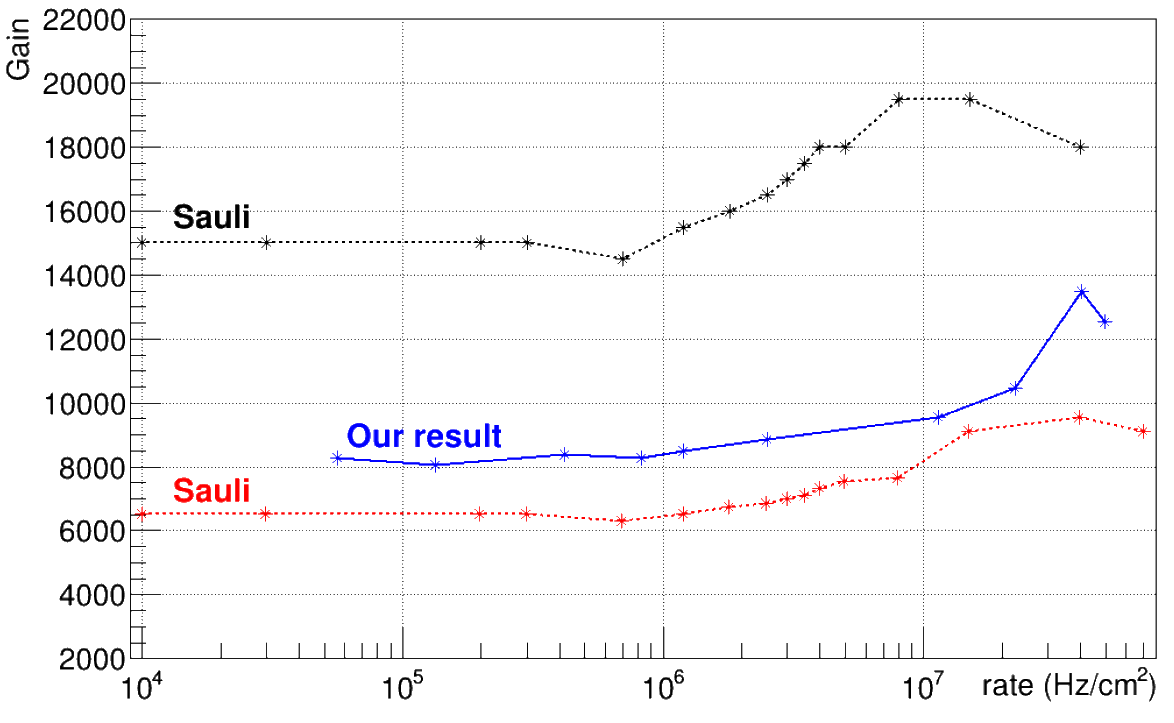}
\end{tabular}
\caption[Cluster charge dependency on beam rate in triple-GEM]{The mean cluster charge in Ar+10\%iC$_4$H$_{10}$ and Ar+30\%CO$_2$ is shown as a function of the beam rate in a triple-GEM with 30$^\circ$ angle between its normal and the beam direction (left). The same points as a function of the gain instead of the charge compared with similar measurement performed by \cite{ref4:sauli_review} (right).}
\label{fig:rate_Q}
\end{figure}

A high rate test has been performed to evaluate the performance of the $\upmu$TPC, In triple-GEM detectors the electrons multiplication occurs inside the holes. Due to the mass difference the electron drift immediately outside the hole while the ions, 1000 times slower, take more time. Some ions are neutralized on the copper of the GEMs while others drift to the cathode. If the flux rate is higher than the recovery rate needed to neutralize the ions then space-charge effects occur and the electric field around the GEM hole is warped. Experimental measurement of the cluster charge shows a dependency as a function of the rate as shown in Fig \ref{fig:rate_Q}. These phenomena has been observed at first by \cite{ref4:sauli_review}: space-charge effect modifies the electric field and increases the transparency\footnote{The transparency is a quantity related to the percentage of electrons that are extracted from a GEM hole. It will be deepen explained in Sect. \ref{sec:digi}} then it increases the gain. At a rate of $10^8 \, \mathrm{Hz/cm^2}$ the gain drops to smaller values.

\begin{figure}[htp]
\centering
\includegraphics[width=0.7\textwidth]{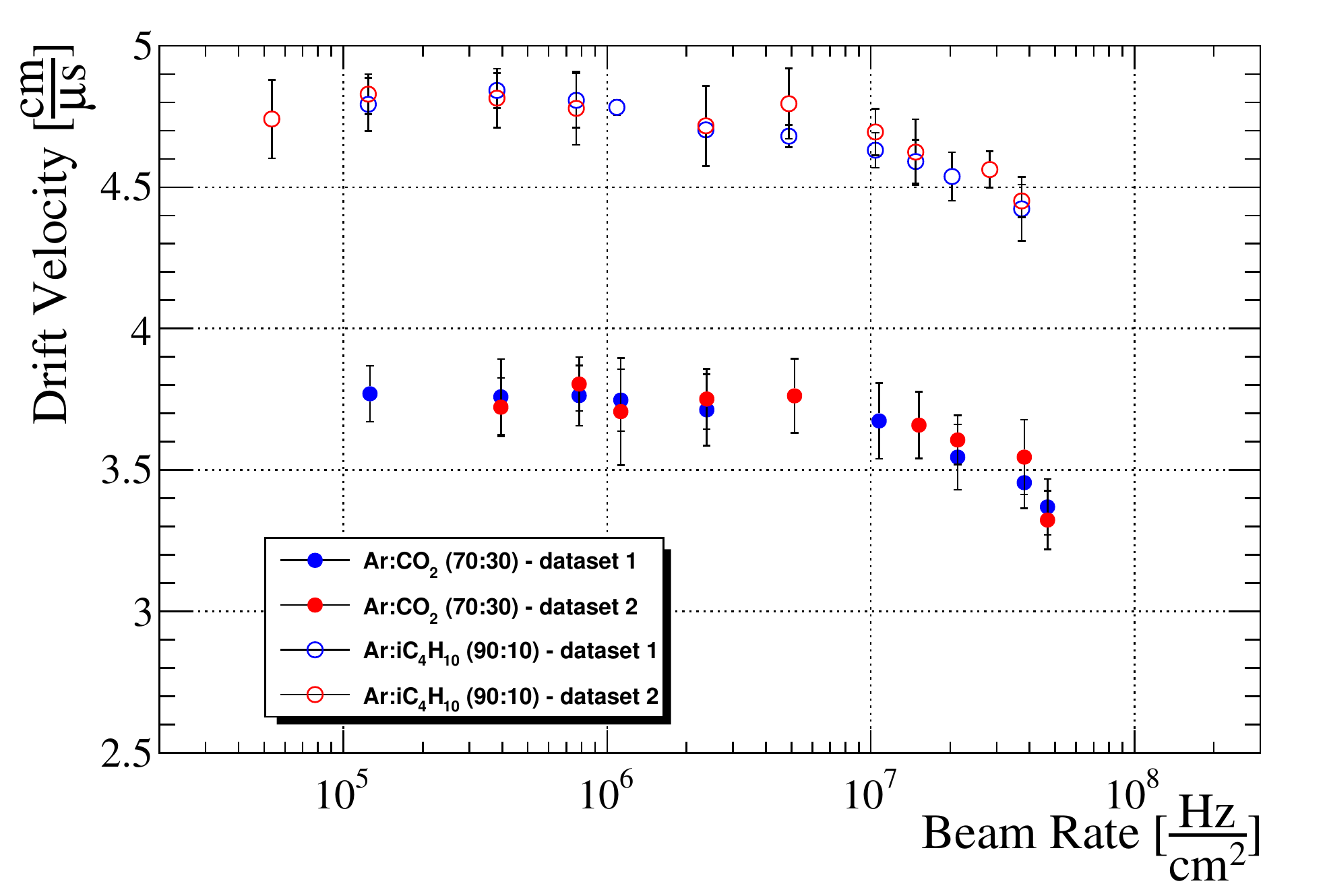}
\caption[Drift velocity dependency on beam rate in triple-GEM]{Measured drift velocity in Ar+10\%iC$_4$H$_{10}$ and Ar+30\%CO$_2$ as a function of the beam rate in a triple-GEM with 30$^\circ$ angle between its normal and the beam direction.}
\label{fig:rate_V}
\end{figure}

The position measured by the $\upmu$TPC depends on the drift velocity. A measurement of this quantity has been performed as described in Sect. \ref{sec:driftvelocity}. Figure \ref{fig:rate_V} shows the drift velocity dependency with respect to the beam rate: a stable value has achieved up to $10^7 \, \mathrm{Hz/cm^2}$  and above this value a significant drop has been observed. The distortion of the electric field is so large that the electrons seem to be slower. The space charge distortion is not homogeneous in the detector and this led to have different drift velocities in different regions of the detector and this set a limit for the $\upmu$TPC application.

\section{Test of Cylindrical GEM}
\begin{figure}[htp]
\centering
\begin{tabular}{cc}
\includegraphics[width=0.5\textwidth]{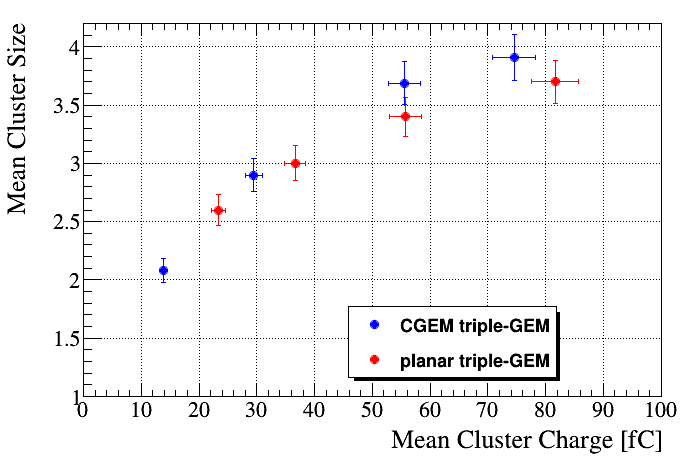} &
\includegraphics[width=0.5\textwidth]{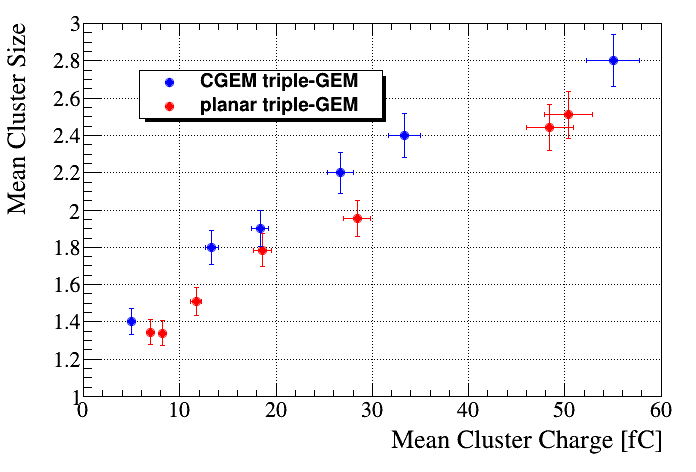}
\end{tabular}
\caption[GEM and CGEM signal comparison]{Mean cluster size as a function of the mean cluster charge is shown for triple-GEM and CGEM with the same  drift gap and gas mixture. Layer1 with $5 \, \mathrm{mm}$ drift gap (left) and proto-Layer2 with $3 \, \mathrm{mm}$ drift gap (right) are shown.}
\label{fig:cgem_signal}
\end{figure}

The geometrical and electrical configuration studied up to now has been used to define the final design of the cylindrical triple-GEM (CGEM) that will be used in the BESIII experiment. Large area CGEMs has been built in Ferrara and Frascati workshops and a test beam has been performed in order to validate the construction, A CGEM with radius $76.9 \, \mathrm{mm}$ and length $532 \, \mathrm{cm}$ has been tested at first at CERN then another one with radius $121.4 \, \mathrm{mm}$ and $690 \, \mathrm{mm}$ length, as labelled $Layer~1$ and $Layer~2$. A study of the collected charge as a function of the cluster size for different gain has been performed to check the behavior of the electrons during the signal formation inside the detector. The results have been compared to the behavior of the planar  triple-GEM and the behavior of the two is compatible, as shown in Fig. \ref{fig:cgem_signal}. This grants the applicability of the CC and $\upmu$TPC algorithms. A spatial resolution of $110 \pm 10 \, \mathrm{\upmu m}$ with Layer 2 and CC algorithm has been reported in agreement with the planar results shown in Fig. \ref{fig:eff_res}. 

\begin{figure}[htp]
\centering
\includegraphics[width=0.7\textwidth]{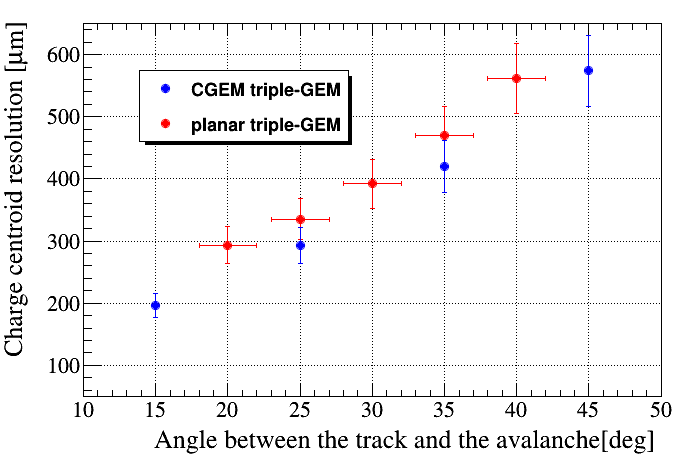}
\caption[CC resolution in CGEM]{CC resolution as a function of the angle between the beam and the avalanche drift direction.}
\label{fig:cgem_b}
\end{figure}

A measurement of the CC performance with Layer1 as a function of the magnetic field has been performed to validate the detector in different condition. Figures \ref{fig:cgem_b} shows the CC as a function of the angle between the avalanche drift direction and the particle beam. This quantity combines the effect of the Lorentz angle and the incident angle. This quantity has been used because the beam was not orthogonal to the CGEM.

\section{TIGER with triple-GEM performance}
\begin{figure}[htp]
\centering
\begin{tabular}{cc}
\includegraphics[width=0.5\textwidth]{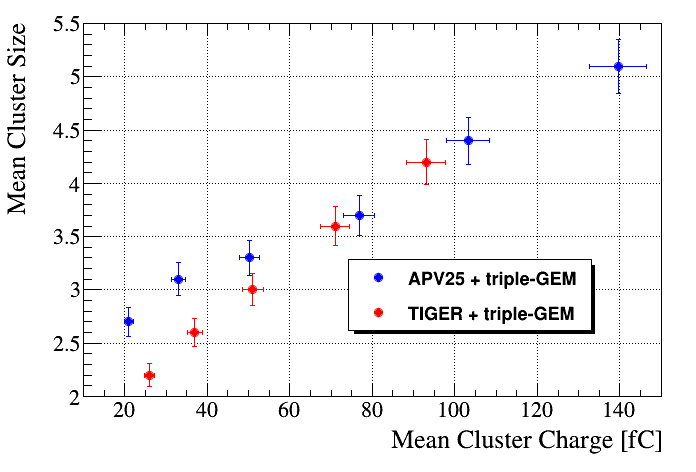} &
\includegraphics[width=0.5\textwidth]{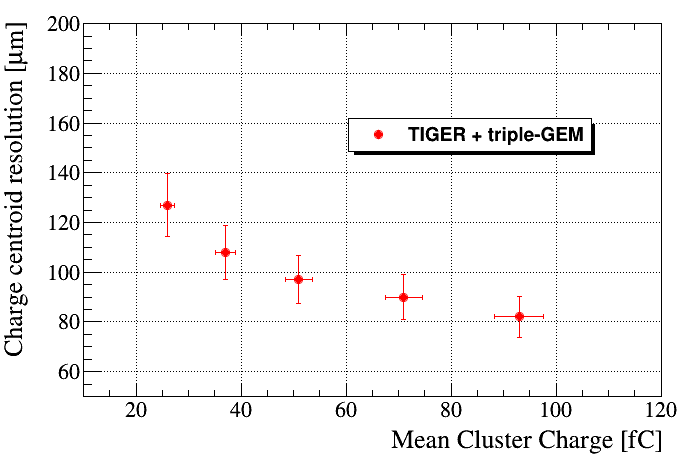}
\end{tabular}
\caption[TIGER with triple-GEM results]{Comparison of the cluster size as a function of the charge for a triple-GEM acquired with two different electronics but similar configuration is shown: APV25 and TIGER (left). The comparison of the CC spatial resolution of a triple-GEM with a TIGER chip as a function of the mean cluster charge is shown (right). Data have been acquired with orthogonal tracks and without magnetic field.}
\label{fig:res_tiger}
\end{figure}

The CGEM-IT will operate inside BESIII but it will not use APV25 electronics but TIGER as mentioned in \ref{sec:tiger}. The research and development of the design and the configuration has been performed mainly with APV25 but to validate the TIGER chip and verify the matching of the triple-GEM performance with this new ASIC, a test beam with triple-GEM and TIGER has been performed. Due to the setup limitation only two triple-GEM has been used then no alignment procedures has been used, neither optimization. The data have been acquired and reconstructed as described in Sect. \ref{sec:tiger_rec} and the clusters within 5 $\sigma$ in $\Delta x_{1,2}$ distribution has been used to measure the signal shape ($Q_{\mathrm{cluster}}$ and $N_{\mathrm{hit}}$ and the CC performance as shown in Fig \ref{fig:res_tiger}.


	\chapter{Triple-GEM simulation}
\label{sec:digi}
A good understanding of a triple-GEM detector and its behavior has been achieved by experimental studies performed in the last few years and by the literature on the subject, as described in previous Chap. A software environment able to reproduce the triple-GEM is needed to extend the knowledge beyond the configurations studied in the test beam and to have a reliable full Monte-Carlo simulation for the new detector to be used by the BESIII collaboration to simulate physics events and to compare to the data acquired by the spectrometer.

Several steps are mandatory to reproduce the signal generation, starting from the gas ionization, to the diffusion of the electrons in the gas volume and their amplification in the GEM holes, up to the induction of the signal on the anode. At present, the most complete software that can reproduce the triple-GEM behavior is Garfield++ \cite{ref5:garfield}. Garfield++ is a toolkit for the detailed computational simulation of detectors which use gases or semi-conductors as sensitive medium. Garfield++ together with Ansys \cite{ref5:ansys} define electric fields, can be used to define the detector geometry and to follow step by step the path and the interaction of each electron to create a complete simulation.
Ansys is a finite element simulator software for static, dynamic and thermal problems. Unfortunately this process is very time consuming: it takes about one day to simulate one event on a modern computer, therefore it is necessary to develop a tool that can reproduce the Garfield++ and the real data results.

In this Chap. the development of a Garfield-based Triplegem Simulator (GTS) will be shown. The basic approach shares the idea shown in \cite{ref5:bonivento} but the method has been extended in a wider range of configurations to take into account the effects of detector gain, incident angle of the track and magnetic field. The simulated events have been reconstructed with the same algorithm described in Chap. \ref{sec:graal} and they provide performance close to the one obtained with the real data.

The idea is to simulate the triple-GEM event through independent processes that can be studied separately:
\begin{itemize}
\item ionization, the interaction between the charged particle and the gas medium and the generation of primary and secondary electrons in space and time;
\item electron drift properties, the electron motion depending on the electro-magnetic field and the gas mixture determines spatial and time distribution of the signal collected at the anode;
\item GEM properties, the electrical configuration of the triple-GEM defines the gain in the GEM holes and their transparency;
\item induction of the signal, the electron motion close to the strips generates a current that is readout by the electronics.
\end{itemize}
\section{Primary ionization simulation}
Heed is an additional tool of Garfield++ that generates ionization patterns of fast charged particles through a model based on photo-absorption and ionization \cite{ref5:heed}. It provides atomic relaxation processes and dissipation of high-energy electrons.

Heed has been used to parametrize the number of primary electrons and their position in the ionization process. If their kinetic energy is sufficiently high then each primary electron can create secondary electrons. The position of the secondary electron is assumed to be the same of the primary. The agglomerate of electrons is called $electron-cluster$.
Ten thousands simulations with Garfield++ have been performed with muons having 150 GeV/c  momentum shooted into $5 \, \mathrm{mm}$ of gas volume filled with Ar+10\%iC$_4$H$_{10}$. This simulation has been used to extract the number of electron-clusters, their relative position and the number of electrons in a cluster.

The number of clusters follows the Poisson distribution, as described in App. \ref{sec:ionization} while the relative distance between two consecutive clusters follows an exponential distribution. The number of the electrons in a cluster depends on the energy loss of the particle: most of the time only one electron is present but some electrons are more energetic and there is a small probability to generate a very high number of secondaries. The electron cluster size, $N_e$, has been limited to 100 to simplify the parametrization of the ionization. 

To reproduce these effects in GTS, an exponential function is used to describe the relative position of the clusters. As a counter-check the number of cluster has to follow a Poisson distribution. 

\begin{figure}[htp]
  \centering
  \begin{tabular}{cc}
    \includegraphics[width=0.4\textwidth]{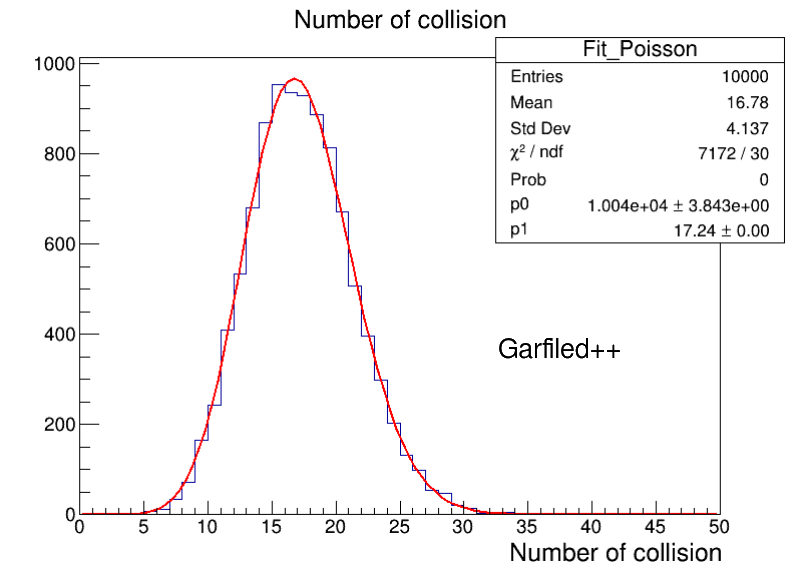}&
    \includegraphics[width=0.4\textwidth]{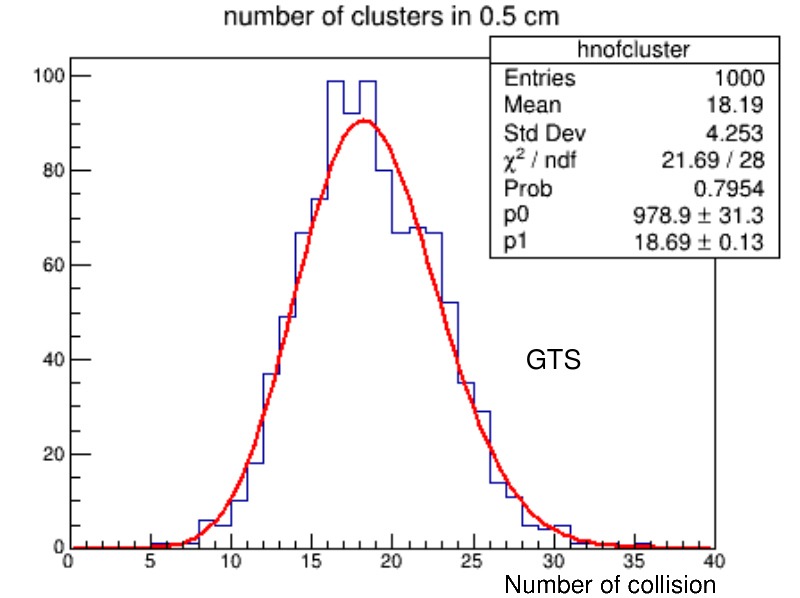}\\
    \includegraphics[width=0.4\textwidth]{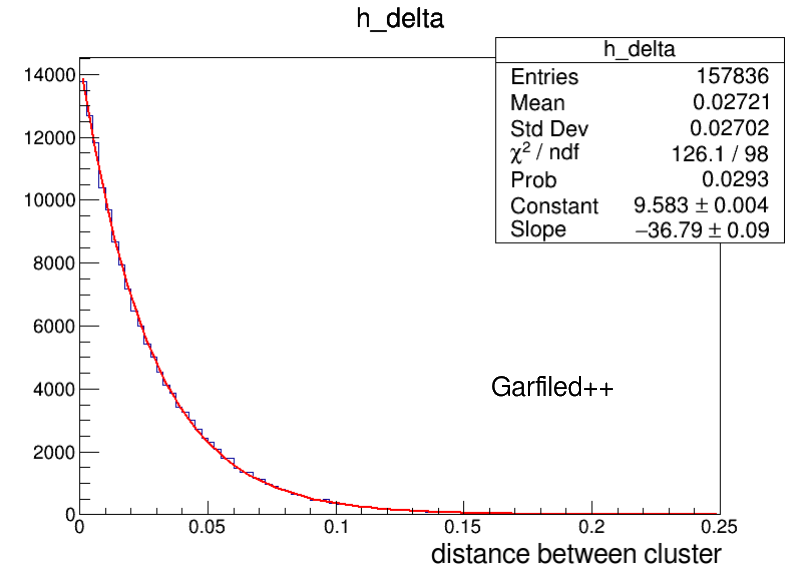}&
    \includegraphics[width=0.4\textwidth]{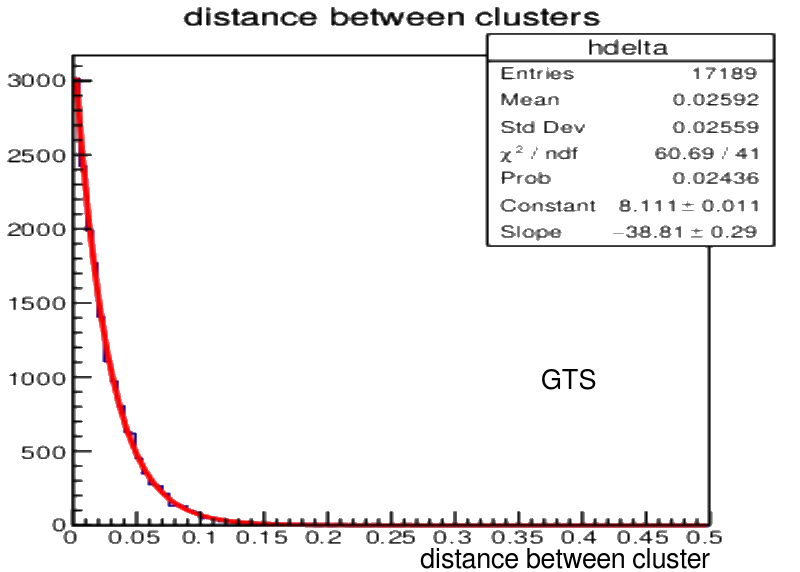}\\
    \includegraphics[width=0.4\textwidth]{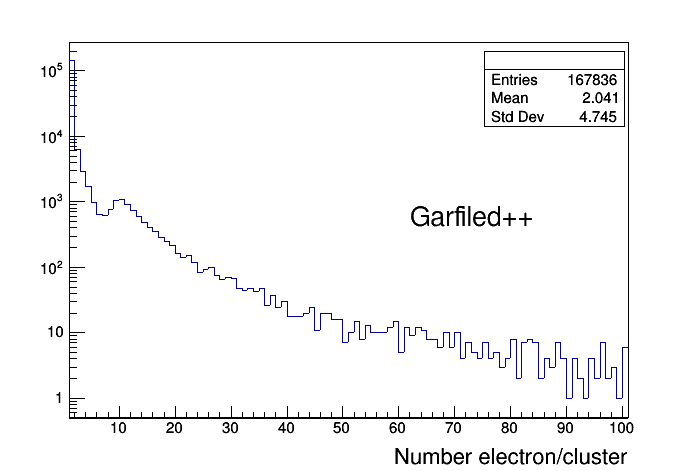}&
    \includegraphics[width=0.4\textwidth]{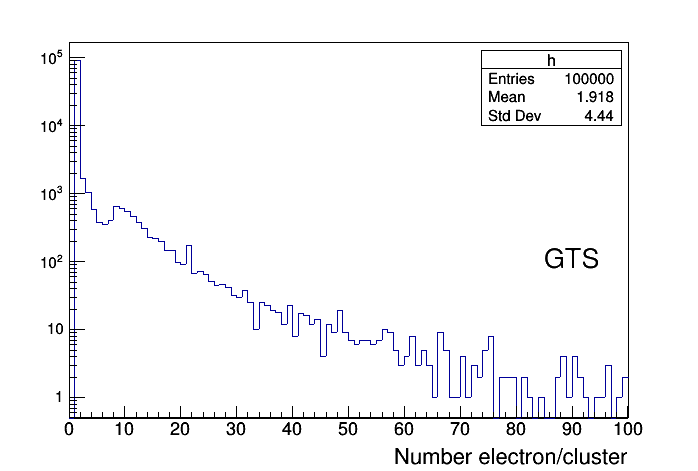}\\
  \end{tabular}
  \caption[Ionization distributions from Garfield++ and GTS]{Distributions of the processes on the ionization from Garfield++ (left) and GTS (right). The first row shows the number of electron-cluster, the second row the distance between each cluster along the ionizing particle path and the third row the number of secondary electrons in the electron cluster.}
\label{fig:digi_ioni}
\end{figure}

The number of primary and secondary electrons is generated randomly from the distribution obtained with Garfield. Figure \ref{fig:digi_ioni} shows the distributions from Garfield++ and their counterpart in GTS.

\section{Electron diffusion}
\label{sec:digi_drift}
The electron diffusion is determined by the gas, the electric and magnetic fields. A description of the electric field inside the detector is needed to measure the spread of the electron avalanche in the detector. This operation can be performed by Ansys. The field maps produced by Ansys can be used as input files by Garfield++.

A triple-GEM detector with bi-conical holes and geometries described in Sect.~\ref{sec:setup} and the high voltage used in the tests beam has been described by Ansys. This output has been used in Garfield++ to determine the diffusion properties of the electrons the electric field. The simulation has been divided in four stages, one for each triple-GEM gas gap. 

\begin{figure}[htp]
  \centering
  \begin{tabular}{cc}
    \includegraphics[width=0.4\textwidth]{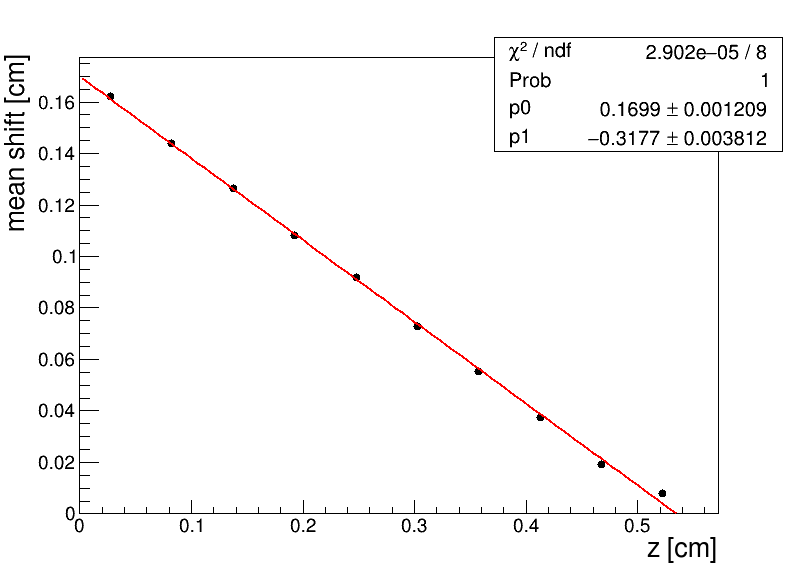}&
    \includegraphics[width=0.4\textwidth]{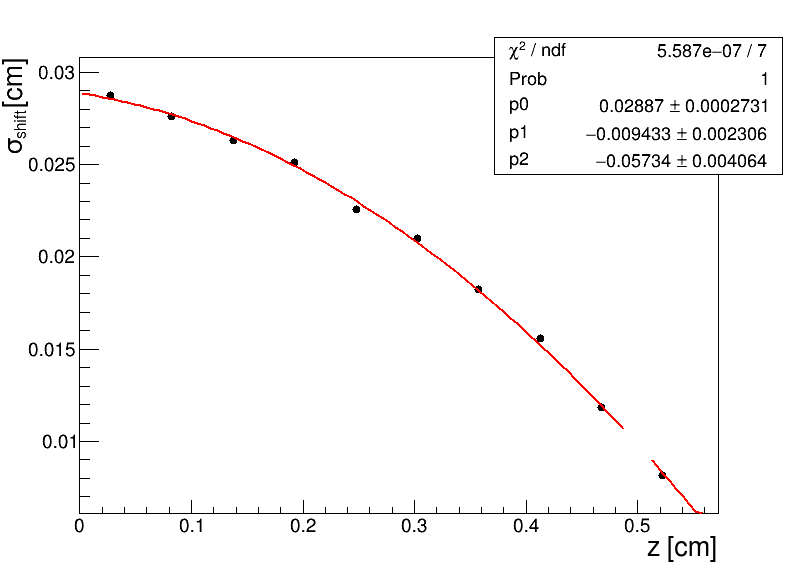}\\
    \includegraphics[width=0.4\textwidth]{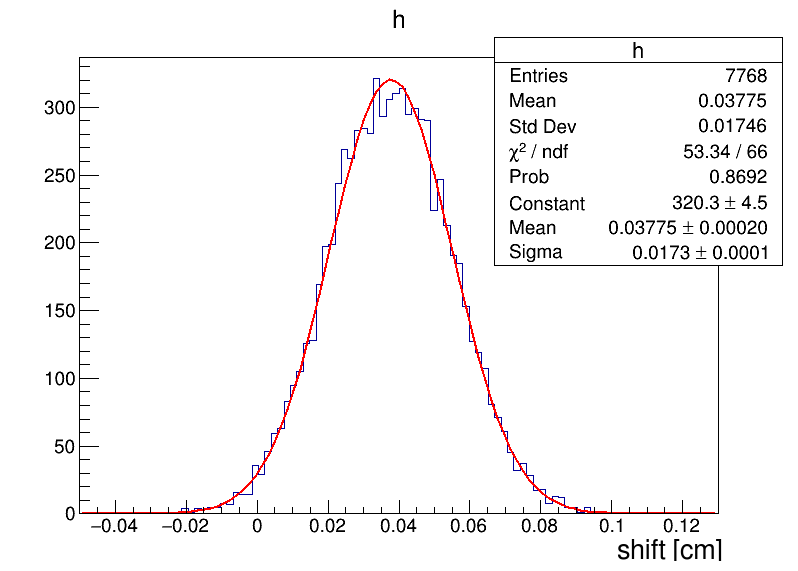}&
    \includegraphics[width=0.4\textwidth]{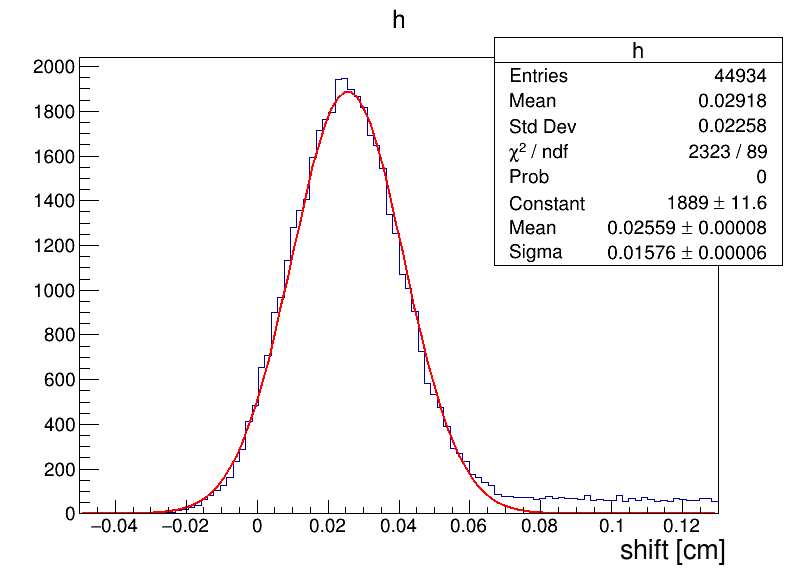}\\
  \end{tabular}
  \caption[Electron drift distributions from Garfield++]{Electron drift properties for the several gaps in a triple-GEM with 1 T magnetic field. In the row it is shown the mean shift (left) and spread (right) in the drift gap as a function of the distance $z$ from the cathode. In the second row the electron distributions in the transfer (left) and in the induction (right) gaps with the Gaussian fit are shown.}
\label{fig:digi_drift}
\end{figure}

In the drift gap a uniform distribution of electrons has been generated between the cathode and a plane placed $150 \, \mathrm{\upmu m}$ before the first GEM. Their spatial and temporal distributions have been studied on the plane to extract the parameters that will be used in GTS. In the transfer gap the electrons have been generated $150 \, \mathrm{\upmu m}$ after the GEM above and the distributions have been evaluated on a plane $150 \, \mathrm{\upmu m}$ before the GEM below. In the induction gap the electrons have been generated $150 \, \mathrm{\upmu m}$ below the GEM and their distributions have been evaluated at the anode plane. Spatial and time distributions in each gap have been fitted with a Gaussian function. In the drift region the distribution has been evaluated as a function of the path length of the electrons because the longer the path the larger the spread. In Fig. \ref{fig:digi_drift} the simulation results for the spatial distribution are shown. Similar analysis have been performed for the time distributions. These simulations have been performed with and without the 1 T magnetic field to parametrize the drift properties in both configurations.
The results are summarized in Tab. \ref{tab:drift}.

\begin{table}[ht]
\begin{center}
\begin{tabular}{l c c}

Spatial shift$_{B=0T}$ &[cm]& $0$ \\
Spatial shift$_{B=1T}$ &[cm]& $-0.317 \, z \, + \, 0.271$ \\
Spatial spread &[cm]& $\sqrt{(-0.057 z^2 \, + \, 0.028 )^2 \, + \, 0.001 }$ \\
\newline\\
Temporal shift &[ns]& $-272 \, z \, + \, 307$ \\
Temporal spread &[ns]& $\sqrt{(-3.34 \, z^2 \, - \, 1.77 \, z  \, + \, 5.00} \, + \, 13.29$ \\

\newline\\
\hline\\
\end{tabular}
\caption[Drift parameters from Garfield++]{Drift parameters from Garfield++ are shown as a function of the $z$, the distance of the generated electron from the cathode (in cm). The magnetic field influences the spatial shift only. The shift and spread are applied to the electron generated in the drift gap and Ar+10\%iC$_4$H$_{10}$ to extrapolate its time and spatial position at the anode.}
\label{tab:drift}
\end{center}
\end{table}	

\section{Transparency and gain of a GEM}
The electric field inside the GEM holes defines the intrinsic gain properties while the electric field outside the holes and the electron diffusion effect determine the GEM transparency $T$. When electric field line stops on the GEM then the transparency is reduced. Some of these lines above the GEM do not enter the GEM holes and some others coming from the holes stop below the GEM. This influence the effective gain of the GEM because a certain fraction of the electrons stops on the GEM foil. In \cite{ref5:bonivento} the dependency of the transparency on the two electric fields outside the GEM foil is shown.

A measurement of the gain and the transparency of the GEM with the electrical configuration used in the test beam has been performed. Ten thousands electrons have been simulated with Garfield++ $150 \, \mathrm{\upmu m}$ above the GEM foil and they have been collected on a plane $150 \, \mathrm{\upmu m}$ below it. For each electron the \textit{collection efficiency} $\varepsilon_{\mathrm{coll}}$, the \textit{extraction efficiency} $\varepsilon_{\mathrm{extr}}$ and the intrinsic gain $G_{\mathrm{intr}}$ has been evaluated. $G_{\mathrm{intr}}$ is number of electron generated in a hole. The \textit{collection efficiency} is the ratio between the number of electrons entering the GEM hole and the number of generated ones; the \textit{extraction efficiency} is the ratio of the extracted ones and the number inside the hole. The number of electron outside the hole $G_{\mathrm{eff}}$ generated by each electron entering the hole is given by:

\begin{equation}		
\begin{tabular}{ccc}
$G_{\mathrm{eff}}=T \, G_{\mathrm{intr}}$ & and & T=$\varepsilon_{\mathrm{coll}} \, \varepsilon_{\mathrm{extr}}$\\
\end{tabular}
\label{eq:trasparency_gain}
\end{equation}

\begin{figure}[htp]
  \centering
  \begin{tabular}{cc}
    \includegraphics[width=0.42\textwidth]{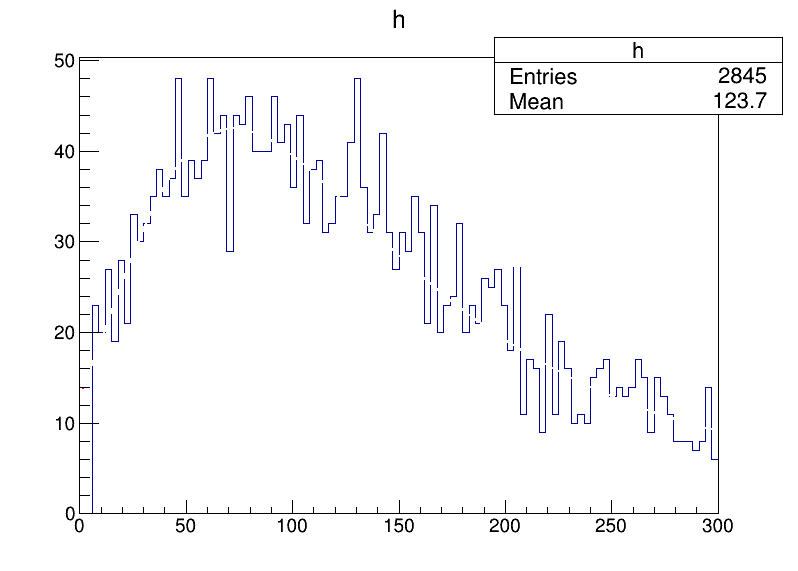}&
    \includegraphics[width=0.3\textwidth]{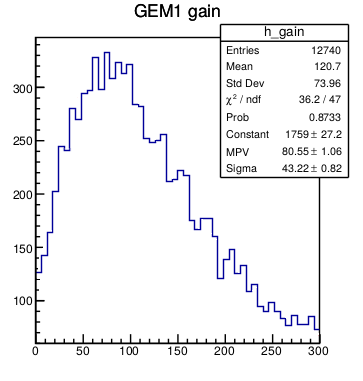}\\
  \end{tabular}
  \caption[GEM gain in Garfield++ and GTS]{GEM gain in Garfield++  (left) and GTS (right).}
\label{fig:GEM_landau}
\end{figure}

A transparency of 37\% and a mean intrinsic gain of about 120 have been measured using in the simulation the standard HV setting from Chapt \ref{sec:graal}. The gain distribution has been fitted with a Polya (see Fig. \ref{fig:GEM_landau}) defined in Eq. \ref{eq:polya}:
\begin{equation}
P(G)= C_0 \frac{(1+\theta)^{1+\theta}}{\Gamma(1+\theta) \left( \frac{G}{\overline{G}} \right) \exp \left[ -(1+\theta)\frac{G}{\overline{G}} \right] }
\label{eq:polya}
\end{equation}

\section{Signal induction on the readout}
The number of multiplication electrons is generated from the Polya function defined above for each electron in the electron-cluster and a shift of the position and the time with respect to the primary electron is attributed following the parametrization in Sect. \ref{sec:digi_drift}. This method allows to generate the spatial and time distributions of the electrons once the avalanche reaches the region below the third GEM. The induction of the signal on the strips takes origin from the electrons motion in the last gap. The electron drift from the last GEM to the anode is sampled each nanosecond and the induced current on the strips is calculated with the Ramo theorem \cite{ref5:ramo}:
\begin{equation}		
I_k=-q \mathbf{v} \cdot \mathbf{E}_k^{\mathrm{weight}}
\label{eq:ramo_current}
\end{equation}
where $k$ is the strip index, $q$ and $\textbf{v}$ the electron charge and velocity, $\mathbf{E}_k^{\mathrm{weight}}$ the weighting field evaluated from the electric field generated by the k-strip if 1 V is applied on it and 0 V on the others. 

\begin{figure}[htp]
  \centering
  \begin{tabular}{cc}
    \includegraphics[width=0.4\textwidth]{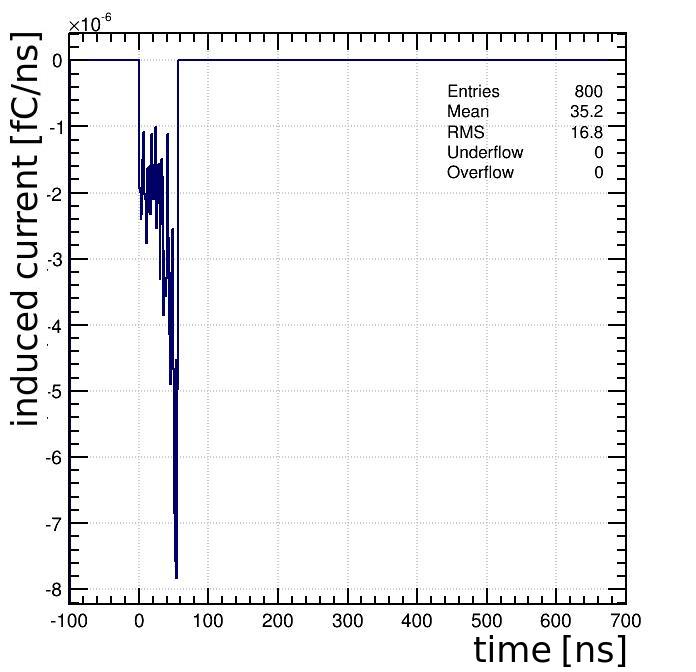}&
    \includegraphics[width=0.4\textwidth]{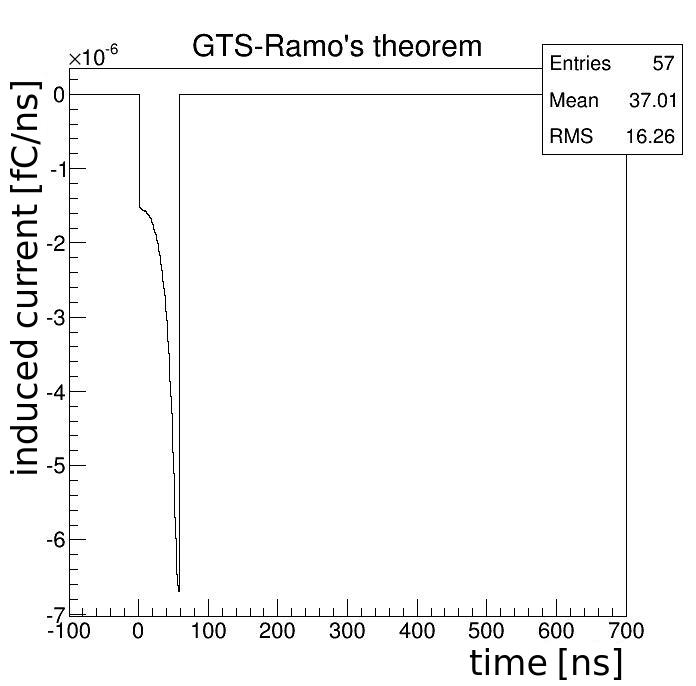}\\
    \includegraphics[width=0.4\textwidth]{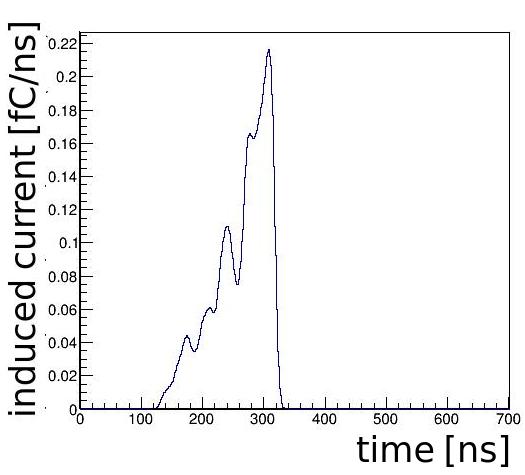}&
    \includegraphics[width=0.4\textwidth]{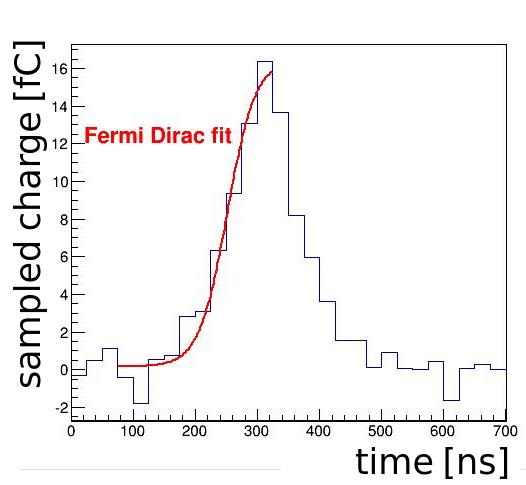}\\
  \end{tabular}
  \caption[Induced current and readout charge in GTS]{In first row the current induced by an electron on a strip as a function of the time is compared between the Garfield++ calculation (left) and the result in GTS using the Ramo's theorem (right). In the second row the integrated current of the event as a function of the time is compared between the Monte-Carlo truth (left) and the one readout by the simulated APV25 (right).}
\label{fig:induced_current}
\end{figure}

The readout electronics plays and important role: a RC integration constant of $50 \, \mathrm{ns}$ simulates the APV25 response to shape the measured current, the charge is integrated every $25 \, \mathrm{ns}$ and it is digitized in 1800 ADC channels to reproduce the pedestal subtraction dynamic range. An example of induced current on a strip is shown in Fig. \ref{fig:induced_current}. The saturation effect is implemented to reproduce better the data. A conversion factor of 30 ADC for each fC is used \cite{ref5:conversion}. A fluctuation of the current induced on the strip is introduced to simulate the electronic noise.

\section{Capacitive effect}
The current induced on the strip can generate a cross talk on its neighbors as a function of the signal amplitude and the inter-strip capacitance. A study on the detector has been performed to evaluate this effect that could not be reproduced by Garfield++ with the current geometry description. The results provides the probability to reproduce this type of noise and its amplitude.

\begin{figure}[htp]
  \centering
  \begin{tabular}{cc}
    \includegraphics[width=0.4\textwidth]{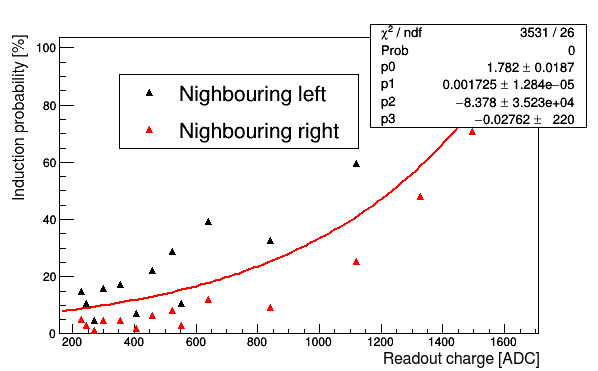}&
    \includegraphics[width=0.4\textwidth]{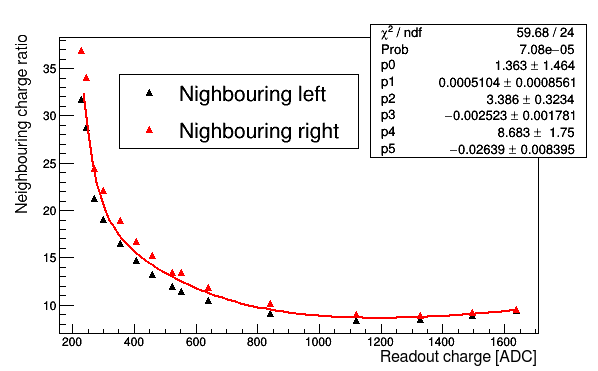}\\
  \end{tabular}
  \begin{tabular}{c}
    \includegraphics[width=0.4\textwidth]{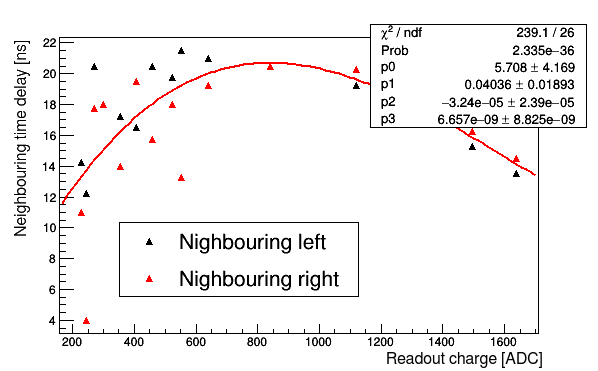}\\
  \end{tabular}
  \caption[Experimental measurement of capacitive effects]{Experimental measurement of capacitive effects as a function of the main strip charge: the probability to induce a charge on the neighboring (top left), the fraction of the induced charge (top right) and the time difference between the signal on the main strip and on its neighboring (bottom).}
\label{fig:capa}
\end{figure}

A signal is injected to the anode strip thanks to a special board inserted between the APV25 and the anode with an attenuator and a 1 pF capacitance. A signal with amplitude between 10 and 400 mV and $40 \, \mathrm{ns}$ length is injected on the strip and readout by the APV25 also on the neighbouring strips. The injected signal is measured mainly on the fired strip and a fraction on its neighbors. The data have been analyzed and the probability to induce the charge on the neighbor strip, the charge induced and the time difference have been measured, their behavior has been fitted and the functions are used in GTS to simulate this effect. Figure \ref{fig:capa} shows the obtained results: higher strip charge generates always a signal of about 5\% of the inducing charge on the neighboring strips after about $15 \, \mathrm{ns}$.

\section{Comparison of the simulation and the test beam data}
The GTS environment described in the previous section can reproduce the real triple-GEM behavior. A validation of the simulation has been performed through the comparison of the Monte-Carlo events with data for different high voltage values on the GEMs, with and without magnetic field and using incident angles of the ionizing particle ranging from -45$^\circ$ to 45$^\circ$. The incident particles are muon having 150 GeV/c momentum like the test beam conditions. Each point of the scan has been simulated with 10000 Monte-Carlo tracks. An agreement better than 30\% has been found between GTS and data.

The high voltage scan has been used to test the behavior of the GEM gain and transparency: the electric field inside the GEM's hole affects the electron multiplication and this influences the charge measured at the readout, the number of fired strips and the charge centroid spatial resolution.

The configurations with different incident angles allow to test the behavior of the detector when primary electrons are spread on a wider region. The charge distribution in each event follows the one showed in Fig. \ref{fig:charge_distr} and the multi-peak trend increases with the incident angle. As a consequence, the spatial resolution measured with the CC algorithm degrades. The time distribution of the readout strips allows to apply correctly the $\upmu$TPC algorithm and it returns good results.

The electronics noise plays an important role: its amplitude influences mainly the CC and cluster size, while the variation of the noise as a function of the time affects the $\upmu$TPC.

\begin{figure}[htp]
  \centering
  \begin{tabular}{cc}
    \includegraphics[width=0.4\textwidth]{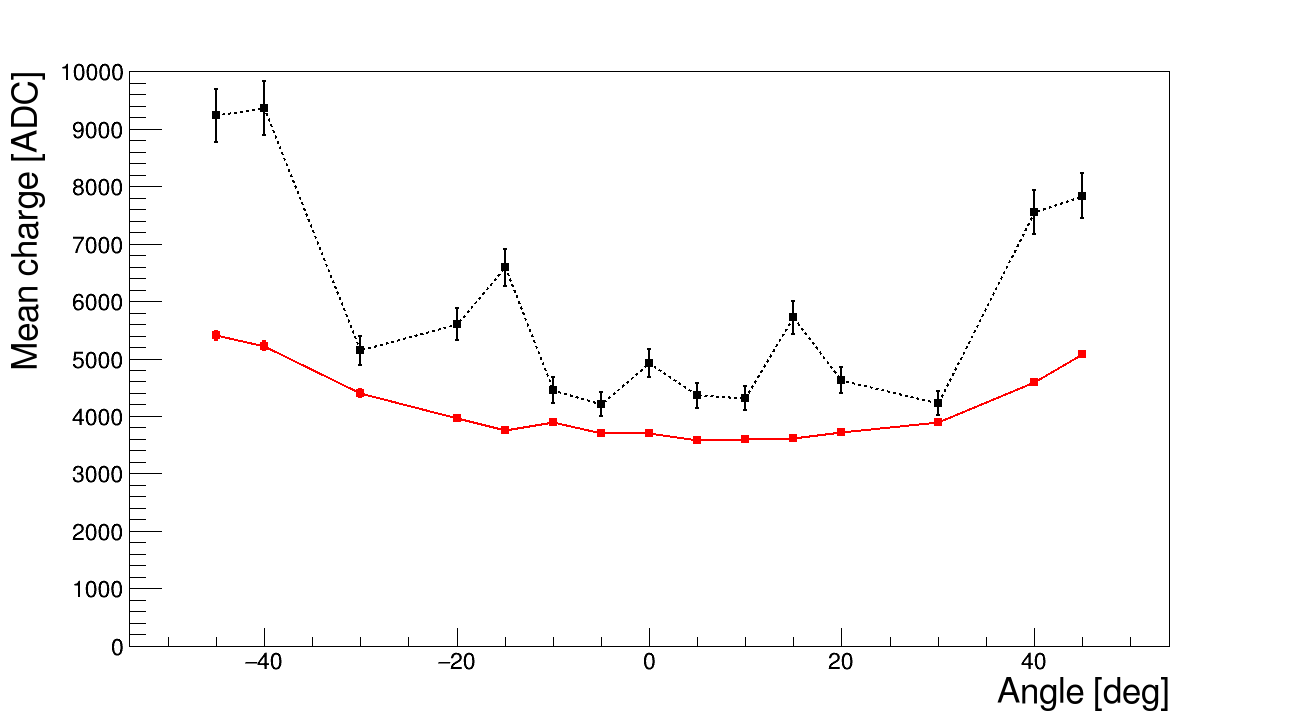}&
    \includegraphics[width=0.4\textwidth]{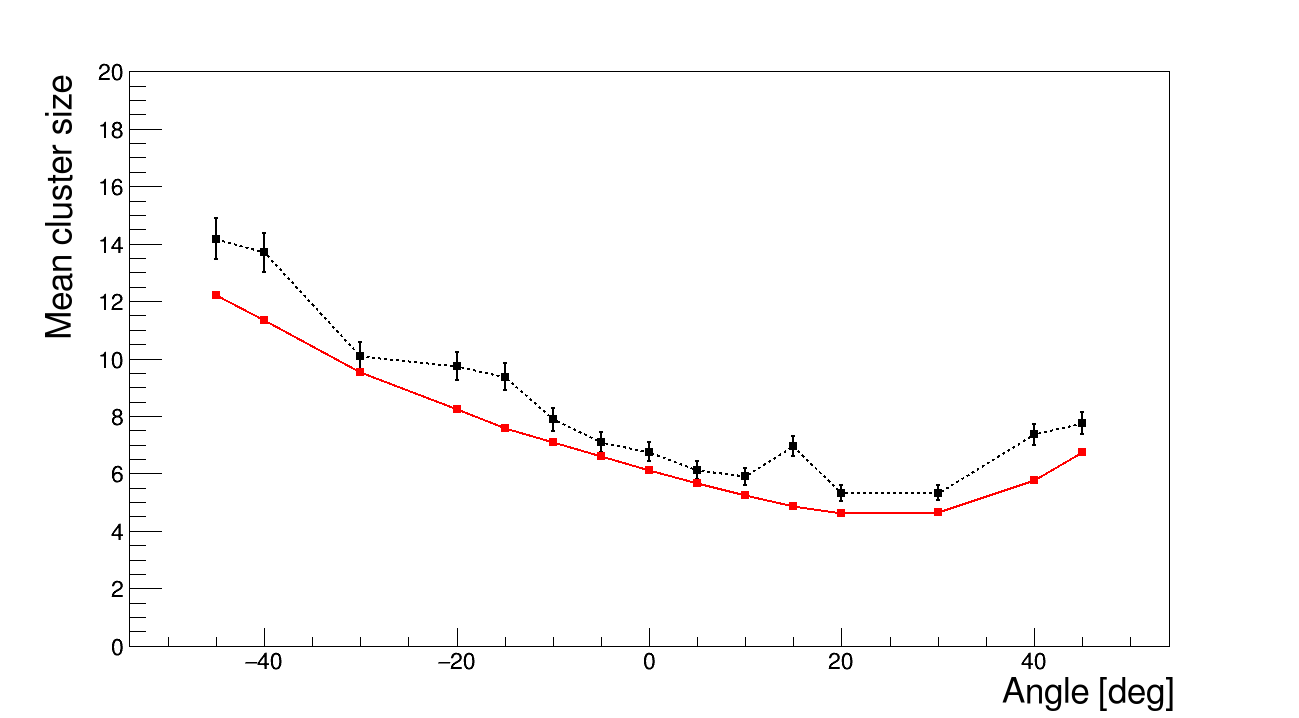}\\
    \includegraphics[width=0.4\textwidth]{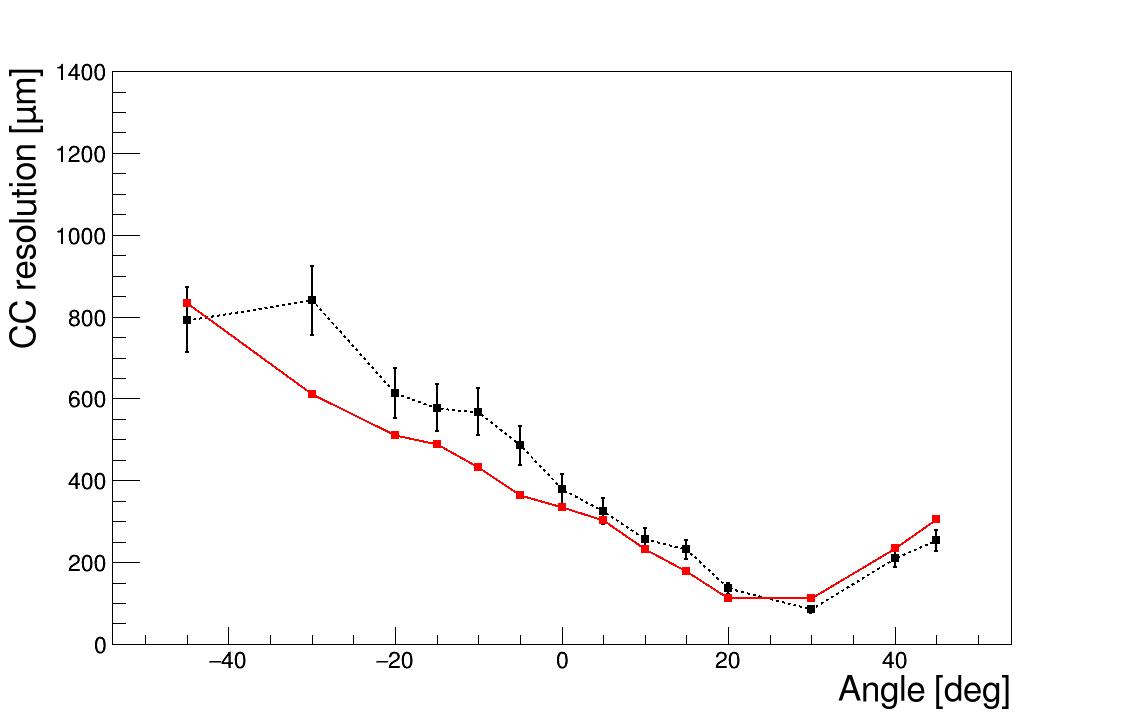}&
    \includegraphics[width=0.4\textwidth]{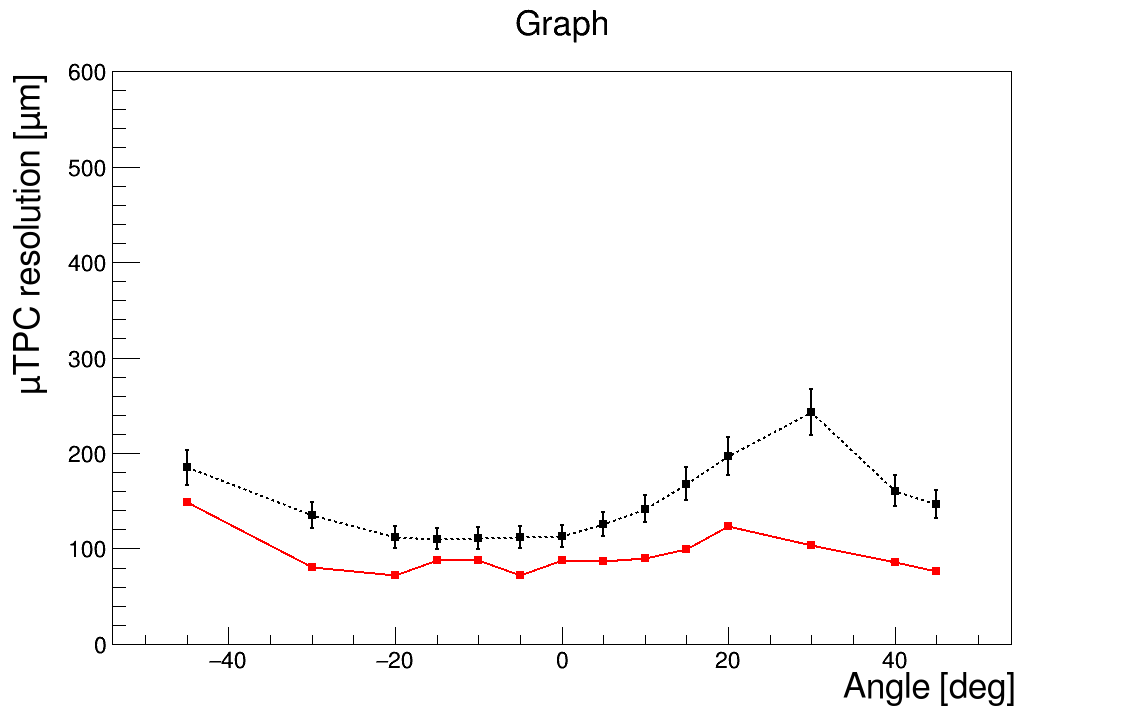}\\
  \end{tabular}
  \caption[GTS and real data comparison]{The comparison between GTS (red line) and real data (black line) is shown for the following variables: the mean cluster charge (top left), the mean cluster size (top right), the CC resolution (bottom left) and the $\upmu$TPC resolution (bottom right).} 
\label{fig:GTS_results_B}
\end{figure}

The magnetic field is introduced and GTS shows compatible results with data: the behavior with orthogonal tracks is transferred in $\theta$=$\theta_{\mathrm{Lorentz}}$. The comparison of GTS and real data with magnetic field as a function of the incident angle is shown in Fig. \ref{fig:GTS_results_B} where cluster size and charge ensure the good shape of the avalanche while CC and $\upmu$TPC spatial resolutions guarantee the correct performances of the triple-GEM in the simulation.

The agreement of cluster charge and size and CC resolution stay within 30 \%. Up to 40\% for the $\upmu$TPC where the simulation does not degrade properly in presence of noise and capacitive effects.

\subsection{Cross checks for $\upmu$TPC}
\begin{figure}[htp]
  \centering
  \begin{tabular}{cc}
    \includegraphics[width=0.4\textwidth]{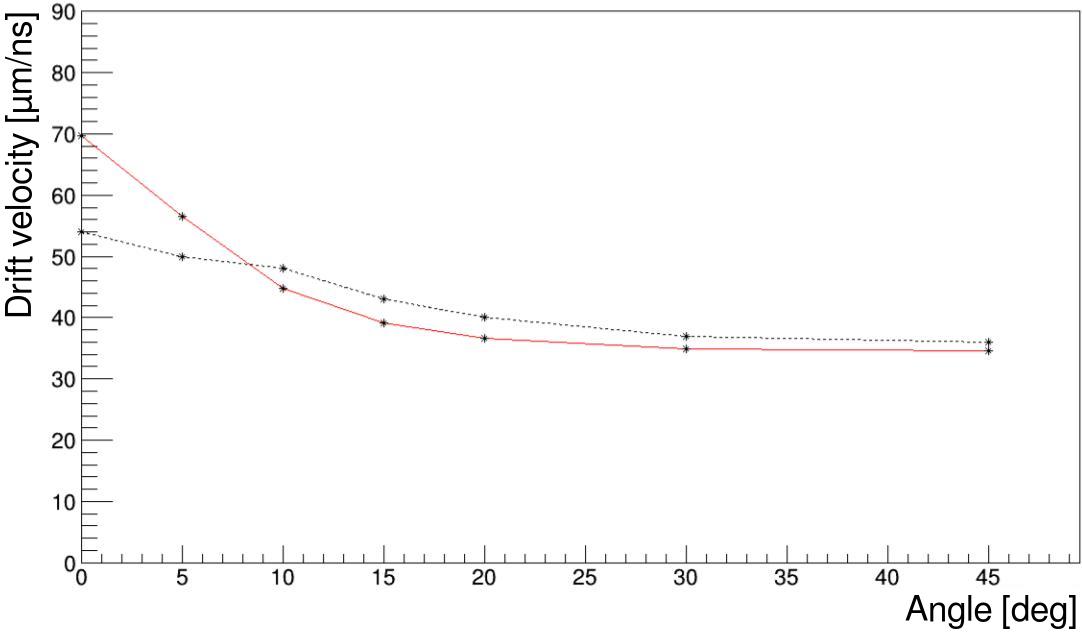}&
    \includegraphics[width=0.4\textwidth]{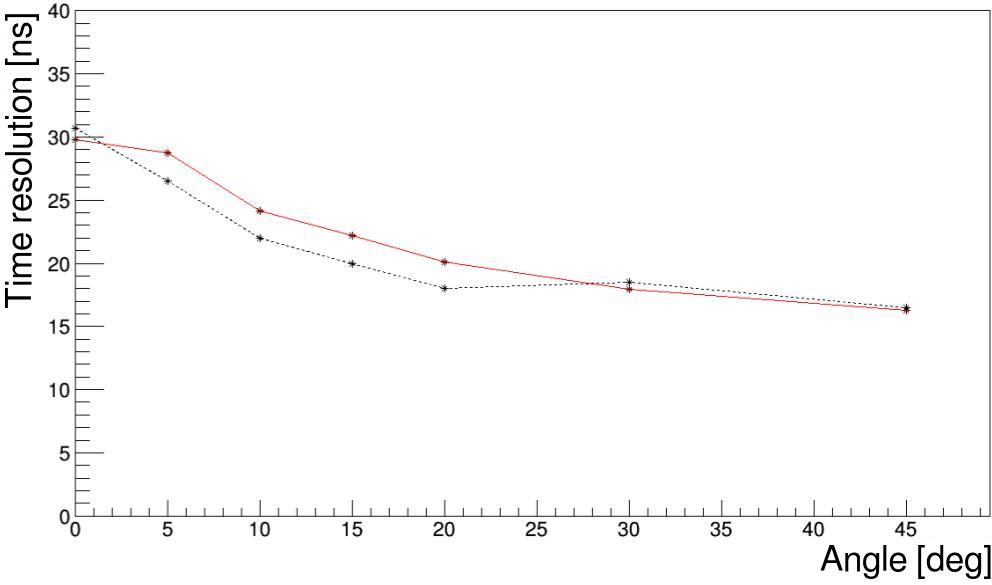}
  \end{tabular}
  \caption[Drift velocity and time resolution comparison]{Comparison between GTS and real data of the measured drift velocity (left) and the triple-GEM time resolution (right) is shown. Red lines are GTS results and black dashed lines the real data.}
\label{fig:time_GTS}
\end{figure}

The $\upmu$TPC performance strictly depends on the signal time. The time fluctuation and its resolution together with the time distribution for each configuration determine the performances of the $\upmu$TPC. Two quantities have been investigated to understand the $\upmu$TPC spatial resolution discrepancy in the focusing region: cluster time resolution and measured drift velocity. The  cluster time resolution is the difference between the fastest strip in the cluster and the trigger time; the measurement of the drift velocity is described in Sect. \ref{sec:driftvelocity}. Those are the variables correlated to the reconstruction algorithm and the comparison between simulation and real data is shown in Fig. \ref{fig:time_GTS}. A good agreement of drift velocity and cluster time resolution has been achieved. 

\section{From GTS to a full simulation}
The reproducibility of a triple-GEM event with a simulation allows to extend the R$\&$D studies to other configurations, $e.g.$ different gap widths, other gas mixtures or electric fields. 

A part from those interesting features, the principal aim of GTS is to provide a digitization algorithm that reproduces all the physical processes to simulate correctly the CGEM-IT inside the BESIII offline software environment. $e^+e^-$ collisions are simulated and the resulting particles interact with the detector. Each detector provides an accurate reconstruction of the process and then the final information used by the physicist for the analysis. An example of a study used to evaluate the impact of the new detector on the BESIII data analysis is shown in the next Chap.

	\chapter{Impact of the CGEM-IT on physics analysis}
In order to check the features and the impact introduced by the new inner tracker (IT), a detailed simulation and reconstruction software have been developed. The new IT is expected to improve the resolution in the z-vertex of charged tracks while the other variables should be as good as in the present IT. This regard several benchmark channels have been studied by the CGEM-IT collaboration, with the standard MDC and with the CGEM-IT. Analysis with large and small track multiplicities, tracks with different momentum ranges, different particle types and short- and long- living particles have been performed. The list of physics channels under study includes: $\psi(3686) \rightarrow \pi^+ \pi^- J/\psi$, $e^+e^- \rightarrow p \overline{p}$, $e^+e^- \rightarrow \pi^+ \pi^- \gamma_{ISR}$, $e^+e^- \rightarrow \pi^+ D^0 D^{\ast -}$, $e^+e^- \rightarrow \Lambda \overline{\Lambda}$ and $D^0 \rightarrow K^0_S K^+ K^-$.  In this Chap. the study of $J/\psi \rightarrow \pi^+ \pi^- \pi^0$ is reported. The event selection has been performed in order to extract a quantitative comparison.

\section{The BOSS and CGEMBOSS environments}
The BESIII Offline Software System (BOSS) is used by the BESIII collaboration for simulation, calibration, reconstruction and physics analysis. The reconstruction algorithms for all sub-detectors, described in Sect.~\ref{sec:bes}, are implemented in BOSS. 
The software is built on the Gaudi architecture and based on C++ language and object-oriented techniques. Moreover, BOSS uses external High Energy Physics libraries, such as CERNLIB, CLHEP, ROOT \cite{ref6:root} and re-uses parts of code from Belle, BaBar, ATLAS and GLAST experiments.
 
A separate package, named CGEMBOSS, has been developed by the Italian, Chinese and Mainz groups to simulate the CGEM-IT within the BESIII environment. The $e^+e^-$ collision is simulated with $KKMC$ event generator \cite{ref6:generator}. A GEANT4 \cite{ref6:geant} model defines the geometries of the three cylindrical detector and the materials of each layer, as described in Sect. \ref{sec:cgem}. A \textit{CgemCluster} is generated from the interaction between the resulting particles and the CGEM-IT. The CgemClusters simulate the response of a strip of the detector and their position is simulated through a smearing of the interaction position of the particle with a spread compatible with the expected spatial resolution of the CGEM-IT. In the last months the CGEM-IT group worked on CGEMBOSS to implement a digitization based on the algorithm shown in Chap. \ref{sec:digi} with some updates and improvements to reach a complete matching between simulations and real data. 

Once the strips information are generated, a track segment finding algorithm defines a tracklet in the CGEM-IT and another tracklet in the Outer Drift Chamber (ODC). For each tracklet in the ODC an extrapolation is performed to the inner region and a match of the two tracklet is applied. The combination between the tracklets with the smaller chi-square is chosen if multiple tracklet are present in the ODC or in the CGEM-IT. Once the information from every BESIII sub-detector is present a fit with Kalman Filter method \cite{ref6:kalman} is used to find the particle helix track parameters.

In the last months the CGEM-IT collaboration worked on the improvement of the tracklet matching between ODC and CGEM-IT. The described approach is inefficient if a particle has a small transverse momentum and it can not reach the ODC. If no tracklet are present in the ODC then this procedure does not uses the CGEM-IT information to measure the particle path. In order to overcome this problem a new approach with the Hough transform track finder method \cite{ref6:hough} is being implemented. Hough transform associates points from the CGEM-IT and the ODC to sinusoidal curves in the Hough space. The interception of those curve defines a point in the Hough space that correspond to a line in the real space: the particle track. The CGEM-IT and ODC points associate to the same track are used instead of the tracklets of the previous approach. This procedure is not complete yet and no results with this procedure will be presented.

\section{Study of the reaction $e^+e^- \rightarrow J/\psi \pi^+ \pi^- \pi^0$ to mark the CGEM-IT improvements}
A data-set of 500000 $e^+e^-$ collisions have been produced at the $J/\psi$ energy and decayed in $\pi^+ \pi^- \pi^0$ and $\pi^0 \rightarrow \gamma \gamma$. This decay channel has been chosen to test the CGEM-IT with multiple-track event, moreover it allows to evaluate the impact of the new IT in presence of decays with photons. The topic has been already studied by the collaboration \cite{ref6:besiii_channel} with measurement of its branching fraction, the di-pion mass spectrum and the Dalitz plot. These measurements are important to shed light on the $\rho \pi$ puzzle and its large branching fraction makes it an important background process for many other studies.

The simulation of those events has been performed with both the CGEM-IT and the present BESIII configuration with the MDC. At least two charged tracks and at least two photons are required to reconstruct the event. The charged tracks are measured with trackers, TOF and EMC. The particle identification of the TOF \cite{ref6:pid} is used to select tracks associated to pions. It is required that both charged pions originate from the same position within $10 \, \mathrm{cm}$ in the beam direction and $10 \, \mathrm{cm}$ in the radial direction. The neutral pion is reconstructed with two photons. Once the three pions are defined then their four-momentum is measured. The event selection requires a kinematic fit with five constrains: zero total tri-momentum, sum of the photons energy to the $\pi^0$ mass and invariant mass of the three pions to the $J/\psi$ mass. A cut on the $\chi^2$ of 200 in the kalman kinematic fit is used to improve the selection of the involved particles: $\pi^+ \pi^- \gamma \gamma$. Events with higher $\chi^2$ are rejected. The analysis cuts are summarized in Tab. \ref{tab:cuts}.

\begin{table}[bht!]
\begin{center}
\begin{tabular}{l l}
Variables & Value \\
\hline
N$^\circ$ Charged tracks & $\ge$ 2\\
N$^\circ$ Photons & $\ge$ 2\\
track vertex distance from \\ the interaction point  & $10 \, \mathrm{cm}$ along z\\
~ & $1 \, \mathrm{cm}$ along r\\
Photon energy & $\ge$ $25 \, \mathrm{MeV}$ in barrel\\
~ & $\ge$ $50 \, \mathrm{MeV}$ in end caps\\
Invariant mass of $\pi^+\pi^-\pi^0$ \\ before 5C & $J/\psi$ mass $\pm$ $150 \, \mathrm{MeV}$\\
Di-photons invariant mass \\ before 5C & $\pi^0$ mass $\pm$ $75 \, \mathrm{MeV}$\\
Invariant mass of $\pi^+\pi^-\pi^0$ \\ after 5C & $J/\psi$ mass\\
Di-photons invariant mass \\ after 5C & $\pi^0$ mass\\
Total tri-momentum \\ after 5C & 0\\
Kalman chi-square cut & $\le$ 200\\
\newline\\
\hline\\
\end{tabular}
\caption[Summary cuts on the benchmark channel]{Summary cuts on the $J/\psi \rightarrow \pi^+ \pi^- \pi^0$ benchmark channel.}
\label{tab:cuts}
\end{center}
\end{table}	

\begin{figure}[ht!]
  \centering
  \begin{tabular}{cc}
    \includegraphics[width=0.5\textwidth]{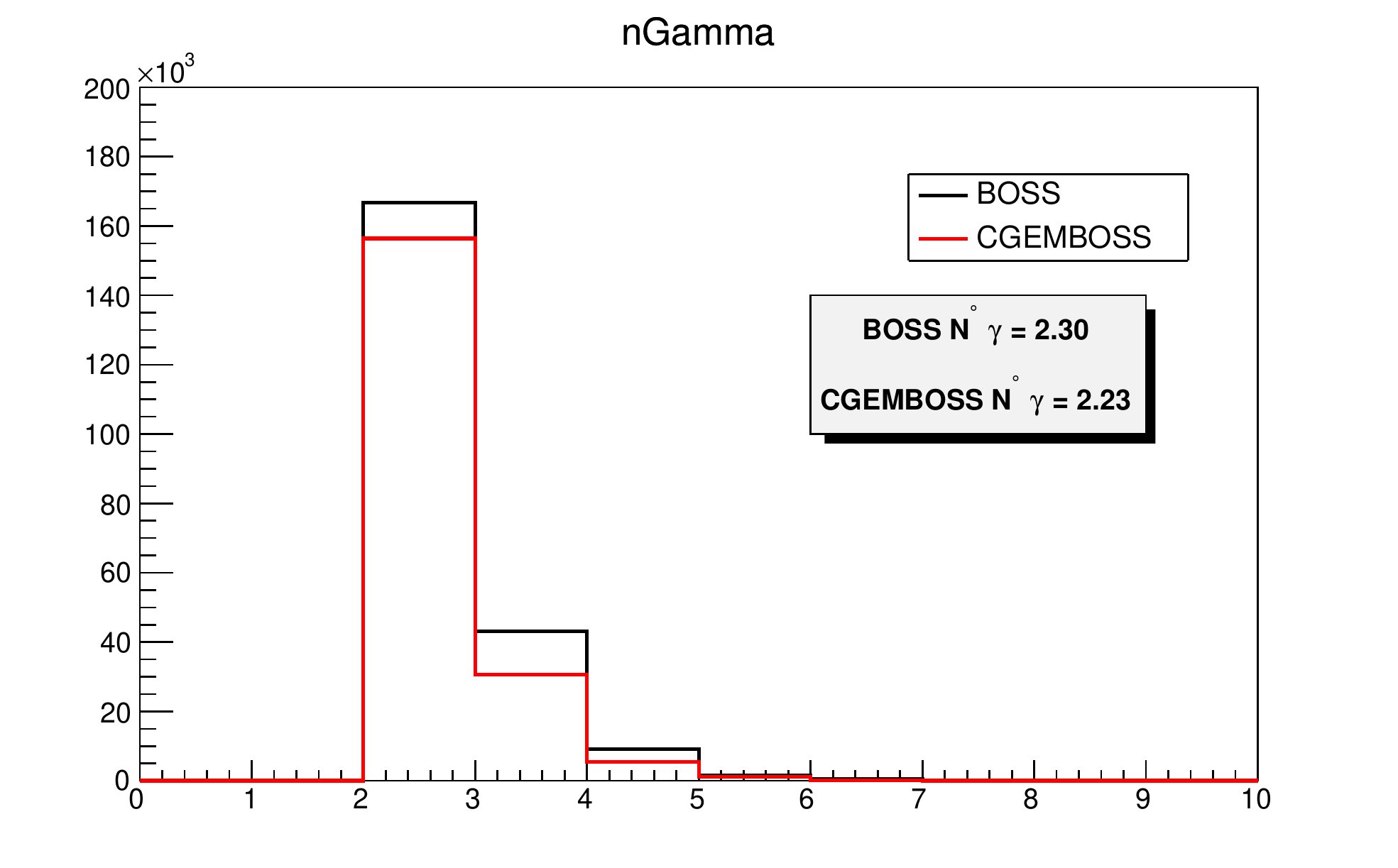}&
    \includegraphics[width=0.5\textwidth]{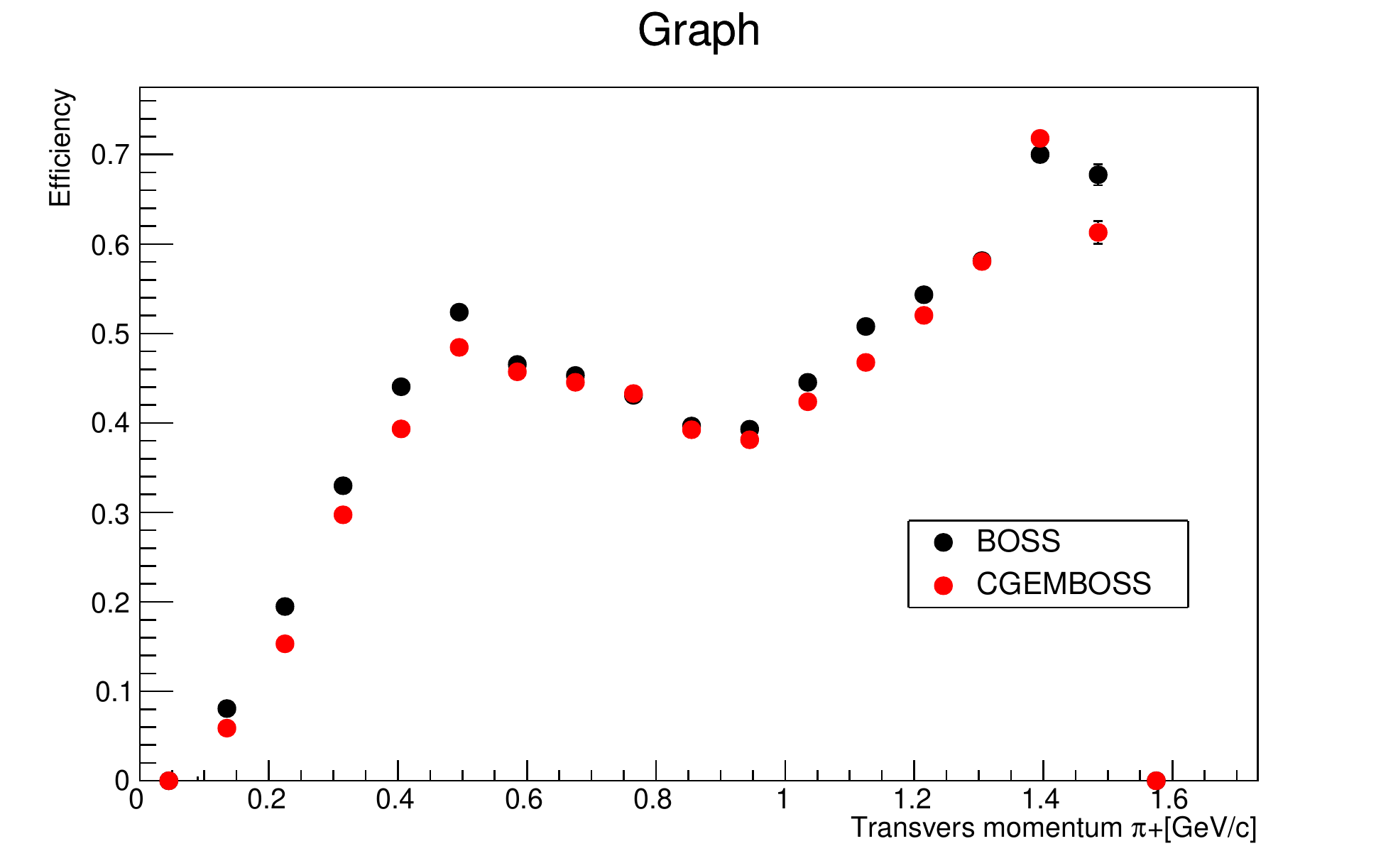}\\
    \includegraphics[width=0.5\textwidth]{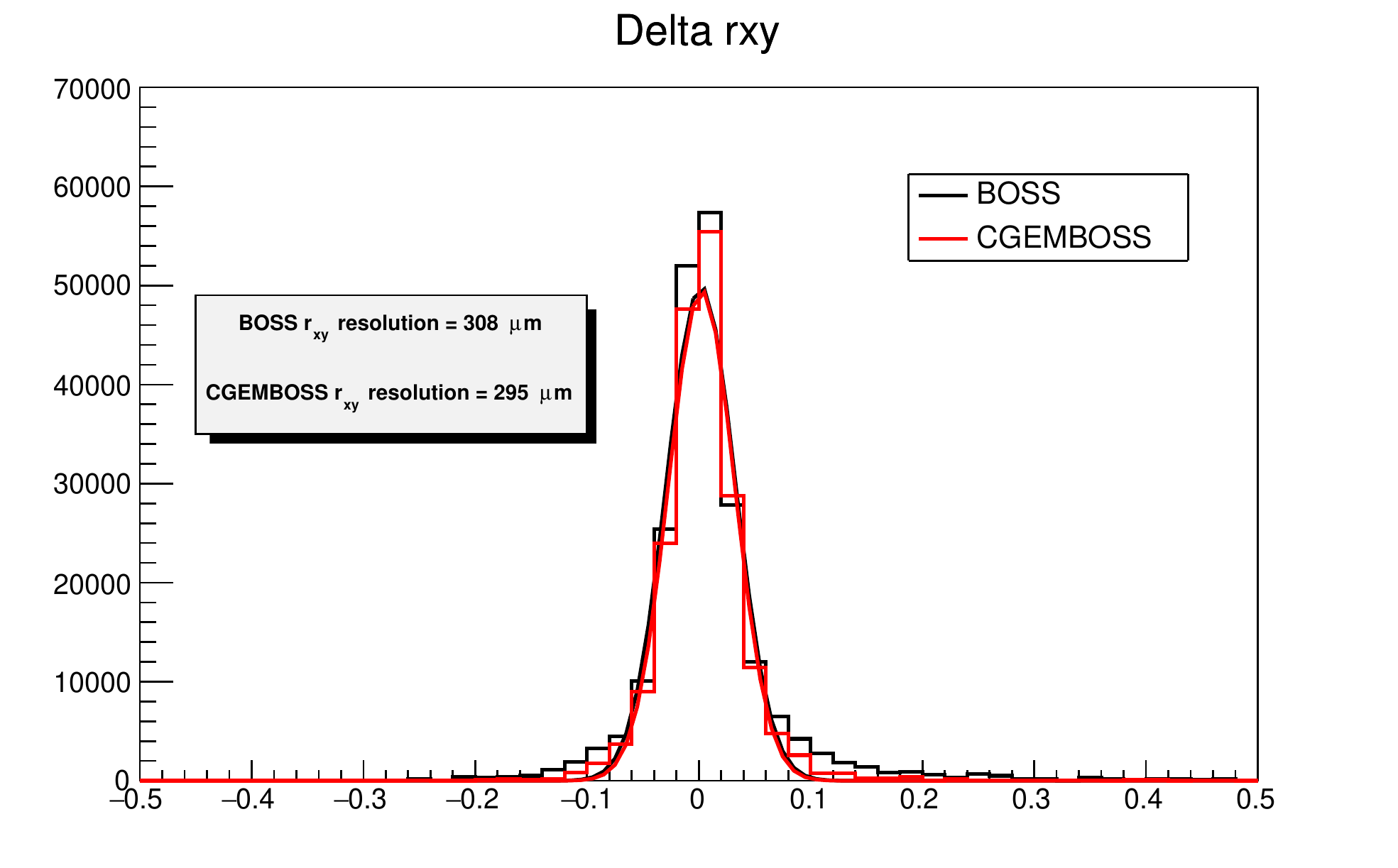}&
    \includegraphics[width=0.5\textwidth]{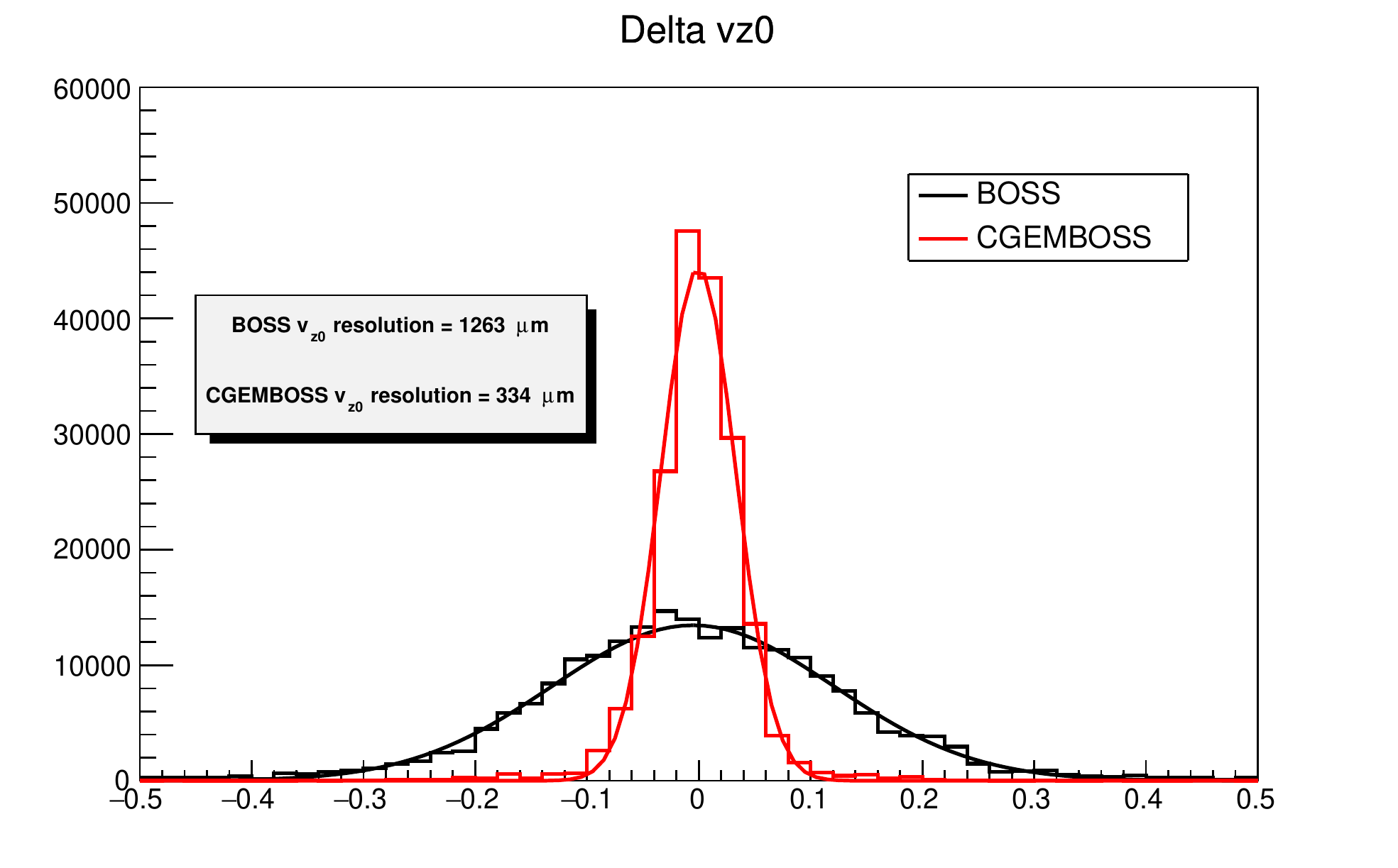}
  \end{tabular}
  \caption[BOSS and CGEMBOSS comparison]{A comparison of some variable of interest measured with BOSS (black bars and dots) and CGEMBOSS (red bars and dots: the number of photons produced in each event (top left), the efficiency as a function of the transverse momentum of the positive pion (top right),  the residual distribution in the $xy$ plane (bottom left) and $z$ plane (bottom right) between the MC-truth $J/\psi$ vertex.}
\label{fig:bench}
\end{figure}

The CGEM-IT grants the same performances of the MDC with three layers of MPGD technology instead of eight layers of wires. The CGEM-IT copies the Inner Drift Chamber (IDC) momentum distribution of the charged pions, reconstruction efficiency and number of photons produced, as is shown in Fig. \ref{fig:bench}. The average number of photons is 2.30 in BOSS and 2.23 in CGEMBOSS. Only $\approx$20\% of the events shows more than the two photons related to the decay channel. The non-increase of the photons number is important for the resolution of the ElectroMagnetic Calorimeter because it could worsen its performances if this number changes. The efficiency of the transverse momentum of the $\pi^+$ (similar to the one of the $\pi^-$) has an efficiency drop for momentum smaller than 0.5 GeV/c due to the algorithm used in the tracklet matching. Further study of this channel will be improved in this momentum region thanks to the new approach with the Hough transform. The spatial resolution of the vertex of the reconstructed $J/\psi$ and the MC-truth return a similar value in the $xy$ plane of about $300 \, \mathrm{\upmu m}$ while in the $z$ direction the improvement of the CGEM-IT is evident: from $1.2 \, \mathrm{mm}$ to $350 \, \mathrm{\upmu m}$. This result is due to a larger stereo angle between the detector layers: in the MDC the angle of the stereo wires ranges from -3.4$^{\circ}$ and 3.9$^{\circ}$ while in the CGEM-IT the stereo strips have an angle of 43.3$^{\circ}$ on layer 1, -31.1$^{\circ}$ on layer 2 and 32.0$^{\circ}$ on the third. A more precise measurement of the vertex allows to improve the analysis with secondary vertexes or some other with long living particles, such as $\Lambda$ and $K^0_S$.

CGEMBOSS simulates the real CGEM-IT geometries and its material. The reconstruction of the signal is based on simulations that reproduce the triple-GEM performance tested in several test beam in similar environments. Therefore the events shown in this benchmark channel are in agreement with the experimental measurement reported in several works of the CGEM-IT group of BESIII \cite{ref6:tipp,ref6:marcello,ref6:lia1,ref6:lia2}.
This ensures that the CGEM-IT fulfill the BESIII requirements and it can be successfully used to substitute the IDC.

	\chapter{Conclusion and outlook}

The R$\&$D of the CGEM-IT is successfully concluded and three layers of cylindrical GEM have been built. Soon they will be installed inside BESIII. Despite this new IT shares many features with KLOE-2 CGEM, a new geometrical and electrical configuration has been developed to maximize the performance keeping a safe working point. A $5 \, \mathrm{mm}$ drift gap has been used to increase the signal amplitude and to improve the $\upmu$TPC reconstruction. Ar+10\%iC$_4$H$_{10}$ as gas mixture has been adopted to reduce the HV on the GEMs without an effective reduction of the gain. Those improvements increased the electrical stability and reduced the discharge probability allowing at the same time a boost of the tracking performance.

Thanks to the charge and time measurements the charge centroid, $\upmu$TPC and the merging algorithms can be deployed to reconstruct the particle position with excellent spatial resolution below $150 \, \mathrm{\upmu m}$ in a large angular range with $1 \, \mathrm{T}$ magnetic field. These studies have been applied both to planar triple-GEM prototypes and to the final CGEM layers. The results are beyond the state of the art for GEM technology in high magnetic field. A paramount importance is given to diffusion and capacitive corrections developed with a data-driven approach.

A parametric simulation code has been developed to reproduce the amplification and induction processes in a triple-GEM and its results have been compared to the experimental data collected in the test beam. The overall agreement between the data and simulation of the variables of interest such as cluster size and charge, CC and $\upmu$TPC spatial resolution is better than 30\%; the simulated $\upmu$TPC differs from the real one in the focusing region and the reasons are under study. 

The reaction $e^+ e^- \rightarrow J/\psi \rightarrow \pi^+ \pi^- \pi^0$ has been simulated with the BESIII offline software to measure quantitatively the performance of the CGEM-IT with respect to the IDC. The CGEM-IT simulations agree with the data collected in the test beam and the benchmark variables show an improvement of the tracking performance of BESIII with the new IT. A new algorithm to combine the CGEM-IT and the ODC based on the Hough transform technique will be implemented soon to improve the efficiency loss of low transverse momentum particles.
	\appendix
	\chapter{Review on the gaseous detectors}
\label{sec:review_gas}
In this Appendix a description of the mechanics inside a gas detector will be reported: ionization, particle motion, amplification and the signal production. Those processes are fundamental to deeply understand a gas detector. Moreover, a brief review of the most important gas detectors is proposed because the GEM technology, focus of this work, comes from the evolution of the previous detectors and this allows to define its background.

\section{Ionization in gaseous detectors}
\label{sec:ionization}
Ionization of charged particles in gas medium allows to detect and measure their flight path. This is the fundamental physical process used in every gaseous detector. Ionization occurs when atoms and molecules in the gas emit an electron if an external element interacts with them, such us a particle that has to be revealed. The average ionization energy per ion pair ranges from 20 to 40 eV for most gases and is the main contribution to the energy loss in matter for charged particles. Tab. \ref{tab:ionization} shows the values of energy loss and ion-pair production for a selection of gases. If the energy transferred is smaller than effective ionization energy (W$_i$) then the resulting effect can be the excitation of the vibrational and rotational levels or a transition between the electronic levels of the atoms or molecules. If the energy transferred is higher that W$_i$ then a photo-electron can be extracted. If the energy is even higher, Compton scattering can occur but this will not be treated because it is not a process of interest for gas detectors. Within a certain probability an electron in the inner shell of the atoms is excited and its disexitation generates the emission of an UV photon and an Auger electron.

\begin{table}[ht]
\begin{center}
\begin{tabular}{c c c c}
Gas & W$_i$ & N$_P$ [cm$^{-1}$] & N$_T$ [cm$^{-1}$]\\
\hline
He & 41 & 5.9 & 7.8\\
Ar & 25 & 25 & 106\\
Xe & 22 & 41 & 312\\
CH$_4$ & 30 & 37 & 54\\
C$_2$H$_6$ & 26 & 48 & 112\\
iC$_4$H$_{10}$ & 26 & 90 & 220\\
CO$_2$ & 34 & 35 & 100\\
CF$_4$ & 54 & 63 & 120\\
\newline\\
\hline\\
\end{tabular}
\caption[Physical constants and approximated values of energy loss and ion pair production]{Physical constants and approximated values of energy loss and ion pair production \cite{appA:sauli77}.}
\label{tab:ionization}
\end{center}
\end{table}	

The ionization processes are totally random and, on average, the path between each interaction $\lambda$ is a function of the ionization cross-section $\sigma_i$ and the electron density of the gas $N_e$ as described in Eq. \ref{eq:path}:
\begin{equation}		
\label{eq:path}
\lambda=\dfrac{1}{N_e\sigma_i}
\end{equation}

The number of primary ionizations $N_i$ expected in a medium along a path length of $d$ is $d/\lambda$ and it follows a Poisson distortion:

\begin{equation}		
\label{eq:poisson}
P^{N_i}_k = \dfrac{(N_i)^k}{k!}e^{-N_i}
\end{equation}

where $k$ is the actual number of pairs. Each interaction creates an electron but several other secondary ionizations can occur and usually the number of total electrons exceed the number of collisions. Moreover, there is a low probability to generate energetic secondary electrons, named $delta$ electrons, that carry a large amount of energy and this determines the peculiar shape of the energy loss distribution. The efficiency to have at least one interaction is an important parameter in the characterization of a detector and it is given by:

\begin{equation}		
\label{eq:det_eff}
\epsilon = 1 - P^{N_i}_0 = 1 - e^{N_i}
\end{equation}

This shows the importance of the gas thickness in the detection efficiency, $e.g.$ to reach an efficiency of 99\% the sensitive gas thickness has to be $d>5\lambda$. The total number of ion pairs expected in a medium where the total energy loss in the gas by the charged particle $\Delta E$ is:

\begin{equation}		
\label{eq:n_ele}
N_T = \dfrac{\Delta E}{W_i}
\end{equation}

In gas mixtures given by different elements it is possible to estimate with a good approximation the average number of expected electrons weighting the values with the gas concentration. $E.g.$ in Ar-iC$_4$H$_{10}$ gas mixture in volumetric proportions 90:10, the average number of total electron can be evaluated for the following expression~\cite{appA:sauli_book}:

\begin{equation}		
\label{eq:n_ele_ariso}
N_T^{Ar:iC_4H_{10}} = 0.9 \cdot N_T^{Ar} + 0.1 \cdot N_T^{iC_4H_{10}} = 117~cm^{-1}
\end{equation}

The energy loss by the charged particles is well described by the Bethe-Block formula in Eq. \ref{eq:bethe}

\begin{equation}		
\label{eq:bethe}
\dfrac{\Delta E}{\Delta x} = -\rho \dfrac{2KZ}{A\beta^2}\left[\ln \dfrac{2mc^2\beta^2}{I(1-\beta^2)}-\beta^2-\dfrac{C}{Z}-\dfrac{\delta}{2}\right] 
\end{equation}

where $\rho$, $Z$ and $A$ are the atomic density, charge and mass, $\beta$ the velocity of the particle, $C/Z$ the inner shell correction needed at low energy and $\delta$/2 the density effect correction used at relativistic velocity. 
The Eq. shows a dependency of the energy loss mainly on the velocity of the particle and the medium nature. The process is dominated by the emission of delta electrons: through an interaction it can carry about the 10\% of the energy loss. The energy loss in low density material can be approximated to a Landau distribution where the mean value is about 30\% higher than the most probable value due to delta electron and the Landau distribution shows the asymmetric tail.


\section{Drift of electrons and ions in gases}
\label{sec:drift}
\begin{figure}[t]
\centering
\includegraphics[width=0.6\textwidth]{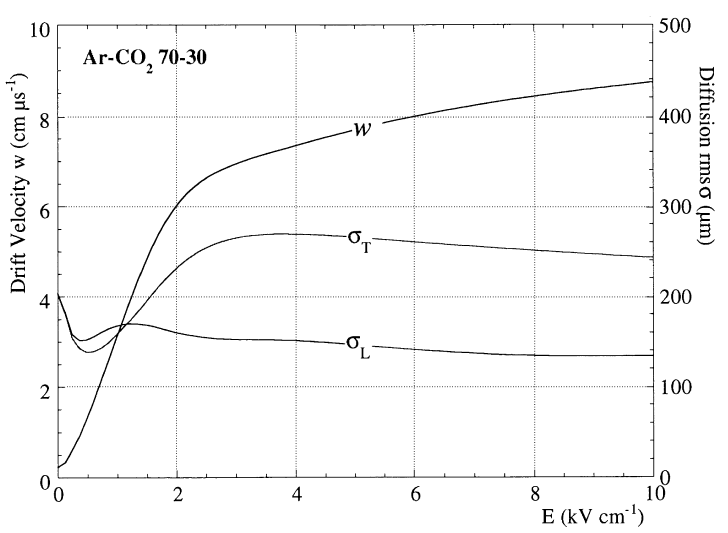}
\caption[Electron diffusion and drift velocity.]{Longitudinal and transverse diffusion and drift velocity \cite{appA:bachmann1999}.}
\label{fig:diffusion_drift}
\end{figure}

Electrons and ions generated by ionization in gas collide with the surrounding molecules and lose their kinetic energy and they are re-absorbed if no external electric field is applied. Otherwise the charged particles move through the gas up to their recombination or to the edges of the gas volume. In this configuration the velocity of the electrons and ions is well above the thermal velocity of a particles in the gas.
The motion of charged particles in electric and magnetic field, $\mathbf{E}$ and $\mathbf{B}$, is described by the motion Eq.:

\begin{equation}		
\label{eq:motion}
m\dfrac{d\textbf{v}}{dt} = q\mathbf{E}+q(\textbf{v} \times \mathbf{B}) - K\textbf{v}
\end{equation}

where $m$, $q$ and $v$ are mass, charge and velocity of the particle, $K$ is the frictional force caused by the interaction of the particle with the gas.  The characteristic time associated to the mean time between collisions is $\tau$ = $m/K$. The solution of the Eq. \ref{eq:motion} is given by:

\begin{equation}		
\label{eq:motion_velocity}
\textbf{v}= \dfrac{q}{m}\tau\dfrac{1}{1+\omega^2\tau^2}\left(\mathbf{E}+\omega\tau[\mathbf{E}\times\mathbf{B}]+\omega^2\tau^2(\mathbf{E} \cdot \mathbf{B})\mathbf{B}\right)
\end{equation}

where $\omega$ is the Larmor frequency equal to $e\mathrm{B}/m$. If no magnetic field is present the expression can be simplified in:
\begin{equation}		
\label{eq:motion_velocity_simpy}
\textbf{v}=\dfrac{q}{m}\tau\mathbf{E}=\upmu\mathbf{E}
\end{equation}

where $\upmu$ is the mobility defined as the ratio of the drift velocity and the electric field in absence of magnetic field. The presence of magnetic field introduces the Lorentz force that modifies the drift properties. If the electric and magnetic fields are orthogonal, the Lorentz angle $\theta_L$ is introduced such as the angle between the drift direction and the electric field in the plane perpendicular to $\mathbf{B}$:
\begin{equation}  
\begin{tabular}{ccc}
$v=\dfrac{\mathrm{E}}{\mathrm{B}}\dfrac{\omega\tau}{\sqrt{1+\omega^2\tau^2}}$ & and & $\tan\theta_L=\omega\tau$
\end{tabular}
\label{eq:motion_velocity_EB}
\end{equation}  
The drift velocity of the electrons and ions is a quantity that depends on the applied electromagnetic field and the nature of the gas. An example is shown in Fig. \ref{fig:diffusion_drift}.
The diffusion is another parameter of interest for electrons and ions drift inside the gas. The scattering along the motion varies the direction of the particles randomly. In the simplest configuration Nerst-Townsend formula relates the diffusion coefficient $D$ to the ion mobility $\upmu$ and the gas temperature $T$:
\begin{equation}		
\label{eq:diffusion}
\dfrac{D}{\upmu}=\dfrac{\varepsilon_k}{q}
\end{equation}
where $\varepsilon_k$ is the characteristic energy, a phenomenological quantity equal to $kT$\footnote{$k$ is the Boltzmann constant} for thermal electrons. The electrons and ions drifting along a distance $d$ diffuse following a Gaussian distribution with a standard deviation:
\begin{equation}		
\label{eq:diffusion_sigma}
\sigma_d=\sqrt{\dfrac{\varepsilon_k d}{qE}}=\sqrt{\dfrac{Dd}{\upmu E}}
\end{equation}
This classical theory assumes a symmetric diffusion described by the diffusion coefficient but in same gases longitudinal and transverse behaviour differs as a function of the electric field as shown in Fig. \ref{fig:diffusion_drift}.\\
The knowledge of the drift velocity of the electron and their diffusion is fundamental to develop gas detector with high performances and accuracy: the drift velocity provides reconstruction stability while the transverse diffusion affect the position resolution and the longitudinal one the time resolution in the detection of the primary ionization.

\section{Amplification of ionization}
\label{sec:amplification}
An electron drifting in a gas medium with a kinetic energy higher that the ionization potential of the gas molecule can interact with it and eject another electron leaving behind a positive ion. An electric field in the proportional regime (few tens of kV/cm) can provide to the electron the needed energy and if the primary and the secondary electrons continue their path, further ionization can occur. The process of avalanche multiplication has been discovered by Townsend \cite{appA:Townsend}. It is possible to define the first Townsend coefficient from the mean free path $\lambda$ as its inverse: $\alpha$~=~$\lambda^{-1}$. It represents the number of ion pairs produced per unit length of drift. It is related to the ionization cross section by the following Eq.:
\begin{equation}		
\label{eq:townsend}
\alpha=N_m\sigma_i
\end{equation}

where $N_m$ is the number of molecules per unit volume. After each mean free path an electron creates another electron and an ion and two electrons will continue the path and the process will be repeated again. The number of electron will increase exponentially. Over a path $d$ the gain is given by:
\begin{equation}		
\label{eq:gain}
G=e^{\alpha d}
\end{equation}

If the gas volume is composed by different species of gas then new channels for radiative or ionizing transitions are possible. Penning effect \cite{appA:pennin} can occur in this configuration: if the ionization potential of one species is lower than the excitation potential of the other, a process of collision transfer that increases the ionized states can take place:
\begin{equation}		
\label{eq:pennin}
A^\ast+B \rightarrow A+B^++e
\end{equation}

\begin{figure}[t]
\centering
\includegraphics[width=0.7\textwidth]{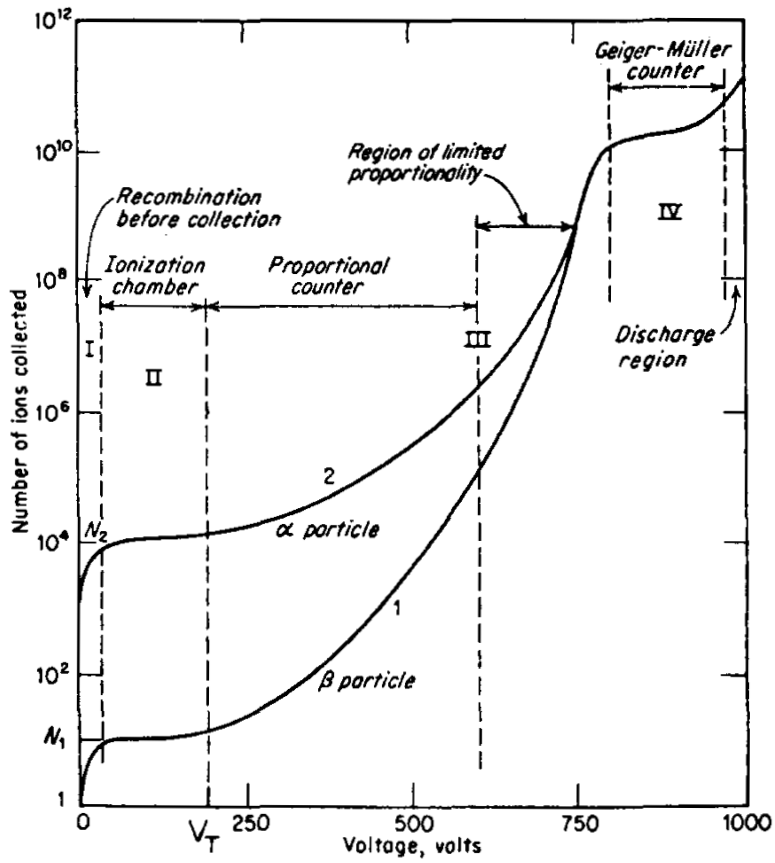}
\caption[HV regime in proportional counter]{Gain-Voltage characteristics for a proportional counter, showing the different region of operation \cite{appA:prince}.}
\label{fig:hvregime}
\end{figure}

If the electric field is increased to higher value and the proportional regime is exceed then there is transition from the avalanche to a streamer: the secondary ionization spread the charge over the volume and strongly increase the space charge distortion of the electric field in front and behind the avalanche. Secondary ionization is triggered by the photons emitted from inelastic collision.  They expand the avalanche and propagate the charge, starting the streamer. Figure \ref{fig:hvregime} shows the different regime as a function of the high voltage.

\section{Gas choice}
The gas mixture is the most important choice in a gas detector. Gas cross section determines the detector properties, as described in the previous Sections. Some gases generate a larger number of ionization with respect to others and this influence the charge collected by the detector. Moreover, the drift velocity of the electron are defined by the gas mixture and this strictly influence the space and time resolution. Also, the multiplication depends on the gas mixture: in noble gases the inelastic cross section is zero below the excitation and ionization threshold \cite{appA:gasgas} and the multiplication occurs at much lower fields than in complex molecules. In principle every gas could be used to amplify the signal but convenience of operation suggests to use noble gas as the main component, such as Argon. A radiative process is needed to de-excite the noble gases and the minimum energy of the emitted photon (11.6~eV) is greater than the ionization potential of the cathode metal (7.7~eV). New electrons can be emitted after the absorption of the photon and those can generate a new avalanche. Poly-atomic gases, such as CO$_2$ or iC$_4$H$_{10}$, can be mixed together the noble gas and this increases the drift velocities of the electrons because of their large inelastic cross sections at moderate energies, which result in $cooling$ electrons into the energy range of the Ramsauer-Townsend minimum. Another role of the poly-atomic gas is to absorb the ultraviolet photons emitted by the excited noble gas atoms in order to limit the avalanche size and constrain it in proportional mode if the streamer mode has to be avoided.

\section{Creation of signal}
The electrons created in the avalanche moving to a grounded metal electrode, $e.g.$ wires, strips or pad, generate a pulse that can be readout by an amplifier. The charge resulting from the multiplication generate a mirror image on the metal surface equal to the inducting charge, if the metal surface is infinite. This can be evaluated from the Gauss law. If the readout plane is segmented into strip of width $w$ and each strip is grounded then the induced charge on the strip depends on the distance $d$:
\begin{equation}		
\label{eq:induction_strip}
Q_{strip}(d)=-\dfrac{2q}{\pi}arctan\left(\dfrac{w}{2d}\right)
\end{equation}

where $q$ is the charge of the particle. 
If the charge is moving then $d=d(t)=d_0-vt$ and thus the induced current is given by the Eq.:
\begin{equation}		
\label{eq:induction_current}
I_{strip}^{ind}(t)=-\dfrac{d}{dt} \{ Q_{strip}[d(t)] \} =\dfrac{4qw}{\pi[4d(t)^2+w^2}v
\end{equation}

Ramo theorem \cite{appA:ramo} can be used to evaluate the induced current. Let's define $\textbf{E}_{strip}(x)$ the electric field created by raising the strip to the potential $V_{strip}$ = 1 V then the induced current can be calculated from:
\begin{equation}		
\label{eq:induction_current_ramo}
I_{strip}^{ind}=\dfrac{-q\textbf{v}(x) \times \textbf{E}_{strip}(x)}{V_{strip}}=-q\textbf{v}(x) \times \textbf{E}_{strip}^{weight}(x)
\end{equation}

where $\textbf{E}_{strip}^{weight}(x)$ is the weighting field. It can be computed once the geometry and the electrostatic configuration is defined. In Eq. \ref{eq:induction_current_ramo} the current depends on the velocity of the electron that is almost constant, the weighting field computed point by point and the angle between the drift direction of the electron and the electric field.
\section{Brief history of the gaseous detectors}
\label{sec:gas_histo}
Gaseous detector uses a gas volume to generate a signal, the primary ionization, in order to detect charged particles and measure their path or the deposited energy in the gas volume; moreover they uses an electric field for amplification and to drift the electrons and ions to the readout plane where a proper electronics with amplifier will measure the signal. In the past century different technologies followed each other with different geometries, process of drift and amplification techniques: from the single wire counters that allows one-dimensional (1D) position measurement to triple-GEM detectors which uses amplification stages and drift region separated and the advantages of the modern production technique to achieve high position resolution in the three-dimensional (3D) space.
\subsection{Single wire chamber}
\begin{figure}[t]
\centering
\includegraphics[width=0.7\textwidth]{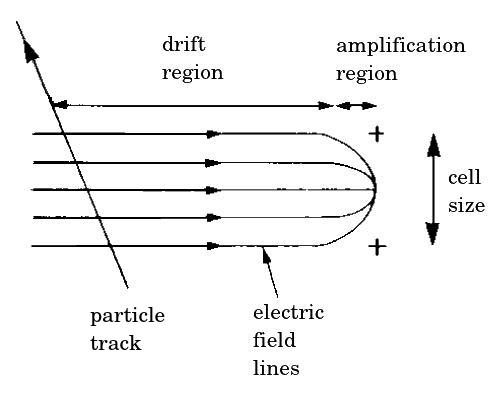}
\caption[Single wire electric field]{Representations of the electric field inside a single-wire chamber.}
\label{fig:singlewire}
\end{figure}

After the Townsend discovery to use the electric field to amplify the charge of the ionization, Rutherford and Geiger \cite{appA:ruthgeig} developed a single-wire counters: an instrument capable to measure the charge deposited by an $\alpha$ particle through a cylinder with a central insulated wire. The volume has been filled by gas at low pressure and an electric field below the sparking value has been applied. The central wire was connected with an electro-meter. In this way the ionization produced by an $\alpha$ could be magnified several times. If the chamber operates with high gas gains ($\sim$ 10$^8$) then all pulses are independent of the primary ionization. Thus is called Geiger mode. Here the electric field allows primary Townsend avalanches generated in the region where the electric field lines concentrate around the wire position. Far away from the wire the electric field is lower and only electrons drift occurs. The advantage of Geiger counter if the amplitude of the signal up to 100~V that does not need any amplification electronics but the disadvantages are corona discharges around the wire that limits the rate capability at 1~kHz and the detector is not charge sensitive because every output has the same amplitude. The electric field of the chamber are shown in Fig. \ref{fig:singlewire}.

\subsection{Proportional counter}
\begin{figure}[t]
\centering
\includegraphics[width=0.8\textwidth]{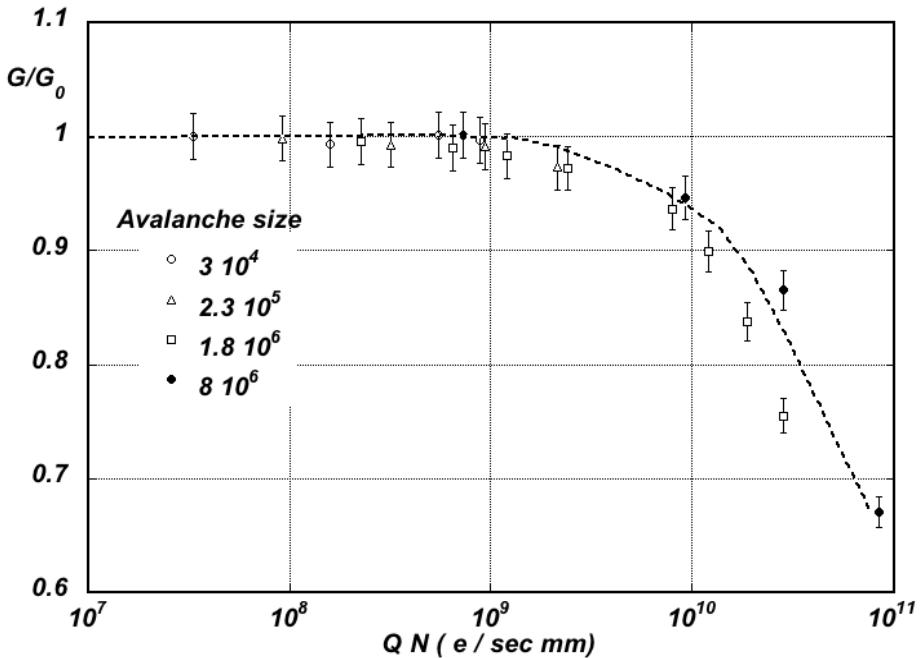}
\caption[Space charge effect in proportional chamber]{Space charge effect on gas amplification. $G/G_0$, gas gain relative to zero counting rate. $Q$: total charge in single avalanche; $N$: particle
rate/wire length \cite{appA:pc_rate}.}
\label{fig:pc_rate}
\end{figure}

The single-wire chamber can operate in proportional mode and the pulse measured is proportional to the charge released in the ionization for voltages smaller than the one used in Geiger mode that correspond to a gain lower than 10$^6$. The detected signal is generated by fast motion of the ions created around the wire and is not due to the collection of the electrons. The pulse spectrum has a Gaussian-like shape and the energy resolution can be evaluated looking at the ration between the Full Width at Half-Maximum (FWHM) and the mean value of the signal amplitude. Resolution between 10\% and 15\% has been achieved with this technology. The typical signal duration is about $10-50 \, \mathrm{\upmu s}$ allowing operations at rates up to $10^5 \, \mathrm{Hz}$. In proportional mode the avalanche in localized in a region around the wire and if the pulse in a single-wire chamber is readout from both edges it is possible to measure the position of the avalanche. Let's define $L$ the wire length and $S_i$ the signal amplitude readout from one edge then the position from the edge $1$ is:
\begin{equation}		
\label{eq:x_pos_wire}
x=\dfrac{S_1}{S_1+S_2}L
\end{equation}
The avalanche process creates electrons and ions, $i.e$ in the wire chamber, around the wire. This effect generates a local drop of the electric field with a consequent reduction of the gain. After several hundred $\upmu$s the ions leave the region and the gain performances are restored. A reduction of the gain has been experimentally observed by \cite{appA:pc_rate} as shown in Fig. \ref{fig:pc_rate}.

\subsection{Multi-wire proportional counter}
\label{sec:MWPC}
\begin{figure}[t]
\centering
\includegraphics[width=0.8\textwidth]{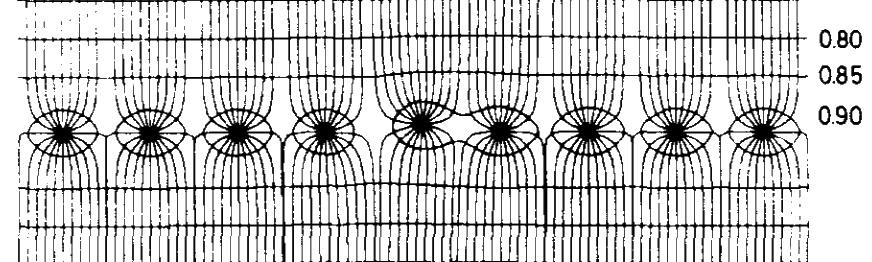}
\caption[MWPC electric field.]{Electric field equipotentials and field lines in a multi-wire proportional chamber. The effect on the field of a small displacement of one wire is also shown \cite{appA:sauli77}.}
\label{fig:MWPC}
\end{figure}

To improve the limitation of the single-wire chamber due to its geometry, several wire counters within the same gas volume can be assembled: a set of wires with a few tens thickness, parallel and equally spaced are placed between two cathode planes \cite{appA:charpak}. The distance between the wires and the cathode is about $10 \, \mathrm{mm}$ while the spacing between the wires between $1$ and $2 \, \mathrm{mm}$. Smaller spacing is difficult to realize and it is not electrically stable. A representations of the electric field in a MWPC is shown in Fig. \ref{fig:MWPC}. The signal collected is of few mV \cite{appA:Bouclier} and it is shared between the wires in the anode and between strips in the cathode. The time spectra of the measured signal show a sharp peak coming from the ionization generated close to the wire then it shows a long tail due the electrons created in the drift region away from the wires. The time resolution achieved is around 30 ns \cite{appA:sauli_book}. Tracks orthogonal to the anode plane generate a signal on a wire most likely, up to 3 depending on the geometry of the detector and the gas properties. This quantity is the cluster size. Once the tracks angle increases (0$^\circ$ correspond to orthogonal tracks) then the cluster size increases too. A signal collected on several wires can be used to measure the position of the avalanche in the anode plane and orthogonality to the wire length. The position is calculated from the center-of-gravity (COG) method:
\begin{equation}		
\label{eq:COG}
\overline{x}=\dfrac{\sum x_i q_i}{\sum q_i}
\end{equation}
where $x_i$ is the central coordinate of the strip i and $q_i$ the charge collected on the strip. Similarly measurement can be provided by the wires of the anode once a proper offset $q_0$ is removed to the measured charge to remove the noise and the cross talk: $q_i \rightarrow q_i - q_0$.

\subsection{Drift chamber}
The signal time measurement in the wire chambers suggested by Charpak \cite{appA:charpak} introduced a new technique to localize the position of the avalanche in the gas volume. The ionization occurs at $t_0$ in the chamber and at the time $t_1$ reaches the anode wire. If the electric field or the space-time correlation function are known then it is possible to associate to t$_1-t_0$ a position along the electric field line. In the simplest situation where the electric field is constant the drift velocity is known and the space coordinate is given by:
\begin{equation}		
\label{eq:spacetime_wire}
z=v(t_1-t_0)
\end{equation}

This technique requires a time reference signal to measure $t_0$ and this can be provided by an external trigger such as a scintillator counter.
Single wire drift chamber can reach up to $50 \, \mathrm{cm}$ drift length with an operating voltage of about 50 kV \cite{appA:saudinos}. To reduce the drift time and cover larger detection area a multi wire geometry is needed. This introduce a new limit because the drift field is not uniform across all the active cell and the space-time correlation is distorted. A modification of the MWPC, named multi-wire drift chamber (MWDC) \cite{appA:mwdc} introduces field wires between anode wires to reduce the low field regions. The position resolution achieved with MWDC depends on the diffusion and the drift properties of the electron in the used gas, the knowledge of the space-time correlation function and the resolution of the electronics used in the time measurement.

\subsection{Time projection chamber}
\begin{figure}[t]
\centering
\includegraphics[width=0.7\textwidth]{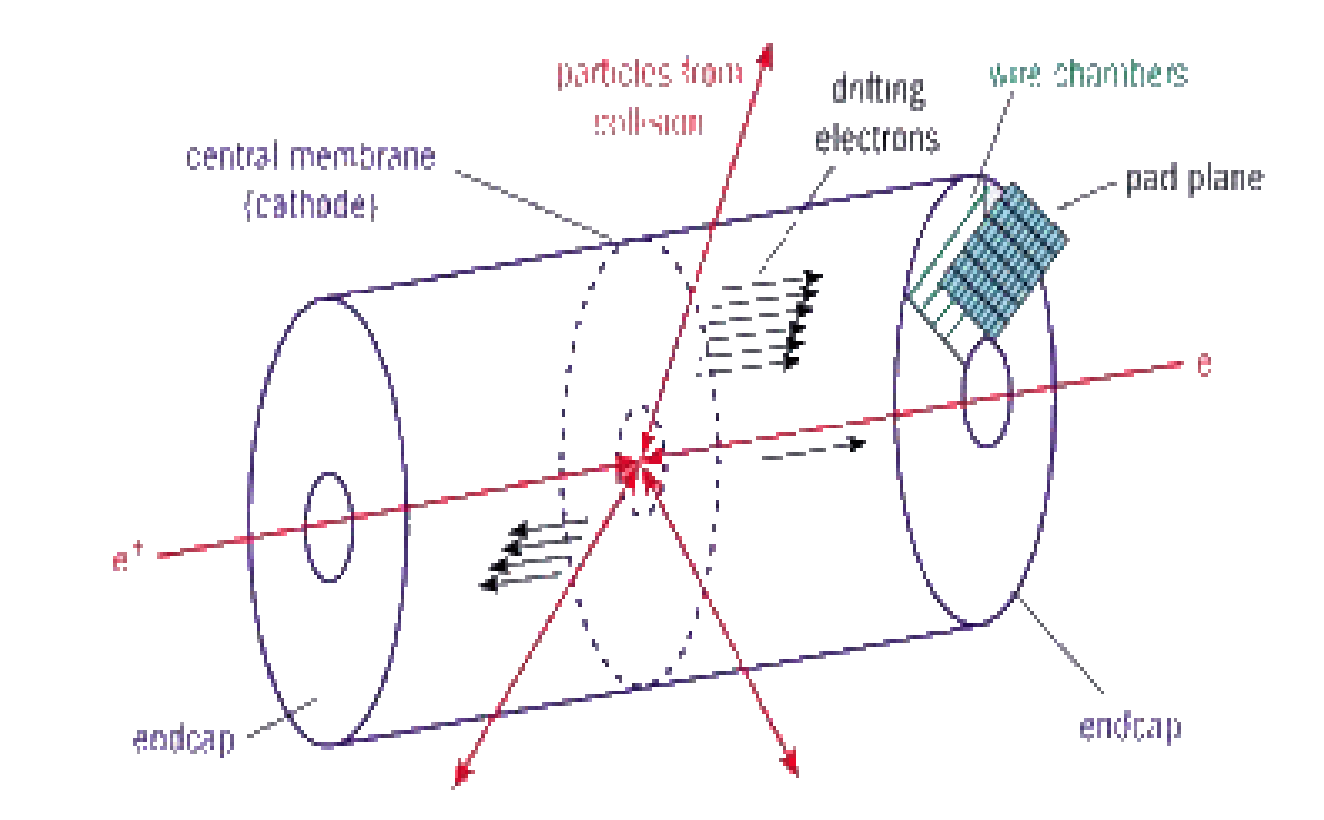}
\caption[Time projection chamber representation]{A schematic representation of the PEP-4 TPC.}
\label{fig:tpc}
\end{figure}

Very large volume drift chamber allows to reach excellent imaging and particle identification capabilities, $e.g.$ in large track multiplicity environment if the particles track is measured with an high segmentation along its path. This measure is affordable with Time Projection Chamber (TPC) technology \cite{appA:allison}. It consist of a large volume of gas from tens of cm to several meters with a field electrode inside the volume and a MWPC in the end cap region with anode and field wires. The cathode of the MWPC consist of a printed circuit with pad. An example is shown in Fig. \ref{fig:tpc}. The drift field is separated to the multiplication field thought a cathode mesh and the drift direction of the electrons generated in the ionization is orthogonal to the MWPC plane. The most important information in a TPC is the drift time measurement provided by the pad rows then the important parameter to optimize are the pad response function and the diffusion of the electrons in the gas. A magnetic field can be used to reduce the transverse diffusion.
\subsection{Resistive plate chamber}
\begin{figure}[t]
\centering
\includegraphics[width=0.9\textwidth]{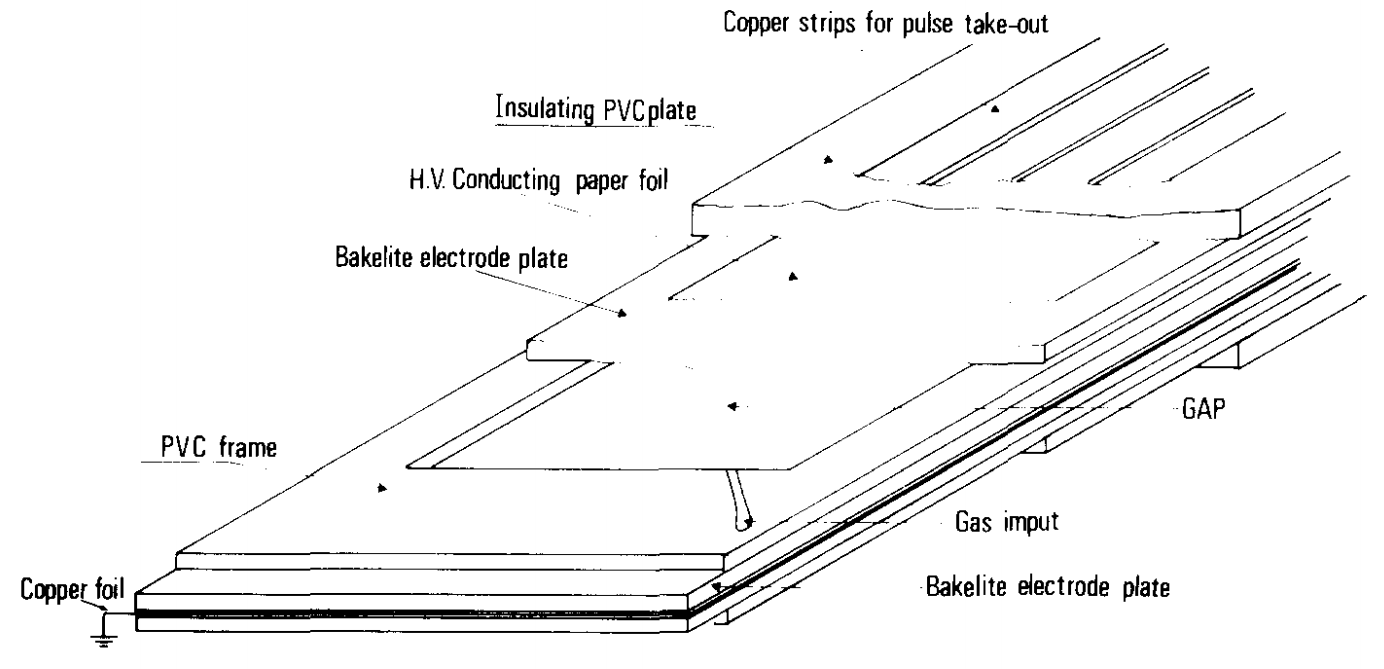}
\caption[Resistive plate chamber]{Sketch of the RPC \cite{appA:santonicco}.}
\label{fig:rpc}
\end{figure}

The resistive plate chamber (RPC) is a technology used in large area particle detection. The design is similar to the parallel plate chamber but it has high resistivity electrodes made of phenolic polymer laminate that provides a volume resistivity of about 10$^{10}$ $\Omega$cm \cite{appA:santonicco}. The sensitive gas is a few mm thick is sandwiched with the high resistivity electrodes and an insulating support frame. Outside there are the readout strips. See representations in Fig. \ref{fig:rpc}. The working gain is around 10$^6$-10$^7$ and it creates a large signal that is dumped by the high resistivity electrodes. A gas mixture photon absorbing is needed to quench the discharge and the photon emission. RPC can work in streamer mode but this leads to an efficiency loss due to the needed recovery time after each avalanche that locally reduce the electric field. This technology can achieve time resolution less thant 100 ps \cite{appA:akindinov} thanks to the direct induction of the signal on the readout strips and the fast amplification process in the thin gap. Several layer of RPC are suitable for particle identification and to measure the particles time of flight in high energy physics experiment.

\section{Micro pattern gaseous detectors}
\label{sec:mpgd}
Modern photolithographic technology and thin-layer polymide depositions led to develop a new design to amplify the ionization charge and to readout the signal: the Micro Pattern Gaseous Detector (MPGD). The granularity of the detector could be improved over wire chamber thanks to pitch size of few hundreds $\upmu$m: this improved the intrinsic rate capability above $10^6 \, \mathrm{Hz/mm^2}$, spatial resolution below $50 \, \mathrm{\upmu m}$, multi particle resolution and single photo-electron time resolution within tens ns \cite{appA:pdg}.
\subsection{Micro Strip Gas Chamber}
\begin{figure}[t]
\centering
\includegraphics[width=0.7\textwidth]{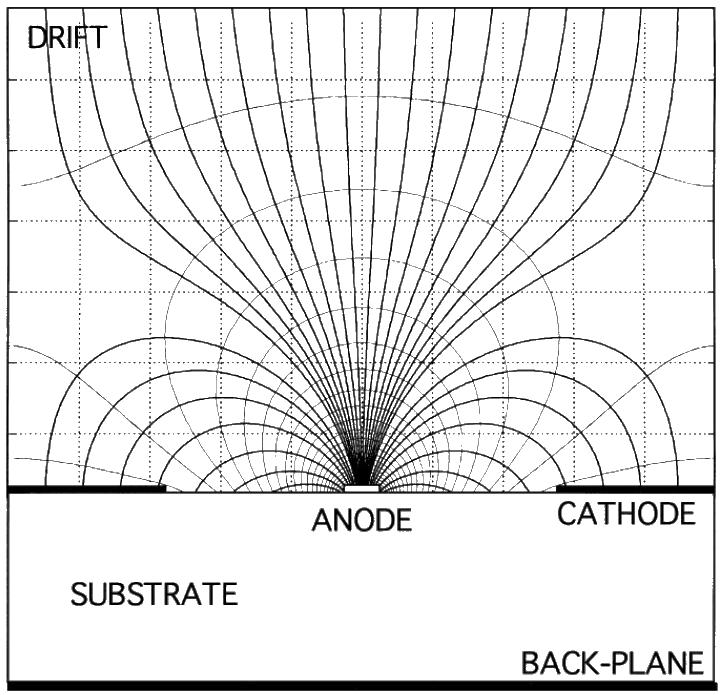}
\caption[Micro Strip Gas Chamber electric field.]{Field and equipotential lines  computed for a back-plane voltage close to the cathode. \cite{appA:saulisharma}.}
\label{fig:msgc}
\end{figure}

The Micro Strip Gas Chamber (MSGC) \cite{appA:oed} is the pioneer in this new detection design and it consist of a plane segment by strips with an alternately large and short width connected respectively to the cathode and the anode; an upper electrode connected to the cathode and within the two plane there is the sensitive gas region where happens the detection and the ionization drift. Figure \ref{fig:msgc} shows the electric field of the chamber. The gas thickness is about few mm while the strip width of the anode strip is $10 \, \mathrm{\upmu m}$ with $200 \, \mathrm{\upmu m}$ pitch: around the strip region the electric field becomes very intense and the avalanche multiplication process takes place. The cathode strips are placed few tens of $\upmu$m away in order to evacuate the ions charge as fast as possible. This design allows to reach higher flux rate because the ions do not have to drift up to the other plane to be neutralized. Due to the capacitance between the strip, a fraction of the induced charge on certain strip is injected into the other and this increases the noise in the chamber. Many problems has been encountered with the MSGC such as gain instabilities and discharges \cite{appA:sauli_book}. This pushed the development of new detection technique that led to the modern MPGD that will be shown in the next sections.
\subsection{Micro-mesh gaseous structure}
\begin{figure}[t]
\centering
\includegraphics[width=0.7\textwidth]{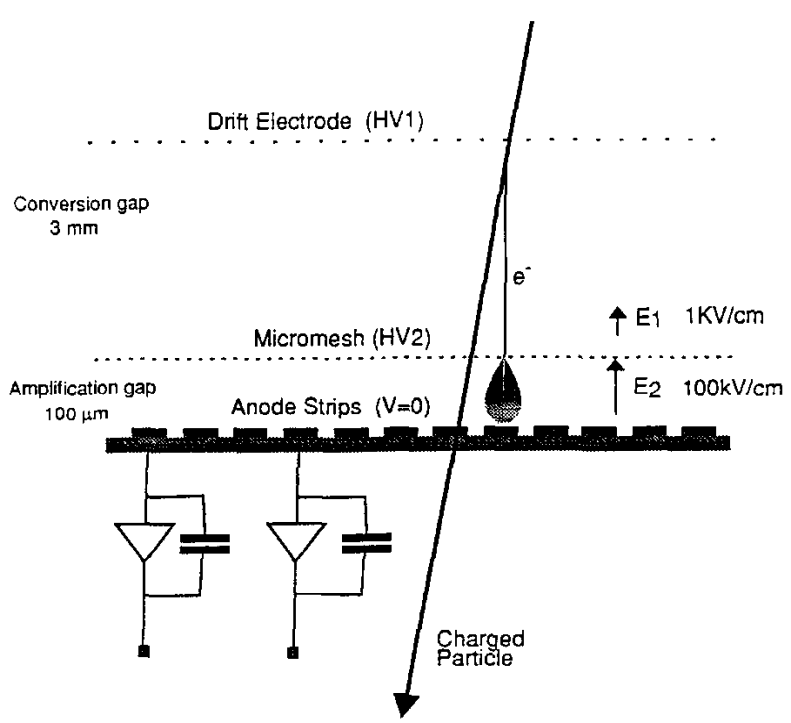}
\caption[MicroMegas.]{A schematic view of MICROMEGAS: the $3 \, \mathrm{m}$ conversion gap and the amplification gap separated by the micro-mesh and the anode strip electrode \cite{appA:saulisharma}.}
\label{fig:micromegas}
\end{figure}

High electric field about $100 \, \mathrm{kV/cm}$ in gaps below the mm saturates the Townsend coefficient during the avalanche and large gains can be attained. This led to the development of the micro-mesh gaseous detector (MicroMegas) \cite{appA:giomataris}: a technology composed by a thin metal grid of about $100 \, \mathrm{\upmu m}$ distance from the anode and above the grid there is the drift region. Thanks to pillar created with photolithographic process the distance between the grid and the anode is kept constant to ensure the gain uniformity. The cathode grid captures most of the ions and this reduces significantly the ions back-flow. The signal is induced by the ions motion in the small gap. The sensitive region is of few mm, here the ionization take place and the electrons drift up to the grid where the amplification of the signal takes place. A sketch of the detector is represented in Fig. \ref{fig:micromegas}. \\
\subsection{Gas Electron Multiplier}
The details of this detector will be described in Sect. \ref{sec:GEM}. The Gas Electron Multiplier (GEM) \cite{appA:sauli_gem} consists of a thin kapton foil with $\upmu$m copper on the faces and it is perforated by a high density holes with a radius of tens of $\upmu$m. The electric field inside the holes of $50 \, \mathrm{\upmu m}$ depth is about $100 \, \mathrm{kV/cm}$. Here the multiplication take place. The GEM foil is placed inside two electrodes, cathode and anode, and divides the gas region in two: the drift gap where the ionization occurs and the electrons are drifted to the GEM; the induction gap where the avalanche electrons drift to the anode and the signal is induced on the strips.
\subsection{Micro Resistive Well}
\begin{figure}[t]
\centering
\includegraphics[width=0.7\textwidth]{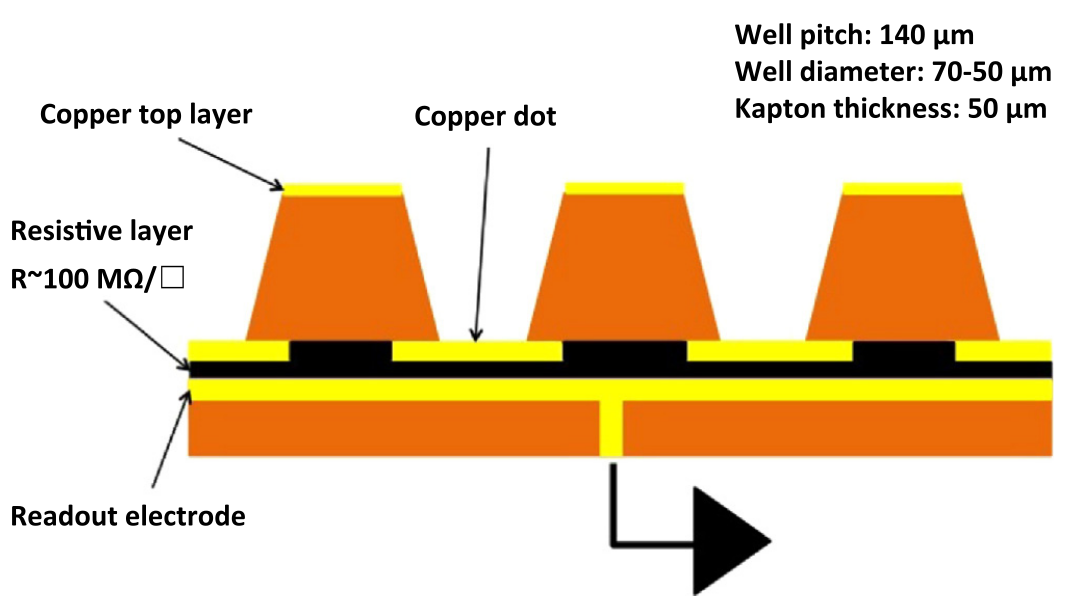}
\caption[$\upmu$-RWELL.]{Schematic drawing of the $\upmu$-RWELL PCB \cite{appA:murwell}.}
\label{fig:murwell}
\end{figure}

The micro Resistive Well ($\upmu$-RWell) \cite{appA:murwell} exploit the GEM amplification technique with a resistive readout plane and it create a single stage amplification detector spark resisting and easily to construct. The $\upmu$-RWell is realized by merging an etched GEM foil with the readout PCB plane coated with a resistive deposition of 50 $\upmu$m thick polymide foil. A cathode foil close the gas volume with a gap of few mm. See the sketch in Fig. \ref{fig:murwell}.

	\chapter{Diffusion and capacitive corrections in $\upmu$TPC}
\label{sec:capacitive}
Beside the $\upmu$TPC development, the ATLAS collaboration performed, with MicroMegas (MM) detectors, simulations and test-beam data studies to detect the source of worsening of the $\upmu$TPC given by the capacitive effects, the $ghost~hits$ effect and the worsening of the hit position measurement on the edges of a cluster \cite{appB:georgios, appB:ntekas}. The third effect is typical of the MicroMegas technology while the other two have to be investigated in triple-GEM detectors.

Due to the capacitance between strips, ATLAS observed an average of 15\% of the inducing charge between a strip and its nearest neighbouring strip. This introduces the need to correct the $\upmu$TPC point position, as a function of the ratio between its charge and the one of the neighbouring strip if the strip considered is at the cluster edge. If the charge measured on those strips is smaller than 300 ADC (10 fC), the charge ratio with its neighbouring strip is 15\% and the time measured is the same, then this is associated to a $ghost~hit$ and it is not generated from electron amplification but by the capacitance between strips. Those hits create a distortion of the $\upmu$TPC reconstruction path and have to be removed to improve the $\upmu$TPC performance in MM detectors.

\section{The correction algorithms}
The studies reported by ATLAS suggest the need to evaluate similar corrections for triple-GEM. When the $\upmu$TPC has been developed in the reconstruction code, the simulation were not yet developed and a study based on the acquired data has been performed to evaluate those corrections. Triple-GEM differs from MicroMegas in the signal amplification: GEMs have a larger gap where the electrons undergo diffusion ($11 \, \mathrm{mm}$) and the amplification of each primary electron is spreaded on several strips. Then the corrections to account for the measured capacitive and diffusion effects were obtained with the following procedures. This allowed a correction of both effects simultaneously.

\begin{figure}[ht]
\centering
\includegraphics[width=0.7\textwidth]{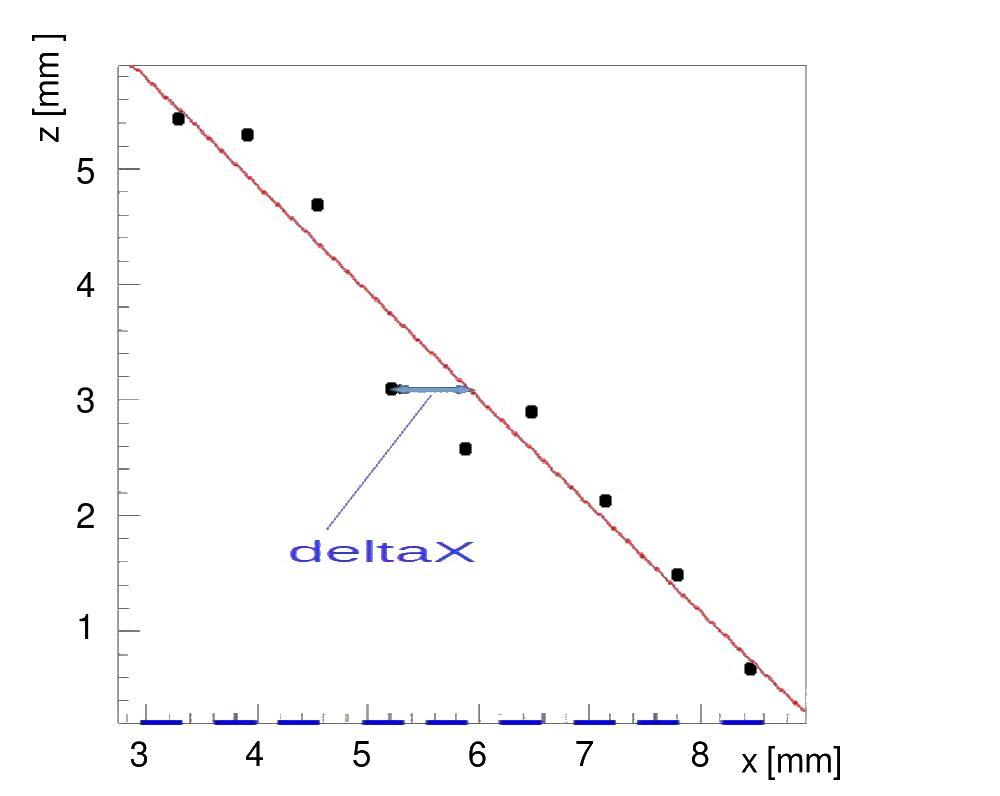}
\caption[The $\upmu$TPC points residual]{The $\upmu$TPC points (black dots) and the $\upmu$TPC fit (red line) are shown. The strip positions are represented by blue bars. For each $\upmu$TPC point the residual with respect to the $\upmu$TPC fitting line is evaluated.}
\label{fig:residual_tpc}
\end{figure}

The residual of each $\upmu$TPC point from the $\upmu$TPC fitting line $\Delta x^{\mathrm{\upmu TPC}}$ has been evaluated as a function of the charge ratio with the neighbouring strip and as a function of the strip index inside the cluster. A representation is shown in Fig. \ref{fig:residual_tpc} with the $\upmu$TPC point (black dots) and the $\upmu$TPC fit (red line).

No residual in the $z$ direction has been evaluated because the $z$ depends on the time measurement and the transverse diffusion or the capacitive effect should not influence that variable. 

The study of those corrections has been performed with a run having an angle between the detector and the beam of 45$^\circ$ and Ar+10\%iC$_4$H$_{10}$ gas mixture. At first a cluster size of nine has been considered then the study has been extended for each cluster size. It must be noted that the correction depends on the gas mixture.

\subsection{Data driven corrections based on the strip index}
\label{sec:corr1}
\begin{figure*}[ht!]
\centering
\begin{tabular}{lr}
\includegraphics[width=0.5\textwidth]{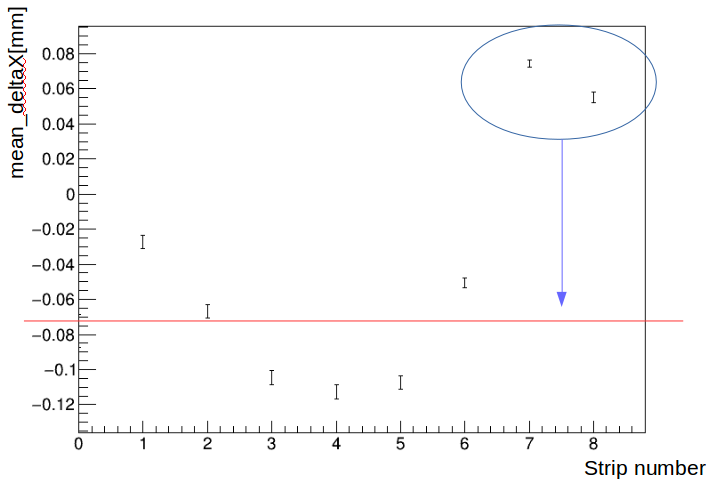} & 
\includegraphics[width=0.5\textwidth]{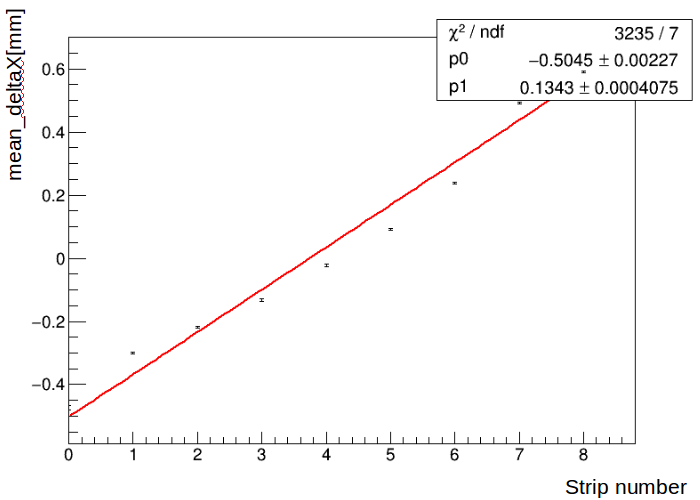} \\
\end{tabular}
\caption[First diffusion and capacitive correction]{Mean $\Delta x^{\mathrm{\upmu TPC}}_i$, where $i$ is the strip position in the cluster, as a function of $i$ is shown in both figures: the first distribution (left) shows an $S$ behaviour {\it before} the correction. The second distribution (right) shows the same plot {\it after} the correction on the last two points: this distribution is fitted with a line and the correction is parametrized, as described in Sect. \ref{sec:corr1}}
\label{fig:deltaX_strip}
\end{figure*}

The $\Delta x^{\mathrm{\upmu TPC}}_i$ residual distribution correlation with the strip index $i$ has been studied and the corrections have been measured in two steps. In Fig. \ref{fig:deltaX_strip} left $\Delta x^{\mathrm{\upmu TPC}}_i$ is shown as a function of the strip index $i$ for a cluster size of nine. The strips on the right show a positive mean value while the others a negative value. This means that the strips on the left tend to be further shifted to the left and the ones to the right tend to be shifted to the right. This is due to the contribution of the peripherical $\upmu$TPC points because they originate from diffusion and capacitive effects, and they bias the $\upmu$TPC fit. 

In a triple-GEM detector those effects affect mainly the first and the last strips but also the second and the second-last. This behavior can be understood by observing the mean cluster size with orthogonal tracks and without magnetic field: it is about four. In these conditions the distribution of the primary electrons is almost point-like in the plane orthogonal to the drift direction. Those four strips are composed by the middle one, where the ionization take place plus one strip and a half (on average) on both sides where the signal is due to diffusion and capacitive effects.

The first correction applied shifts the last strip to the mean value of the others (about 180 $\upmu$m) then the distribution of the residuals as a function of the strip number is evaluated again. Due to this shift now the distribution shows a linear behavior as a function of the strip index (see Fig. \ref{fig:deltaX_strip} right) and can be fitted with a line with $p0_{\mathrm{shift}}$ and $p1_{\mathrm{shift}}$ parameters. The corrections introduce a shift of the $\upmu$TPC point as a function of the strip position in the cluster. 

\begin{figure*}[ht!]
\centering
\begin{tabular}{lr}
\includegraphics[width=0.5\textwidth]{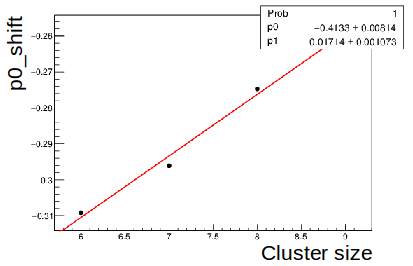} & 
\includegraphics[width=0.5\textwidth]{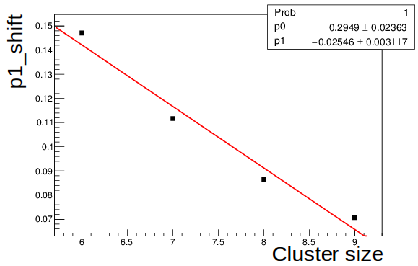} \\
\end{tabular}
\caption[First diffusion and capacitive corrections parameters]{The distribution shown in Fig. \ref{fig:deltaX_strip} on the right is fitted with a line. The line parameters change as a function of the cluster size and their values are shown in those plots: the intercept (left) and the slope (right). Those distributions are fitted too to be used in the correction algorithms.}
\label{fig:m_p_shift}
\end{figure*}

The next step of this correction is to evaluate the $p0_{\mathrm{shift}}$ and $p1_{\mathrm{shift}}$ coefficients for every cluster size while the first correction of $180 \, \mathrm{\upmu m}$ on the last strip is kept constant. The procedure is the same and the line parameters behaviors are shown in Fig. \ref{fig:m_p_shift}.

\subsection{Data driven corrections based on the charge ratio}
The corrections described in the previous paragraph shift the residual mean value around zero. The second part of the correction, described in this paragraph, evaluates the dependency of the $\Delta x^{\mathrm{\upmu TPC}}$ on the ratio of the charge strip and its neighbouring strip. If the track incident angle is positive then the left neighbouring strip is considered because it is the one reached by earlier electrons, otherwise the right neighbouring strip is considered.

\begin{figure*}[ht!]
\centering
\begin{tabular}{lr}
\includegraphics[width=0.4\textwidth]{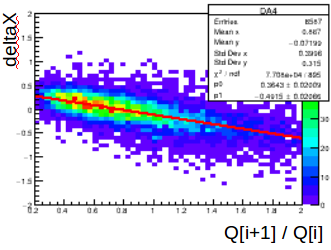}  & 
\includegraphics[width=0.4\textwidth]{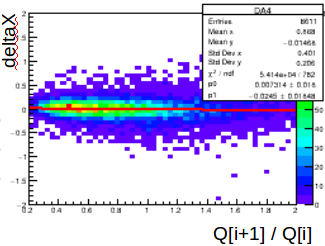} \\
\includegraphics[width=0.4\textwidth]{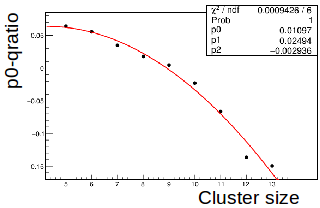}  &
\includegraphics[width=0.4\textwidth]{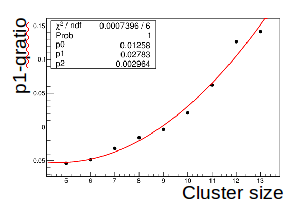}  
\end{tabular}
\caption[Second diffusion and capacitive correction and its parameters]{In the first row the $\Delta x^{\mathrm{\upmu TPC}}$ distributions as a function of the charge ratio of neighbouring strips is shown: before (top left) and after the correction (top right). This distribution is fitted with a line and its parameters depend on the cluster size. In the second row the intercept (bottom left) and the slope (bottom right) as a function of the cluster size are shown.}
\label{fig:corr_qratio}
\end{figure*}

On the top left in Fig. \ref{fig:corr_qratio} the dependency of $\Delta x^{\mathrm{\upmu TPC}}$ on the charge ratio before the correction is shown: the distribution is fitted with a line with parameters $p0_{\mathrm{q-ratio}}$ and $p1_{\mathrm{q-ratio}}$. The corrections introduce a shift of the $\upmu$TPC position as a function of the charge ratio. On the top right the same distribution after the corrections is shown.

The same $p0_{\mathrm{q-ratio}}$ and $p1_{\mathrm{q-ratio}}$ are used for all the strips of clusters with the same cluster size, contrarily to the ATLAS approach that applies this kind of correction at the edges only. In triple-GEM detectors the charge distribution at the anode is very variable and those kinds of effect affect both the edges and the strips in the middle of the cluster. Charge distributions with multi-peak trend are very common.

This kind of correction has been evaluated for different cluster sizes and the different $p0_{\mathrm{q-ratio}}$ and $p1_{\mathrm{q-ratio}}$ are reported on the bottom in Fig. \ref{fig:corr_qratio}. The behaviour of those variables shows a parabolic dependency as a function of the cluster size.

\subsection{Data driven corrections based on the mean position}
\begin{figure}[ht]
\centering
\includegraphics[width=0.9\textwidth]{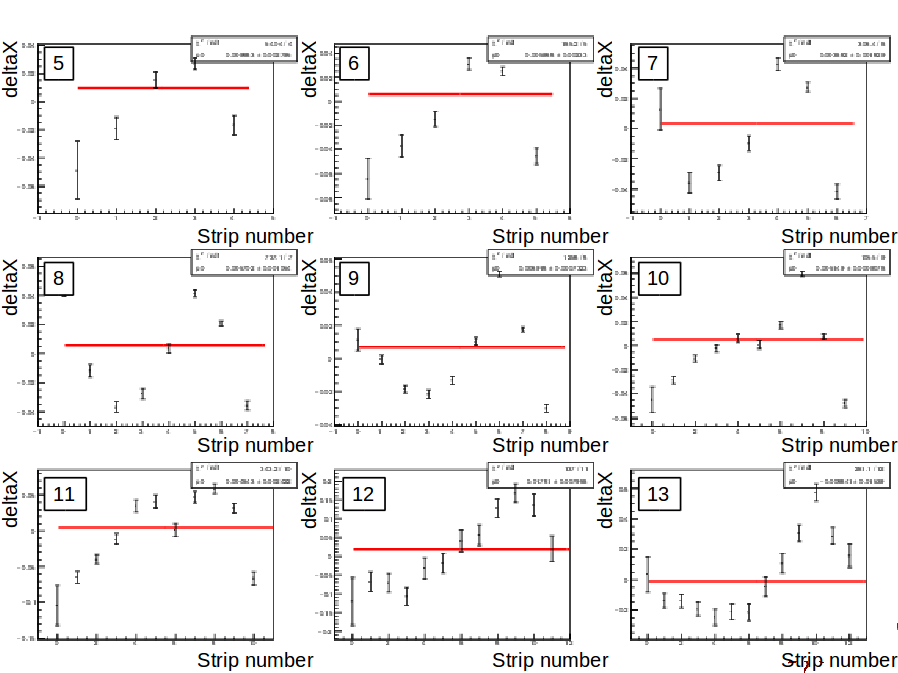}
\caption[Third diffusion and capacitive corrections]{The mean $\Delta x^{\mathrm{\upmu TPC}}_i$, where $i$ is the strip position in the cluster, as a function of $i$ for different cluster size: from 5 to 13. The distributions are fitted with a constant as last correction. The $S$ shape is an indication that some worsening effects are still present and the $\upmu$TPC could be improved more.}
\label{fig:corr_mean}
\end{figure}

The first group of corrections is applied to the strip as a function of the strip index and the cluster size. The second of corrections depends on the charge ratio between neighbouring strips and cluster size. The third kind of corrections reported in this section depends on the cluster size only. The mean $\Delta x^{\mathrm{\upmu TPC}}_i$ as a function of the $i$ strip index and the cluster size is evaluated, as shown in Fig. \ref{fig:corr_mean}, and the mean value from each plot is taken to correct the position of the $\upmu$TPC point of each strip in the cluster as a function of the cluster size. It varies from $-50$ to $150 \, \mathrm{\upmu m}$. Figure \ref{fig:corr_mean} reports the behavior of the point after the application of the first two groups of corrections. On average the distribution is almost around zero but an $S$ shape of the mean $\Delta x^{\mathrm{\upmu TPC}}_i$ is still present. This behavior should be investigated more but so far those corrections already return a significant improvement on the $\upmu$TPC spatial resolution.

\section{Diffusion and capacitive correction on real and simulated data}
\begin{figure}[htp]
  \centering
  \begin{tabular}{cc}
    \includegraphics[width=0.5\textwidth]{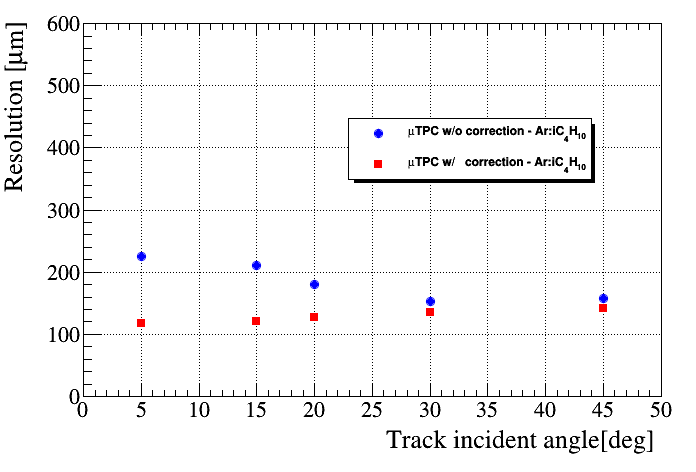}&
    \includegraphics[width=0.5\textwidth]{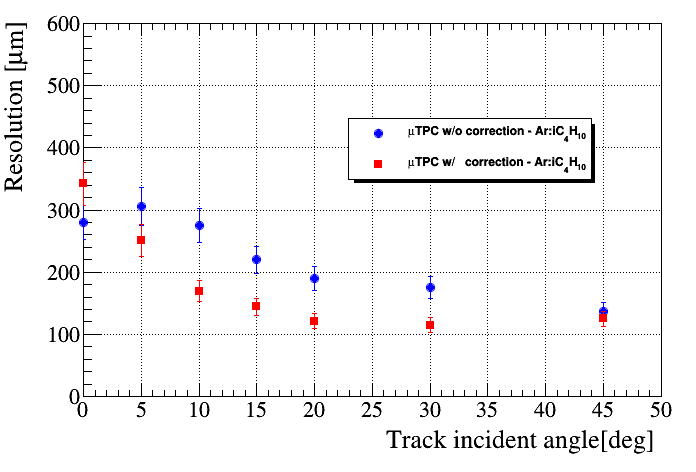}\\
  \end{tabular}
  \caption[$\upmu$TPC resolution with and without corrections]{$\upmu$TPC spatial resolution with (red squares) and without (blue dots) correction algorithms from real data (right) and simulation (left). As explained in Sect. \ref{sec:digi} the data-simulation agreement is good for angles greater than 10$^\circ$.}
\label{fig:w_wo_corr}
\end{figure}

The measured corrections have been extracted from a certain data-set and they have been applied to every other run with Ar+10\%iC$_4$H$_{10}$. Those remove, on average, the dependency on diffusion and capacitive effects from the residual $\Delta x$ as well as the dependency on the charge. The $\upmu$TPC spatial resolution improves with those corrections as shown in Fig. \ref{fig:w_wo_corr}. These algorithms have been applied both to real data and simulations and in both cases the improvement is significant. Despite the $\upmu$TPC simulation still does not agree perfectly with real data, the effectiveness of the algorithm is clear. The results of $\upmu$TPC algorithm shown in Chap. \ref{sec:graal} for real data and Chap. \ref{sec:digi} for simulations already contain these corrections in the reconstruction code. 

---

	\chapter*{Aknowledgements}

My PhD thesis is the result of a fruitful collaboration with an active comunity and I feel obliged to acknowledge the support receive from several people.\\
\newline
To Dr. Gianluigi Cibinetto goes my gratitude for the sustenance given on these years through a constant comparison on the research line and its faith on the PhD activities. His push let me to grow up on several fields and to have a complete instruction.\\
\newline
A special thanks goes to Dr. Lia Lavezzi, I am grateful for our successful collaboration on the activities of these years and for her presence in the time of need.\\
\newline
The acknoledgements on this theses go to each researcher who let me learn more and more about physics and beyond. Dr. Giovanni Bencivenni and Dr. Marco Poli Lener are two splendid person that crossed my walk in Physics and I am happy to have met.\\
\newline
I would also like to thank Dr. Giulio Mezzadri and the entire BESIII Italian collaboration for the efficient cooperation and the time we've spent together.\\
\newline
Finally I want to aknowledge the support of several people I have met: Prof. Diego Bettoni and Prof. Mauro Savri\'e, Dr. Alessandro Calcaterra and Prof. Marco Maggiora.
	\chapter*{}

\footnotetext{The game.}

\end{document}